\pdfoutput=1
\documentclass[a4paper,fleqn,usenatbib]{mnras}

\usepackage{newtxtext,newtxmath}

\usepackage[T1]{fontenc}
\usepackage{ae,aecompl}

\usepackage{graphicx}	
\usepackage{amsmath}	
\usepackage{amssymb}	
\usepackage{deluxetable}

\newcommand{\angstrom}{\mbox{\normalfont\AA}}

\newcommand{\ergs}{erg s\ensuremath{^{-1}}}

\def\gtrsim{\mathrel{\hbox{\rlap{\hbox{\lower4pt\hbox{$\sim$}}}\hbox{\raise2pt\hbox{$>$}}}}}

\newcommand{\halpha}{H\ensuremath{\alpha}}

\newcommand{\hbeta}{H\ensuremath{\beta}}

\newcommand{\kms}{km~s\ensuremath{^{-1}}}

\newcommand{\lbol}{\ensuremath{L_{\mathrm{bol}}}}
\newcommand{\loiii}{\ensuremath{L_{\mathrm{[O{\tiny III}]}}}}

\newcommand{\lsun}{\ensuremath{L_{\odot}}}

\newcommand{\nii}{[N~{\small II}]}

\def\lax{{$\mathrel{\hbox{\rlap{\hbox{\lower4pt\hbox{$\sim$}}}\hbox{$<$}}}$}}
\def\gax{{$\mathrel{\hbox{\rlap{\hbox{\lower4pt\hbox{$\sim$}}}\hbox{$>$}}}$}}

\newcommand{\oiiil}{\oiii$\lambda$5007}
\newcommand{\oiiill}{\oiii$\lambda$4959}

\newcommand{\lbolfifteen}{\ensuremath{L_{\mathrm{bol,~15 ~ \mu m}}}}

\newcommand{\oii}{[O~{\small II}]}
\newcommand{\oiii}{[O~{\small III}]}
\newcommand{\sii}{[S~{\small II}]}
\newcommand{\nev}{[Ne~{\small V}]}
\newcommand{\uintensity}{$\times10^{-15}$ erg s$^{-1}$ cm$^{-2}$ arcsec$^{-2}$}

\title[EELR in Obscured AGN with HSC]{Imaging Extended Emission-Line Regions of Obscured AGN with the Subaru Hyper Suprime-Cam Survey}

\author[A.-L. Sun et al.]{Ai-Lei Sun,$^{1,2}$\thanks{E-mail: asun27@jhu.edu}
Jenny E. Greene,$^{3}$
Nadia L. Zakamska,$^{2}$
Andy Goulding,$^{3}$
\newauthor
Michael A. Strauss,$^{3}$
Song Huang,$^{4}$
Sean Johnson,$^{3}$
Toshihiro Kawaguchi,$^{5}$
\newauthor
Yoshiki Matsuoka,$^{6}$
Alisabeth A. Marsteller,$^{3}$
Tohru Nagao$^{6}$
and Yoshiki Toba$^{1}$
\\
$^{1}$Academia Sinica, Institute of Astronomy and Astrophysics, No.1, Sec. 4, Roosevelt Rd., Taipei 10617, Taiwan, R.O.C. \\
$^{2}$Department of Physics and Astronomy, Bloomberg Center, Johns Hopkins University, Baltimore, MD 21218, USA \\
$^{3}$Department of Astrophysical Sciences, Princeton University, Princeton, NJ 08544, USA \\
$^{4}$Department of Astronomy and Astrophysics, University of California, Santa Cruz, 1156 High Street, Santa Cruz, CA 95064, USA \\
$^{5}$Department of Economics, Management and Information Science, Onomichi City University, Hisayamada 1600-2, Onomichi, Hiroshima 722-8506, Japan \\
$^{6}$Research Center for Space and Cosmic Evolution, Ehime University, 2-5 Bunkyo-cho, Matsuyama, Ehime 790-8577, Japan
}

\date{Accepted XXX. Received YYY; in original form ZZZ}

\pubyear{2018}

\begin{document}
\label{firstpage}
\pagerange{\pageref{firstpage}--\pageref{lastpage}}
\maketitle

\begin{abstract}
Narrow-line regions excited by active galactic nuclei (AGN) are important for studying AGN photoionization and feedback. 
Their strong \oiii{} lines can be detected with broadband images, allowing morphological studies of these systems with large-area imaging surveys. 
We develop a new technique to reconstruct the \oiii{} images using the Subaru Hyper Suprime-Cam (HSC) Survey aided with spectra from the Sloan Digital Sky Survey (SDSS). 
The technique involves a careful subtraction of the galactic continuum to isolate emission from the \oiiil{} and \oiiill{} lines. 
Compared to traditional targeted observations, this technique is more efficient at covering larger samples with less dedicated observational resources. 
We apply this technique to an SDSS spectroscopically selected sample of 300 obscured AGN at redshifts 0.1 - 0.7, uncovering extended emission-line region candidates with sizes up to tens of kpc. 
With the largest sample of uniformly derived narrow-line region sizes, we revisit the narrow-line region size -- luminosity relation. 
The area and radii of the \oiii{} emission-line regions are strongly correlated with the AGN luminosity inferred from the mid-infrared  (15 \micron{} rest-frame) with a power-law slope of $0.62^{+0.05}_{-0.06}\pm0.10$ (statistical and systemic errors), consistent with previous spectroscopic findings. 
We discuss the implications for the physics of AGN emission-line region and future applications of this technique, which should be useful for current and next-generation imaging surveys to study AGN photoionization and feedback with large statistical samples. 
\end{abstract}




\section{Introduction}

Feedback from active galactic nuclei (AGN) is often invoked to explain a number of phenomena in galaxy evolution, including the lack of massive star-forming galaxies in the recent universe \citep[e.g.,][]{Bower2006,Croton2006}, and the tight correlation between the mass of supermassive black holes and the gravitational potential of their host galaxies \citep[e.g.,][]{Gebhardt2000,McConnell2013,Kormendy2013}. 
A number of mechanisms have been proposed to inject the energy released by AGN accretion to the galactic interstellar medium. These include radiation pressure, radiative heating, radiative-driven wind, and relativistic jets, but it remains unclear how these mechanisms work and which are most important. 
However, further progress can be made by observing feedback in action, by exploring the relation between the AGN and the interstellar medium of their host galaxies. 

The narrow-line region (NLR) is one of the defining features of active galactic nuclei. NLR were discovered early in the history of the studies of quasi-stellar objects \citep[quasars, e.g.,][]{Wampler1975}. 
They represent the influence of the AGN ionizing photons on the galactic interstellar medium on 0.1 - 10 kpc scales. The NLR is more extended and prominent in more luminous AGN
 \citep[e.g.,][]{Stockton1987, Mulchaey1996, Mulchaey1996a,Liu2013b}. 
 The most extended NLRs hve sizes of tens of kpc; we refer them as extended emission-line regions (EELR). 

Their large physical extent make narrow emission lines, in particular, \oiiil{}, excellent tracers for a number of AGN studies, including feedback, photo-ionization, variability, and obscuration. 
As a kinematic tracer of ionized gas, \oiiil{} has been used to identify AGN outflows, which has led to the discovery of strong ionized outflows of $\gtrsim$ 10 kpc in size, as evidences of AGN feedback \citep[e.g.,][]{Greene2011,Greene2012,Harrison2014,Liu2013,Harrison2015,Sun2017}. 
Extended emission-line regions also provide a unique probe of AGN variability on a time scale of $\sim 10^{4-5}$ years. 
Comparisons of narrow-lines from the EELR, which reflect past AGN luminosity, to instantaneous AGN luminosity tracers, such as X-ray or mid-infrared, suggests that AGN can fade by orders of magnitude on such time scales, giving rise to AGN light echoes \citep[e.g.,][]{Lintott2009,Schirmer2016,Keel2012a,Keel2017}. 
Extended emission-line regions can also constrain the geometry of AGN obscuration. Cone-like emission-line regions are evidence for an obscuring torus \citep{Netzer2015} on smaller scales than can be resolved directly, and its opening angle provides constraints on the torus geometry \citep{Keel2017a,He2018}. 

All of these studies require spatially resolved observations of extended emission-line regions. 
Traditionally, these rely on targeted narrow-band imaging, long-slit, or integral-field-unit spectroscopy, all of which are intensive in observational and data-reduction resources. 
Thus, the samples studied in this way have been limited to a few to a few dozens of objects. 
On the other hand, broadband imaging surveys, such as the Sloan Digital Sky Survey (SDSS), the Panoramic Survey Telescope and Rapid Response System (Pan-STARRS), the Dark Energy Survey (DES), and the Subaru Hyper Suprime-Cam Survey (HSC), have accumulated deep data covering large areas of the sky and have been used for automated identification of various populations of galaxies. 
It is interesting to explore whether these broadband imaging datasets could be useful for selecting extended emission-line regions and resolving their spatial structure. 

In fact, some of the most spectacular extended emission-line regions known were identified with broadband images. 
The strong emission lines, in particular, \oiiill{} and \oiiil{}, create flux excess in the broadband images that gives them distinct colors from normal galaxies. 
The famous quasar light echo -- Hanny's Voorwerp -- was discovered by a citizen scientist who spotted a region with peculiar colors in the SDSS images \citep{Lintott2009}. Dozens of other candidates have been identified with the same method by others in the Galaxy Zoo project \citep{Keel2012}. 
Automated and more systematic searches have been performed to look for emission line systems based on SDSS catalog photometry, such as the Green Peas \citep{Cardamone2009} and Green Beans \citep{Schirmer2013} projects. 
But as the catalog photometry is optimal for galaxies that have regular morphology, such samples are limited to emission-line regions that are compact or concentrated. 
Nonetheless, these works confirm that broadband images are able to detect or even resolve line emission from AGN. 

In this paper, we develop a new technique to image the \oiii{} emission line in AGN with broadband images, allowing an automated search for extended emission-line regions. 
The key of the method is a careful continuum subtraction with the aid of SDSS spectrophotometry. 
To maximize our sensitivity to diffuse faint emission and to explore systems at higher redshifts, we use the deep high-resolution images of the Subaru Hyper Suprime-Cam survey. 
The sample consists of optically selected type 2 AGN, which do not have the glaring AGN continuum and broad lines that would interfere with the narrow-line measurements. 
We also revisit the relation between the size of the EELR and the luminosity of the AGN and compare our results with previous works. 

This paper is organized as follows: 
Section \ref{sec:data} describes the sample and the data. The methodology of imaging and measuring the \oiii{} region is detailed in Sec. \ref{sec:method}. 
The results of the size -- luminosity relation and extended emission-line region candidates are presented in Sec. \ref{sec:res}. We discuss the results in Sec. \ref{sec:discussion}. The paper is summarized in Sec. \ref{sec:summary}. 
Isophotes are quoted at rest-frame surface brightness levels in units of 1\uintensity{}, which corresponds to 1.33 \lsun{} pc$^{-2}$, unless otherwise stated.  
We use an $h=0.7, \Omega_m=0.3, \Omega_{\Lambda}=0.7$ cosmology throughout this paper. 
Vacuum wavelengths are used for the analysis but we keep the conventional notation of air wavelengths for lines, e.g. \oiiil. All error bars represent 1-sigma errors.

\section{Sample and Data} \label{sec:data}
\subsection{Subaru HSC Images} \label{sec:data:hsc}

Our high-quality broadband images are taken from the Subaru Hyper Suprime-Cam (HSC) Survey, which is part of the Subaru Strategic Program \citep[SSP;][]{Aihara2018a}. 
This survey takes advantage of the wide field capability (1.77 deg$^2$ field-of-view) of the Hyper Suprime-Cam \citep{Miyazaki2012,Miyazaki2018,Komiyama2018,Furusawa2018} on the Subaru 8.2m telescope to deliver high-quality $g$, $r$, $i$, $z$, $y$ broadband images over a total survey area of $\sim$ 1400 deg$^2$. 
The survey contains three nested layers -- the ultra-deep, deep, and wide layers, of progressively larger areas and shallower depths. The median seeing in $i$-band is 0\farcs6, compared to a typical seeing of 1\farcs4 in SDSS $i$-band. 

The data used throughout this paper is based on the S16A internal release taken between March 2014 and April 2016. The wide layer, which this paper focuses on, contains 174 deg$^2$ of data with full depth in all five bands. 
Some of the data were reprocessed to give a more accurate estimate of the point-spread-function (PSF). 
The depth of the wide layer expressed as 5-sigma detection limit in 2 arcsec diameter apertures is 26.1 and 25.9 mag in the $r$ and $i$-band, respectively.  Basic data processing, including bias and background subtraction, flat-fielding, astrometric calibration, individual exposure co-addition, and object detection was performed using hscPipe 4.0.1 and 4.0.2, which is an HSC-specific derivative of the Large Synoptic Survey Telescope (LSST) processing pipeline \citep{Ivezic2008,Axelrod2010,Juric2015}. For further details regarding hscPipe and the HSC-SSP data releases, see \citet{Bosch2018,Aihara2018}; and \citet{Huang2018}. 
For our analyses, we use the co-add images that are registered to a common World Coordinate System (WCS) and have a pixel scale of 0.168 arcsec.

\subsection{Sample} \label{sec:data:t2sample}

Our targets are selected from four SDSS specroscopically identified obscured type 2 AGN samples -- \citet{Zakamska2003,Reyes2008,Mullaney2013}; and \citet{Yuan2016}. 
In general, these type 2 AGN candidates are selected to have no broadline features (\halpha{}, \hbeta{}, or $\mathrm{[Mg {\tiny II}]\lambda 2800}$) but only narrow emission lines with line ratios characterized by AGN-like photoionization, e.g., \oiii{}/\hbeta{}, \nii{}/\halpha{}, \sii{}/\halpha{} \citep{Baldwin1981,Veilleux1987,Kauffmann2003a}. At high redshifts where the \nii{} and \halpha{} lines move out of the spectral coverage, the detection of \nev{} lines \citep{Gilli2010} is also used to separate type 2 AGN from star-forming galaxies. 
Each of the samples differs in their exact criteria for broad lines, their spectrum signal-to-noise ratio cut, and their \oiii{} luminosity and equivalent width cuts, which introduces heterogeneity in the sample. 
We refer to each individual paper for their specific selection criteria. 
The first three samples \citep{Zakamska2003,Reyes2008,Mullaney2013} are selected based on spectra from the Sloan Digital Sky Survey \citep[SDSS;][]{York2000}, which are observed through fibers of diameter 3\arcsec{}, and the \citet{Yuan2016} sample is selected from the Baryon Oscillation Spectroscopic Survey \citep[BOSS;][]{Dawson2013} of SDSS-III \citep{Eisenstein2011}, which has a smaller aperture of 2\arcsec{}. 
\citet{Mullaney2013} targets low redshift objects ($z\lesssim0.3$), while \citet{Zakamska2003} and \citet{Yuan2016} primarily focuse on higher redshifts ($z\sim0.3-0.7$), and \citet{Reyes2008} have both. 
The typical \oiiil{} luminosities increase with redshifts from a median of $\loiii \sim 10^{41.5}$ \ergs{} in the \citet{Mullaney2013} sample to a median of $\loiii \sim 10^{42.5}$ \ergs{} in the \citet{Yuan2016} sample. 
There are overlaps between these samples (525 duplicates and 38 triplicates), resulting in 17051 unique objects in the combined sample; see Tab. \ref{tab:sample_source}. 

\begin{figure*}
	\centering
	\vbox{
	\includegraphics[width=3.in]{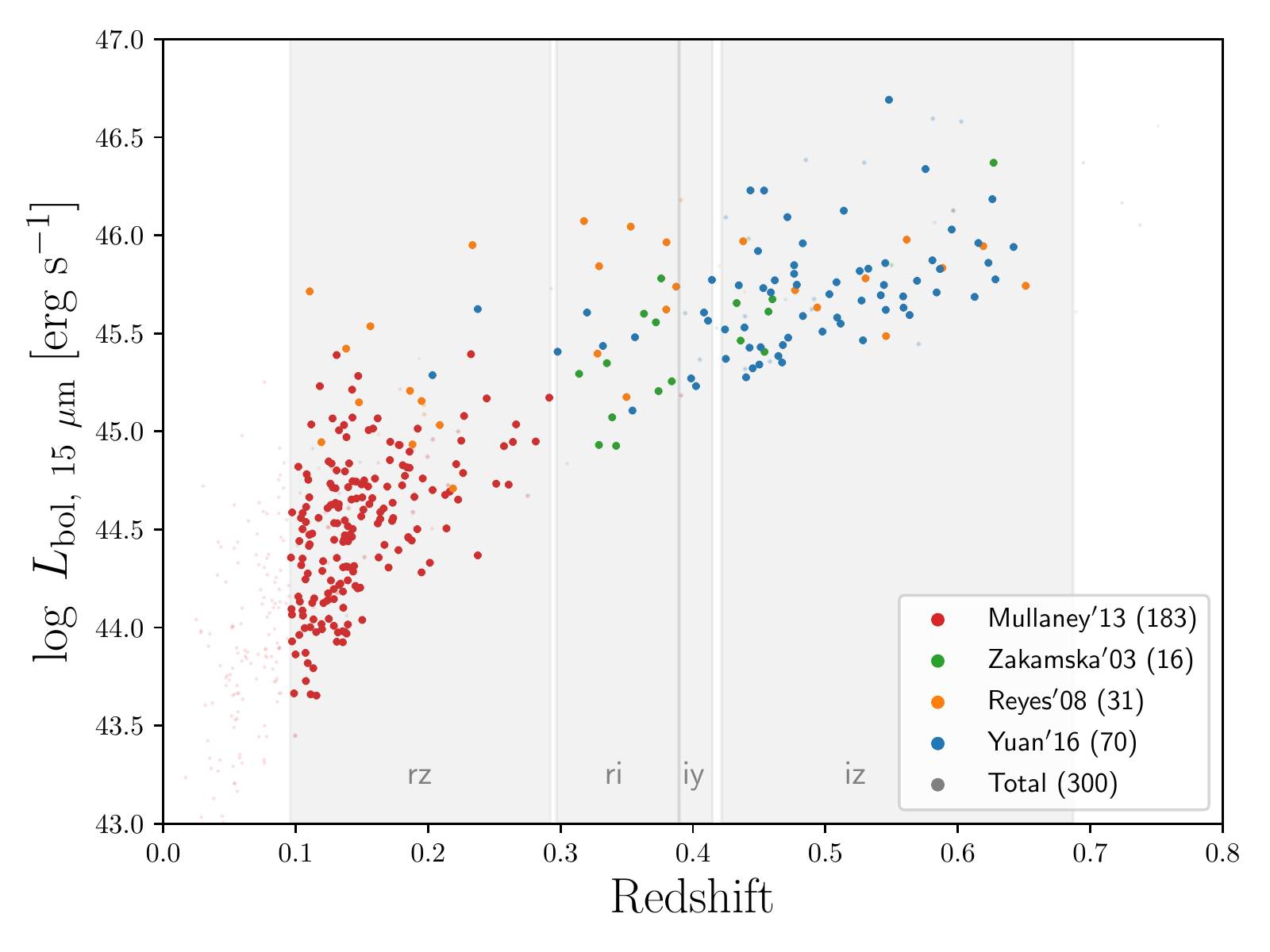}
	\includegraphics[width=3.in]{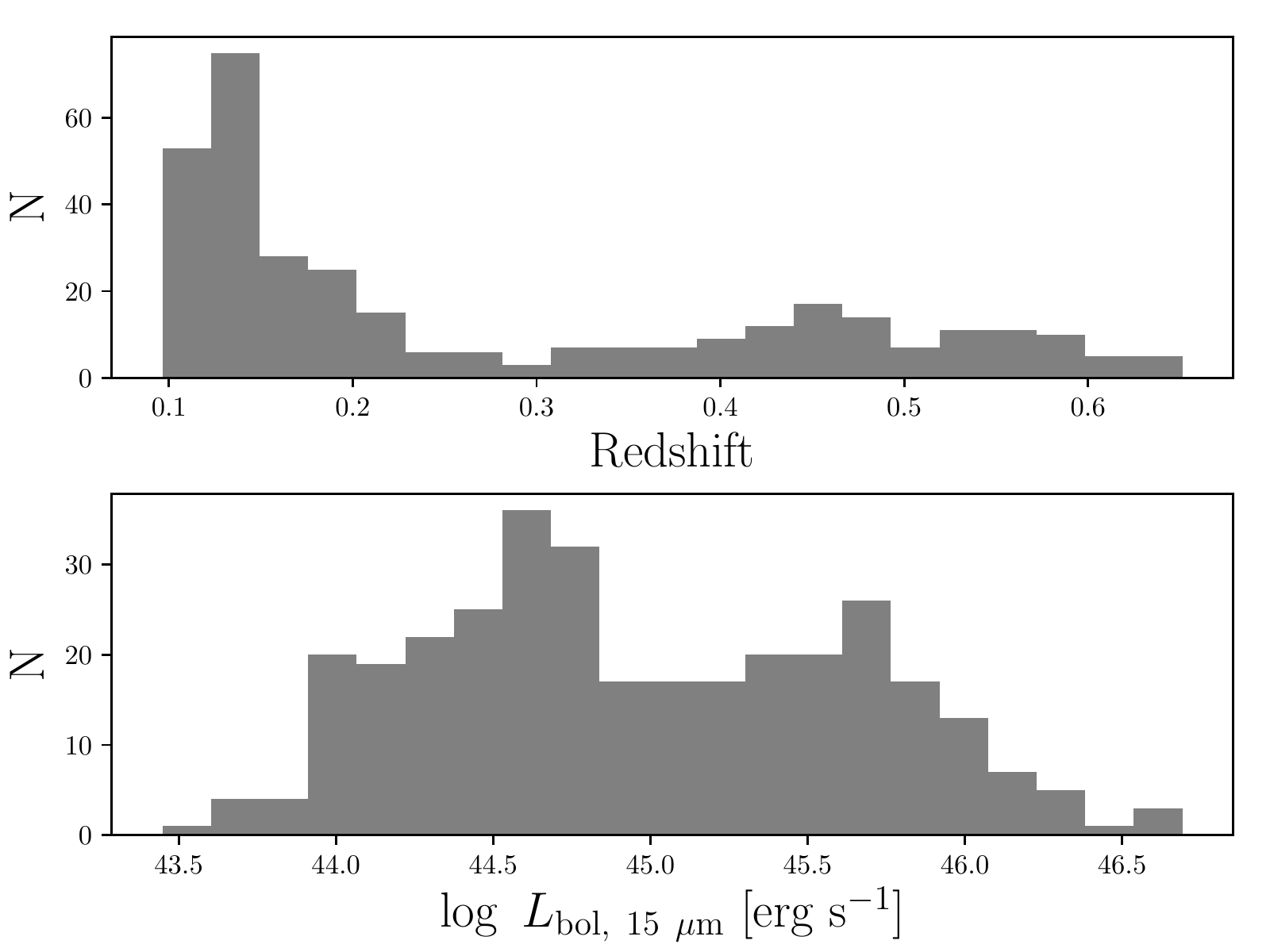}
	}
	\caption{
	{\it Left:}
	The AGN luminosities and redshifts of our type 2 AGN sample. 
	The large dots are the primary sample as defined in Sec. \ref{sec:method:linemap} with the colors indicating the source of the target. The sample sizes are shown in the brackets. 
	The small transparent dots in the background are the type 2 AGN within the HSC footprint but not included in the primary sample due to their redshift or data quality issues. 
	The shaded areas are the redshift ranges where the \oiiil{} line can be imaged with one of the band configurations listed in Tab. \ref{tab:sample_batch}. 
	{\it Right:} The redshift and AGN luminosity (\lbolfifteen{}) distributions of the primary sample. 
	}
	\label{fig:sample}
\end{figure*}

We crossmatch this SDSS sample with the sample of galaxies with $i$-band Kron magnitudes $<$ 24 in the HSC S16A-wide catalog with a matching radius of 2 arcsec. 
The matched sample contains 527 objects, as the HSC survey is much smaller in area than SDSS. 
We then focus on the objects where the \oiiil{} line falls in either the HSC $r$-band ($0.10 < z < 0.39$) or the $i$-band ($0.39 < z < 0.69$; see Sec. \ref{sec:method:overview}). We require that the two HSC bands required for the \oiii{} image reconstruction (which depends on the redshift, see below) are available for the objects. 
In addition, two AGN from the \citet{Mullaney2013} sample are excluded at this stage because our visual inspection finds that their spectra show type-1-like blue continuum and broad lines.
The resulting sample contains 333 systems covering a redshift range from 0.1 to 0.7 and an AGN bolometric luminosity of $10^{42-46.5}$ \ergs{} (see below), see Tab. \ref{tab:sample_source}. 
In what follows, we will drop 33 objects whose image quality is inferior, leaving us with a primary sample of 300 objects on which to do statistical studies. 
Their redshift and luminosity distributions are shown in Fig. \ref{fig:sample}.

\begin{table*}
\begin{center}
\caption{Type 2 AGN Sample Size}
\begin{tabular}{cccccc}
\hline
\hline
 & Yuan et al. 2016 & Reyes et al. 2008 & Zakamska et al. 2003 & Mullaney et al. 2013 & Total \\
\hline
Original Sample & 2758 & 887 & 291 & 13716 & \nodata \\
Unique Instances & 2758 & 842 & 145 & 13306 & 17051 \\
\oiii{} Imaged & 84 & 36 & 18 & 195 & 333 \\
Primary Sample & 70 & 31 & 16 & 183 & 300 \\
\hline
\end{tabular} \\
\textit{Note.}
Type 2 AGN sample size from each of the parent SDSS spectroscopic samples. The first row is the size of the original sample. The second row is the size of the sample after resolving duplicated sources between samples. The third row shows the number of type 2 AGN that are imaged with HSC. 
These objects are within the HSC footprint, have their \oiiil{} line covered by either the HSC $r$- or $i$-band, and have HSC images available in the both the line-band and continuum-band. 
The last row shows the primary sample size used in this study, which includes objects that pass the image and spectrum quality criteria as discussed in Sec. \ref{sec:method:linemap}. 
\label{tab:sample_source}
\end{center}
\end{table*}

\subsection{AGN Luminosity from WISE} \label{sec:data:wise}

We use the mid-infrared luminosities from the Wide-field Infrared Survey Explorer \citep[\emph{WISE};][]{Wright2010} \textit{ALLWISE} catalog \citep{Cutri2013} as a proxy for the AGN bolometric luminosity.  
The mid-infrared captures emission from AGN heated hot dust and has been used as an AGN luminosity indicator in previous studies of NLR size -- AGN luminosity relations \citep[e.g.,][]{Hainline2013}. 
Unlike optically-based AGN luminosity indicators such as the \oiiil{} line, the mid-infrared is independent from the properties of the emission-line regions and is thus arguably a cleaner tracer for the AGN luminosity. 
However, differences in mid-infrared colors have been found between type 1 and type 2 AGN \citep{Yan2013}, such that the inferred AGN luminosity could depend on the AGN type or the inclination of the system. 

As discussed in \citet{Sun2017}, mid-infrared luminosities at longer wavelengths, such as rest-frame 15 \micron{}, are more robust to such type dependences than shorter wavelengths. 
We adopt the same treatment as \citet{Sun2017} to interpolate the \emph{WISE} W2, W3, and W4 bands with a second order spline to obtain the rest-frame 15 \micron{} luminosity ($L_{\mathrm{15~\mu m}}$). We adopt a bolometric correction factor of 9 to infer the AGN bolometric luminosity $L_{\mathrm{bol,~15~\mu m}} = 9\times L_{\mathrm{15~\mu m}}$ \citep{Richards2006}. 
Six of the 300 objects in the primary sample, which is defined in Sec. \ref{sec:method:linemap}, do not have the \emph{WISE} detections to infer the 15 \micron{} luminosity, and we do not use them in the analysis of the mid-infrared-based size -- luminosity relations, see Sec. \ref{sec:res:sizelum}.

\section{Methodology} \label{sec:method}

In this section, we describe our methodology to image and measure the \oiii{} emitting region of type 2 AGN. Section \ref{sec:method:overview} gives an overview of the spectral characteristics of type 2 AGN and how to observe them with broadband imaging. Section \ref{sec:method:conti} describes the continuum subtraction technique. 
The conversion from broadband image to line surface brightness is detailed in Sec. \ref{sec:method:linemap}. 
A discussion on the choice of the isophotal threshold is in Sec. \ref{sec:method:isolevel} and
the isophotal measurements are described in Sec. \ref{sec:method:iso}. The Python code that automates the analysis described in this section is made public\footnote{http://github.com/aileisun/bubbleimg}.

\subsection{Overview} \label{sec:method:overview}

\begin{figure*}
	\centering
	\includegraphics[width=5in]{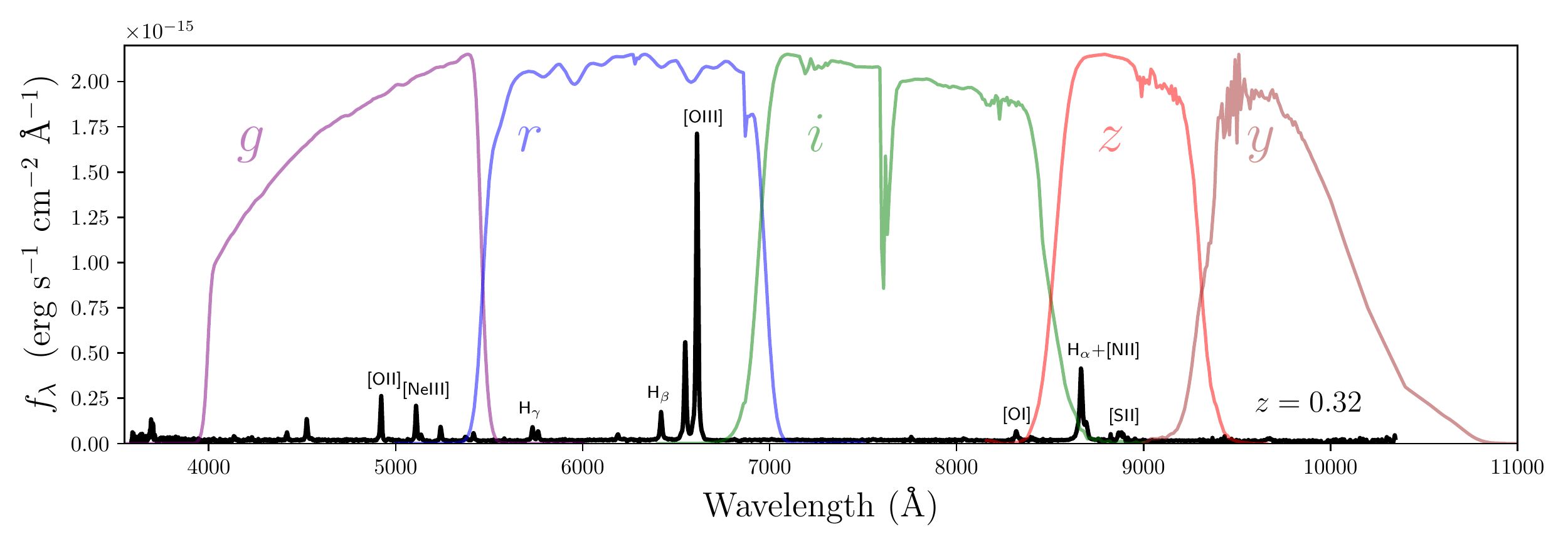}
	\caption{The SDSS spectrum of an example luminous obscured AGN in our samle, together with the filter response functions of the 5 HSC bands \citep{Kawanomoto2018}. Strong narrow emission lines, including the \oiii{} doublet and the \halpha{}+\nii{} lines, stand out from the stellar continuum. The part of the spectrum between the \oiiil{} and the \halpha{} lines as well as the part redward of the \halpha{} line are extended wavelength ranges without strong emission lines for the imaging of the stellar component. 
	This AGN is SDSS J160550+440540 at $z=0.32$ with \loiii{}~$=10^{42.94}$ \ergs{} from the sample of \citet{Yuan2016}. The spectrum is taken by the BOSS spectrograph. 
	}
	\label{fig:spec}
\end{figure*}

As discussed in the introduction, the strong emission lines in type 2 AGN, in particular, the \oiiill{} and \oiiil{} lines, can create a flux excess in broadband images, see Fig. \ref{fig:spec}, allowing reconstruction of emission line images from broadband images. 

The flux excess depends on the observed equivalent width (EW) of the emission line and the band width (BW) of the photometric band, according to 
\begin{equation}
\Delta m_{AB} = -\frac{5}{2} \log_{10} \left( 1+\frac{EW}{BW} \right). 
\end{equation}
With a typical broadband width of $1000$ \angstrom{}, even a moderately luminous obscured type 2 AGN (\lbol{} $\sim 10^{45.5}$ \ergs{}) that has an \oiiil{} $EW \sim 100$ \angstrom{} can create a measurable flux excess of $\Delta m_{AB} \sim 0.1$. The flux excess can be even more noticeable at the outskirts of galaxies where the continuum is faint. 
The emission-line regions in unobscured AGN may be harder to identify because the strong AGN continuum lowers the line equivalent widths. 

The broadband images contain the contribution from both the emission lines and the galactic stellar continuum. 
Thus, to map the narrow emission-lines, one needs to subtract out the stellar continuum. 
Fortunately, there are large continuum-only wavelength ranges in the spectrum without strong emission lines, in particular, between the \oiiil{} line and the \halpha{}/\nii{} lines, as well as redward of \halpha{}/\nii{}. 
The image of this line-free wavelength range serves as a model for the continuum subtraction, as described in Sec. \ref{sec:method:conti}. 

In this paper, we focus on the type 2 AGN in which the \oiiil{} line falls in the $r$-band or the $i$-band, i.e., at redshifts between 0.10 and 0.69. The choice of line-band and continuum-band depends on the redshifts. In total, there are four different line-band and continuum-band combinations for the \oiii{} image reconstruction, i.e., $r-z$, $r-i$, $i-y$, and $i-z$, covering redshifts from 0.10 to 0.69, as plotted in Fig. \ref{fig:sample} and tabulated in Tab. \ref{tab:sample_batch}. 
The redshift boundaries of these four bins are defined as where the \oiiil{} line is within the line-bands where the filter is above 60\% of its peak sensitivity, and where the \halpha{}/\nii{} lines are outside the continuum-bands where the filter is above 40\% of its peak sensitivity. 

\begin{table*}
\begin{center}
\caption{Band-Configurations for \oiiil{} Imaging}
\begin{tabular}{cccccc}
\hline
\hline
Line- \& Conti-band & $r-z$ & $r-i$ & $i-y$ & $i-z$ & Total\\
(Redshifts) & (0.096 - 0.292) & (0.297 - 0.390) & (0.389 - 0.415) & (0.422 - 0.687) & \\
\hline
Primary Sample & 197 & 23 & 5 & 75 & 300 \\
\hline
\end{tabular} \\
\textit{Note.}
The four band-configurations used for \oiii{} line imaging. 
The redshift ranges within which the line-band captures the \oiiil{} line and the continuum-band captures line-free continuum are indicated in brackets. 
The numbers are the primary sample size of each of the configurations. 
\label{tab:sample_batch}
\end{center}
\end{table*}

\subsection{Continuum Subtraction} \label{sec:method:conti}

The image of the continuum-band is used as a continuum subtraction model after it is scaled by a constant factor to match the continuum flux in the line-band. This scaling factor depends on the color of the stellar continuum, which can be determined from the SDSS spectrum. The continuum-band and line-band images are matched in point-spread-function (PSF) before subtraction. This continuum subtraction should be suitable as long as there is no significant color gradient in the continuum. 

To estimate the scaling factor, we use the science primary spectra, which are taken with either the SDSS or BOSS spectrograph, from the SDSS data release 12 \citep{Alam2015}. 
To obtain the continuum spectrum, we mask the emission lines and fit the spectrum with a stellar population synthesis model. 
The models are taken from \citet{Bruzual2003} with 156 spectra covering ages from 0.1 to 14 billion years and metallicities from $Z=$0.004 to 0.05. 
For simplicity, the original model spectrum is used without reddening applied and no power-law continuum (e.g., from type 1 AGN light) is used. 
To reduce over-fitting, linear Lasso regression is used to find a small subset of these templates to best fit the data. 
The weighting of each model spectrum is non-negative. 
We evaluate the performance of the continuum fitting and find that, in the wavelength range where there are no strong lines, the best-fit model gives broad-band averaged flux that is consistent with the observed spectrum within an uncertainty of 10\% (1-$\sigma$). This includes uncertainties that may arise from any mismatch between the actual continuum and the assumed template. 

The observed spectra do not cover the redmost 20-25\% of the HSC $z$-band (SDSS spectrum) or $y$-band (BOSS spectrum). In the galaxies where the $z$- or $y$-band is needed for continuum subtraction, this corresponds to a missing coverage of 100 - 170 \AA{} at rest-frame wavelengths 5600 - 6600 \AA{} or 7100 - 8500 \AA{}. 
When tested with $z < 0.2$ SDSS and BOSS spectra that cover these wavelength regions, we find that the best-fit model can reproduce the continuum spectrum in this missing range accurately with an uncertainty of only 6$\%$.  

We convolve the best-fit continuum with the HSC filter transmission functions, which include the sky transmission and detector efficiency, to calculate the corresponding flux density in a given band, taking into account that the CCD integrates over the photon number instead of energy. 
The ratio of the flux densities between the line-band and the continuum-band is used to scale the continuum-band image to match the continuum level in the line-band. 
This scaled image is then taken as the model for continuum subtraction in the line-band based on the assumption that the galaxy's continuum color does not vary significantly across the image.  

As the line-band and the continuum-band image may have different PSFs, PSF matching is required before subtracting the two images.  The PSF models are taken from the output of the HSC pipeline.  
We convolve the smaller PSF with the 2D Moffat profile that allows us to reproduce the larger PSF. The Moffat function has been found to be suitable for modeling the PSF in ground based images and has been used as a PSF homogenizing kernel \citep{Chan2015,Liu2015}. 

\begin{figure*}
    \centering
    \includegraphics[width=6in]{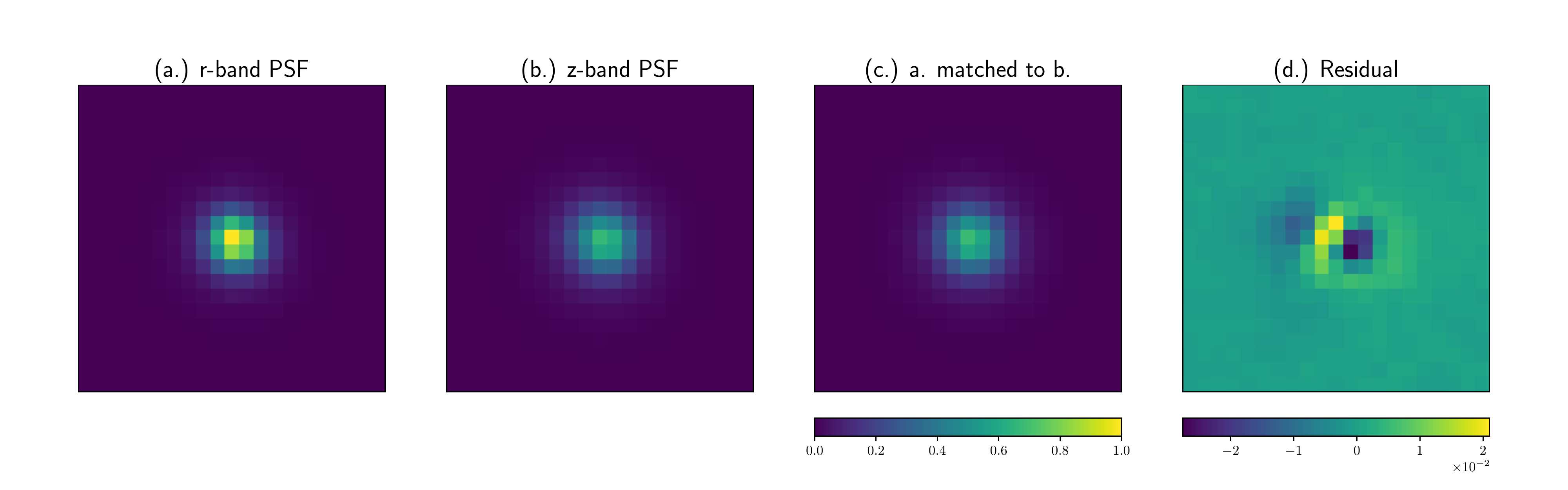}
    \caption{Example of PSF matching. Panel (a.) shows the $r$-band PSF (FWHM = 0$\farcs$49), which is sharper than the $z$-band PSF (FWHM = 0$\farcs$59) shown in (b.). 
	Panel (c.) shows the degraded $r$-band PSF that is matched to the $z$-band, and panel (d.) shows their difference. In this example, the maximum residual amplitude is 4\% of the peak of the $z$-band PSF, which is typical for our sample. 
    }
    \label{fig:psf}
\end{figure*}

\begin{figure*}
    \centering
    \hbox{
    \includegraphics[width=3.in]{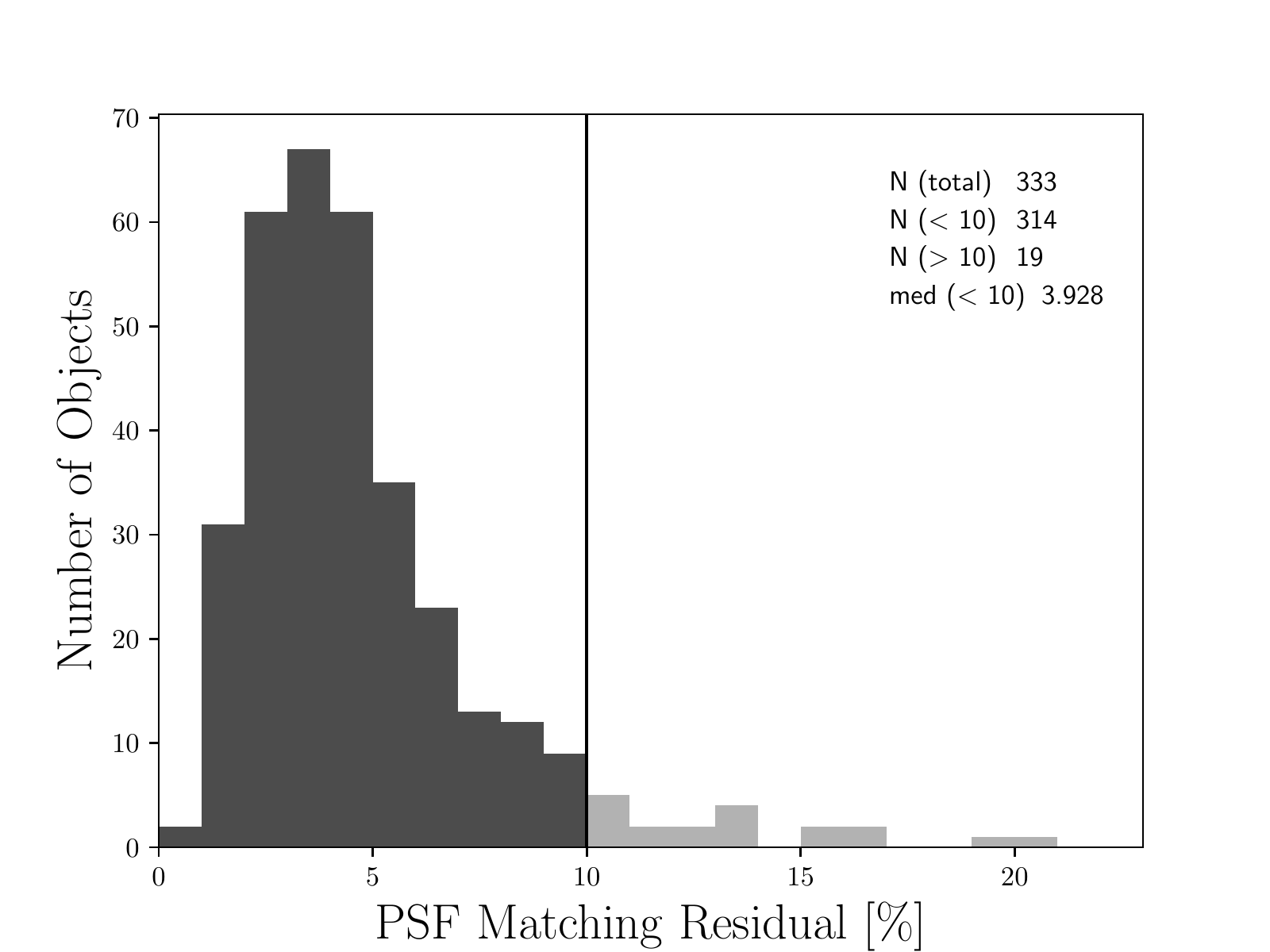}
    \includegraphics[width=2.8in]{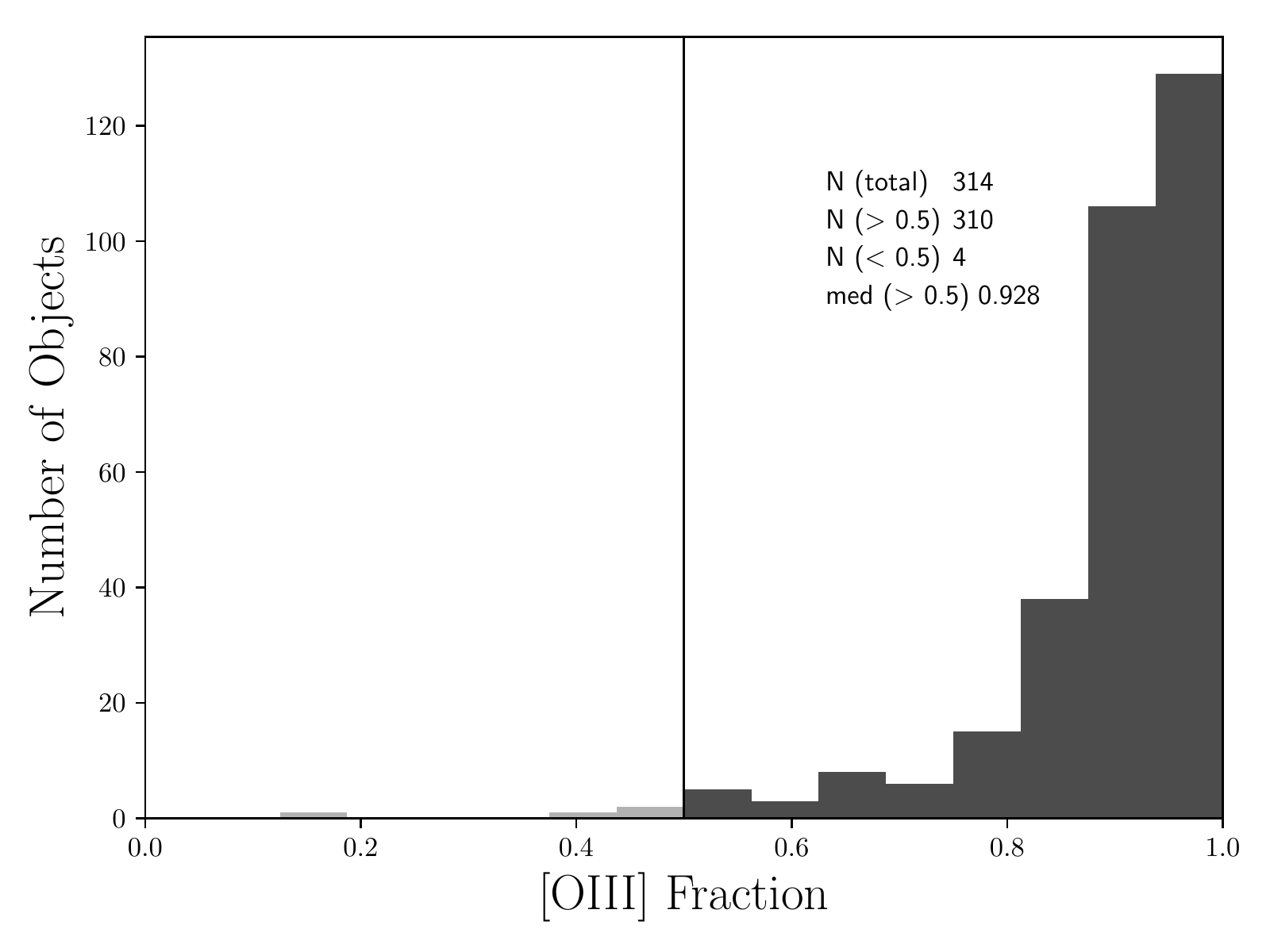}
    }
    \hbox{
    \includegraphics[width=3.in]{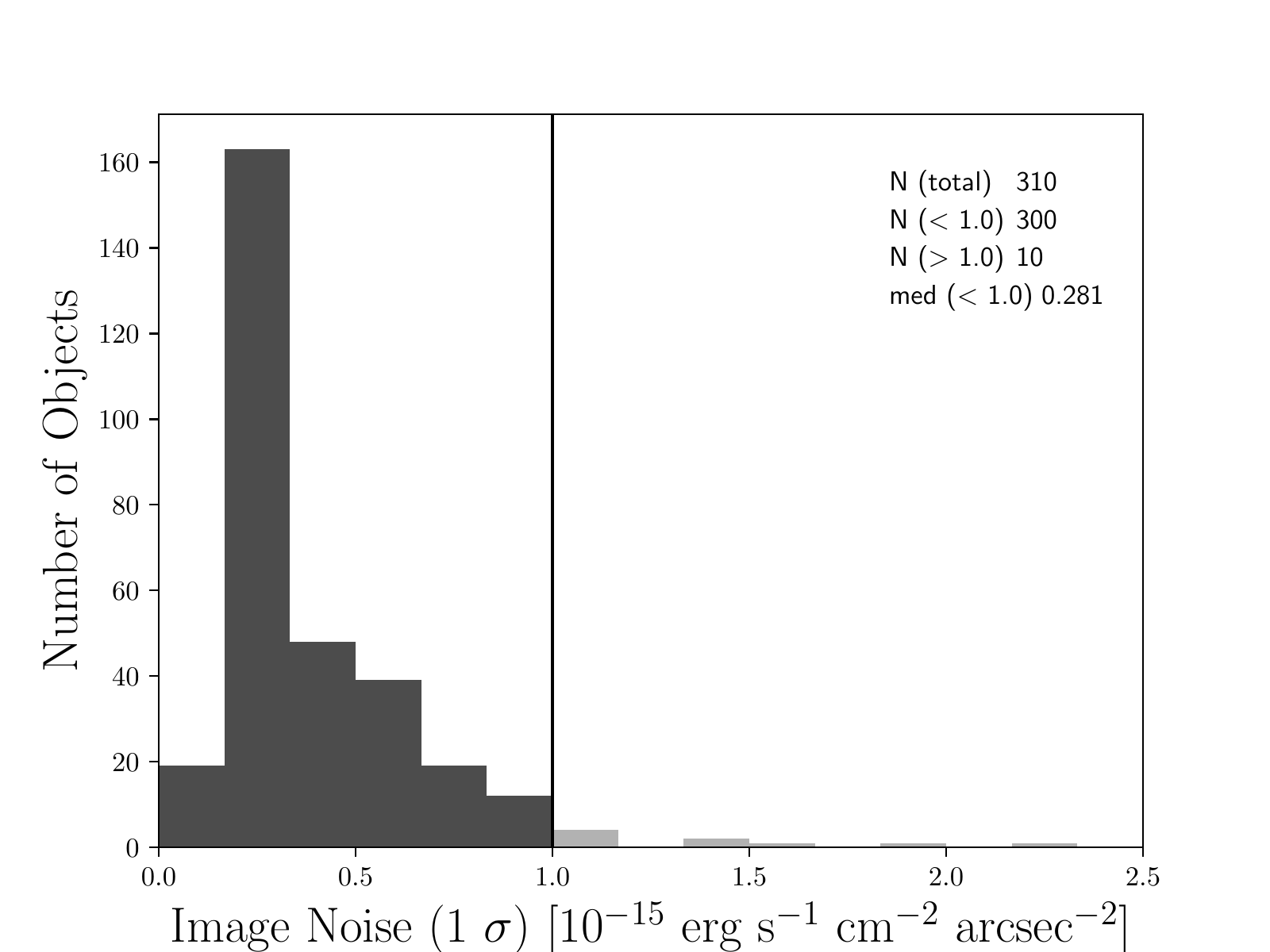}
    \includegraphics[width=3.in]{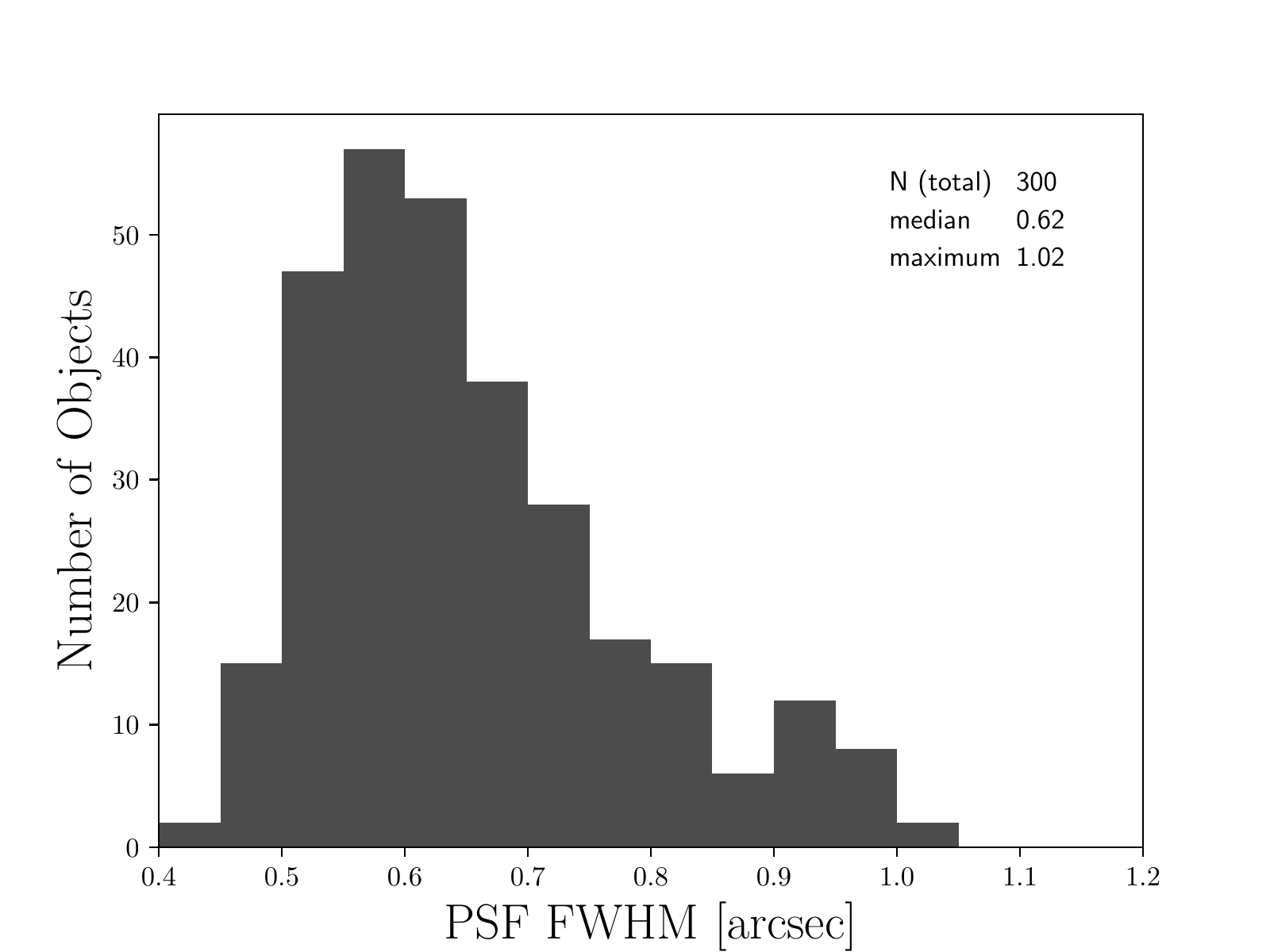}
    }
    \caption{The image and spectrum quality criteria for the primary sample as described in Sec. \ref{sec:method}. 
    {\it Upper left}: the histogram of the PSF matching residual, 
    the ratio of the maximum pixel value of the residual to the peak of the original PSF. The sample contains objects from all four line-band -- continuum-band combinations.  Objects with residuals more than 10\% are excluded from our primary sample. 
    {\it Upper right}: the distribution of the fraction of the emission line flux in the line-band attributed to the \oiiill{} and \oiiil{} lines as calculated from the SDSS spectrum. Objects with \oiii{} contributing to less than 50\% are excluded from our primary sample. 
    {\it Lower left}: the distribution of the image noise level (1-$\sigma$) of the \oiiil{} map. Objects with noise higher than 1 \uintensity{} are excluded from our primary sample. 
    The resulting sample of 300 objects is the primary sample used in the analysis. Their distribution of PSF FWHM is shown on the 
    {\it lower right}, which has a median of 0\farcs62. 
    The texts on the upper right corners of the first three panels show the total number of objects, the number of objects below and above the threshold, and the median value of the plotted quantify of the selected sample highlighted in dark shade, respectively. 
    The texts in the {\it lower right} panel show the total number of objects, the median, and the maximum value of the PSF FWHM of the primary sample, respectively. 
    }
    \label{fig:qc}
\end{figure*}

Fig. \ref{fig:psf} shows an example of the PSF matching. The residuals are dominated by the asymmetry in the PSF shape, which is not captured by the circularly symmetric Moffat profile. The residuals are typically a few percent of peak of the original PSF. 
However, there are 19 objects that have PSF matching residuals larger than 10\%, see the upper left panel of Fig. \ref{fig:qc}. To avoid artifacts, these objects are excluded for the following analysis.


\subsection{\oiiil{} Line Intensity Map} \label{sec:method:linemap}

We use the observed SDSS spectrum to quantify the contamination from emission lines other than \oiiill{} and \oiiil{} in the line-band. 
The continuum-subtracted image that we obtain in Sec. \ref{sec:method:conti} is a linear combination of all the emission lines in the band, weighted by the transmission function of the band, that is,
\begin{align}
	I_{\nu}^{\mathrm{contsub}} = \frac{\sum_i I_i T(\lambda_i) \lambda_i / c}{\int{T(\lambda) d\lambda/\lambda}},
\end{align}
where $I_{\nu}^{\mathrm{contsub}}$ is the specific intensity of the continuum subtracted image, $I_i$ and $\lambda_i$ are the intensity and the wavelength of the line $i$, $T$ is the filter transmission function, and $c$ is the speed of light. 
The relative intensities of the lines are estimated from the SDSS spectrum and are integrated over a width of $\pm 1400$ \kms{}. 
In addition to the two \oiii{} lines, we consider the \hbeta{}, [Ne~{\small III}]3870, [Ne~{\small III}]3969, H$\gamma$, [O~{\small I}]6302, and [O~{\small I}]6366 lines. 

The \oiii{} doublets -- \oiiill{} and \oiiil{}, which have a fixed theoretical flux ratio of 2.98 \citep{Storey2000} -- typically together contribute $> 80\%$ of the flux in the continuum-subtracted image for the objects in our sample, see the upper right panel of Fig. \ref{fig:qc}. 
Thus, the reconstructed \oiiil{} image is not sensitive to intensity variations of the other lines, such as \hbeta{}, across the galaxy.  
The \oiiil{} to \hbeta{} ratio is typically constant across the galaxy for luminous type 2 AGN \citep[e.g.,][]{Greene2012}. 
But even if the fractional flux contribution from \hbeta{} varied by a factor of 2, for example in cases where star formation dominates the line intensity, it would introduce less than 30\% uncertainty in the inferred \oiiil{} flux. 
There are four galaxies for which the \oiii{} doublets contribute less than half of the flux in the line-band. We exclude these objects for the following analysis. 

The continuum-subtracted image ($I_{\nu}^{\mathrm{contsub}}$) is then translated to a \oiiil{} intensity map ($I_{k}$) given the fractional contributions of the lines. Expressing the \oiiil{} line as line $k$, the \oiiil{} image is
\begin{align}
	I_{k} = \frac{1}{T(\lambda_{k}) \lambda_{k}} \gamma_{k} \times \tau \times c \times I_{\nu}^{\mathrm{contsub}}, 
\end{align}
where
$\gamma_{k} = f_{k} T(\lambda_{k}) \lambda_{k} / \sum_i f_i T(\lambda_i) \lambda_i$
is the fraction of the flux contributed by the \oiiil{} line, and $\tau = \int{T(\lambda) d\lambda/\lambda}$ is a normalizing factor. 
The ratio between the \oiiil{} and \oiiill{} line is fixed at the theoretical value of 2.98. The final \oiiil{} image is expressed in rest-frame intensity after a cosmological dimming correction of $I_{\mathrm{rest}} = I_{\mathrm{obs}} \times (1+z)^4$. 
The intensity maps shown below are in units of 1\uintensity{}, which is 1.33 \lsun{} pc$^{-2}$ in surface brightness. 

\begin{figure*}
    \centering
    \vbox{
    \includegraphics[width=5in]{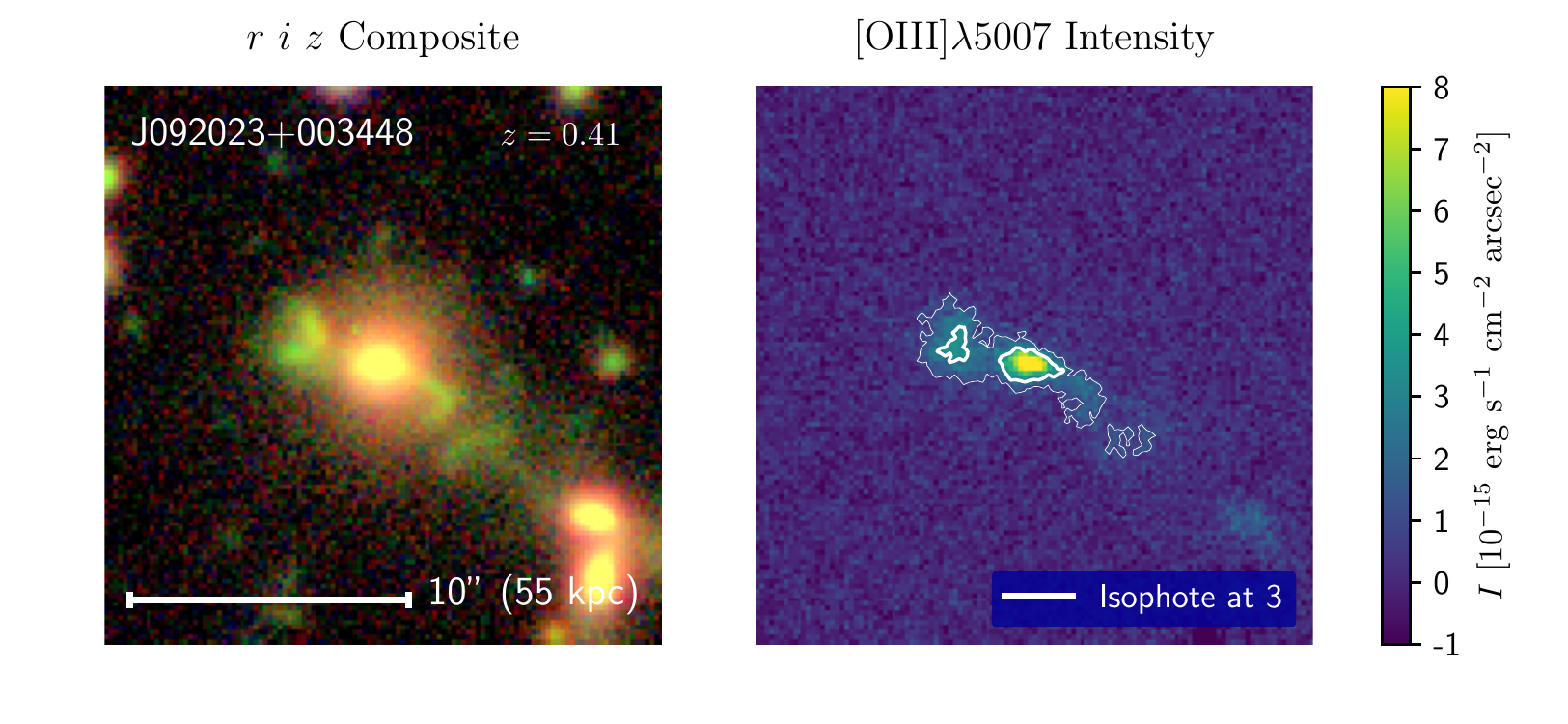}
    }
    \caption{Example of \oiii{} intensity map reconstruction. 
    {\it Left}: The HSC $r$, $i$, $z$-band composite image of type 2 AGN SDSS J092023+003448. The green color shows the $i$-band flux excess due to the strong \oiii{} emission. {\it Right}: 
    The reconstructed \oiiil{} rest-frame intensity map in units of 1\uintensity{}. 
    The thick contour is at an intensity of 3\uintensity{}. This is what we use for the isophotal area measurement. The thin contour is at 1\uintensity{} to outline the extended emission. 
    This map captures the morphology of the $i$-band excess and is dissimilar from the bulge-like galactic continuum. 
    }
    \label{fig:map_example}
\end{figure*}

Figure \ref{fig:map_example} shows the reconstructed \oiiil{} rest-frame intensity map for SDSS J092023+003448.  The scaled $y$-band image is subtracted as a continuum from the $i$-band image that contains the \oiiil{} line. The line emission is successfully separated from the stellar continuum, and shows an elongated morphology that is distinct from the rounder and smoother galaxy shape. 

As shown in the lower left panel of Fig. \ref{fig:qc}, the median 1-$\sigma$ noise level per pixel of the \oiiil{} image is 0.3 \uintensity{}, which is measured by fitting a Gaussian function to the distribution of the image pixel values. 
This is deep enough to allow detection of the extended \oiii{} emission. 
It is three times lower than the isophotal cut adopted in previous spectroscopy studies of AGN emission-line regions \citep[1 \uintensity{}, e.g.,][]{Liu2013b,Hainline2013,Sun2017}. 
As discussed in Sec. \ref{sec:method:isolevel}, we adopt a higher isophote cut to avoid other galactic contaminants. 
There are 10 objects, predominantly at higher redshifts, that have significantly higher noise and they are excluded from the following analysis, see Fig. \ref{fig:qc}. 

In total, from the sample of 333 type 2 AGN in our sample, 33 are excluded due to data quality issues regarding PSF, \oiii{} fraction, or image noise, resulting in a final sample of 300 systems, which we define as the primary sample of this paper for the following analysis. 
This primary sample has a median \oiii{} fraction ($\gamma_{k}$) of 92\%, median PSF FWHM of 0\farcs62, and a median image noise per pixel of 0.3 \uintensity{} in the \oiii{} intensity map. 

\subsection{Isophotal Threshold} \label{sec:method:isolevel}

To quantify the size and area of the \oiiil{} emission-line regions, we adopt an isophotal measurement similar to previous studies \citep[e.g.][]{Liu2013b,Hainline2013,Sun2017}. 
Unlike parametrized model fitting, isophotal measurements are more adaptive to irregular morphologies that are commonly seen in AGN emission-line regions, showing filamentary morphology \citep{Greene2014}, outflow driven bubbles \citep{Harrison2015}, ring-like or isolated light-echoes \citep{Lintott2009,Schweizer2013,Keel2015}, among other shapes. 

Previous spectroscopic studies of the narrow-line region size -- luminosity relation often use a faint isophote level of 1 \uintensity{} rest-frame \citep[][]{Liu2013b,Hainline2013,Sun2017}. This threshold is chosen to be sensitive to the faint outskirts of the emission-line regions in luminous AGN. Indeed, the measured radii are often larger than the typical size of galaxies ($>$ 10 kpc) and in some cases as large as 30 kpc when ionized tidal features are included. 
However, such a faint isophote level may not be suitable for lower luminosity AGN, where the emission-line regions are smaller (hundreds of pc). 
In such cases, the emission (or absorption) coming from various galactic structures such as the spiral arms, dust lanes and star-forming regions could affect the isophotal measurements. 
For example, the \oiiil{} line emitted by star formation regions by itself can reach a surface brightness of $1$ \uintensity{} in starburst systems, if we adopt typical \oiii{} luminosities \citep{Moustakas2006} and star formation surface densities \citep{Kennicutt2012}. 
The \hbeta{} emission could also be comparable or sometimes even stronger than \oiiil{}. 
Indeed, Seyfert galaxies sometimes exhibit a transition in emission-line ratios from AGN to HII regions at a certain radius, when observed with spatially resolved spectral diagnostics \citep{Bennert2006a,Wylezalek2018}. 
Therefore, it is important to determine a robust isophote threshold that is faint enough to trace the extended part of the emission-line regions and at the same time bright enough to be free from various galactic contaminations such as star formation. 

\begin{figure*}
\centering
	\hbox{
    \includegraphics[width=3.in, trim={1cm 1cm 0 0}, clip]{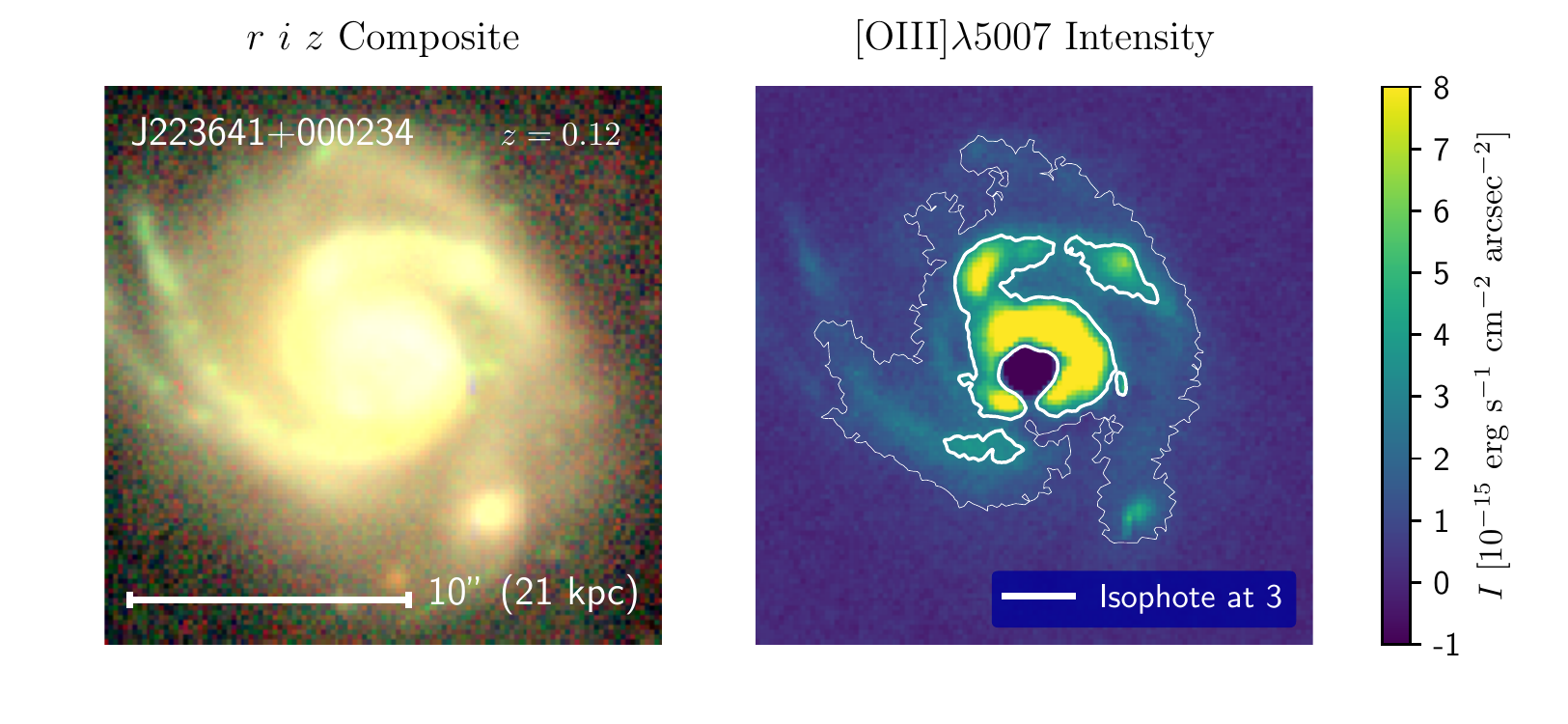}
    \includegraphics[width=3.75in]{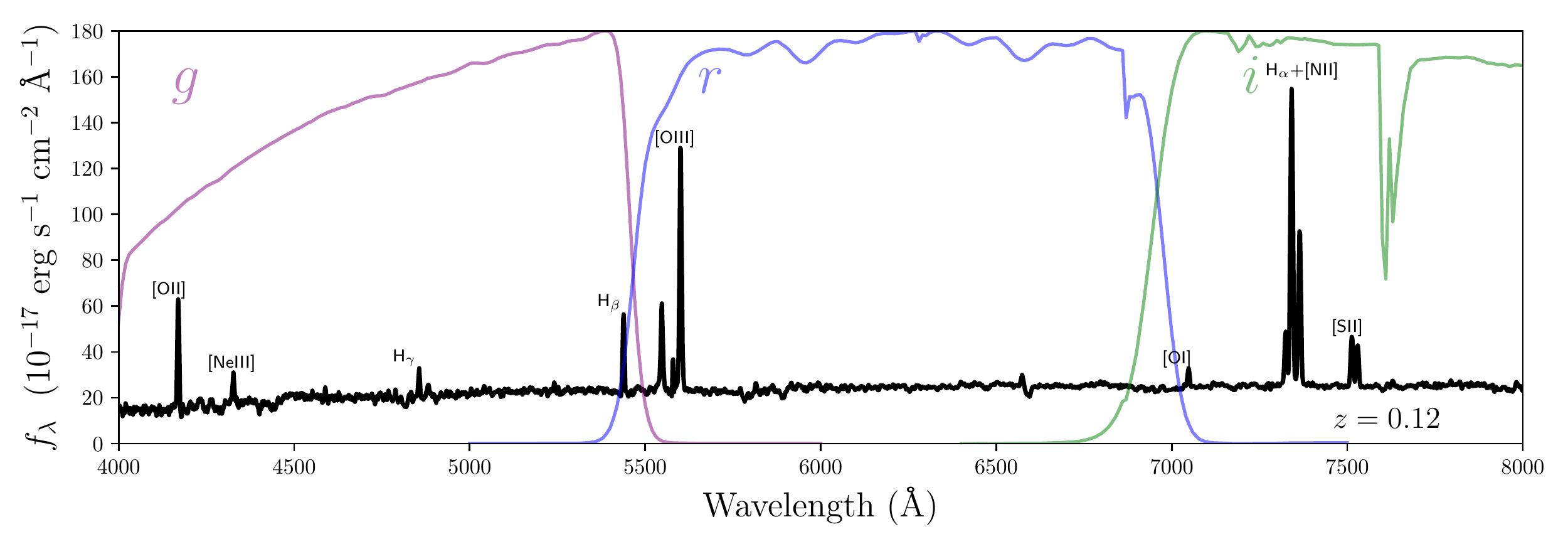}
    }
\caption{Example of isophotal measurements at a faint isophote of 1 \uintensity{}. 
{\it Left}: the HSC composite image and the \oiiil{} emission line map of an AGN in a star forming galaxy. {\it Right}: its SDSS spectrum. The isophote, shown in white thin contours, captures the galactic star-forming spiral arms. The thicker isophote at 3\uintensity{} also seems to trace galactic structures. The hole at the center of the \oiii{} map is due to image saturation of the nucleus in the $r$-band. 
}
\label{fig:lowcut}
\end{figure*}

We visually inspect the isophotes of our \oiiil{} emission-line map and find that, particularly for low luminosity AGN, the isophote at 1 \uintensity{} often outlines galactic structures. 
The left panel of Fig. \ref{fig:lowcut} shows one such example. 
SDSS J223641+000234 is a moderate luminosity AGN (\lbolfifteen{} = $10^{45.2}$ \ergs{}) with star formation. 
The thin white contour at the isophote of $1$ \uintensity{} traces the star-forming spiral arms. 
Although it is not entirely certain whether such galactic structures are due to star formation or continuum color variations, it is clear that the 1 \uintensity{} isophote may not be a robust tracer of the AGN emission-line region, at least with our broadband imaging methodology. 

We experiment with higher isophote cuts and find that 3 \uintensity{} is faint enough that it still captures extended emission-line regions, but is significantly less affected by galactic structures. 
As will be discussed in Sec. \ref{sec:res:sizelum}, there is a strong correlation between the isophotal area at 3 \uintensity{} and the luminosity of the AGN, suggesting that the \oiiil{} emission above this surface brightness is AGN dominated, while an isophote threshold of 1 \uintensity{} gives much a weaker correlation. 
Given the trade-off between sensitivity and robustness, we believe 3 \uintensity{} is a sensible choice for uniform measurements among all AGN in our sample and therefore adopt it for the primary measurements in this paper. 

\subsection{Area and Radius Measurements} \label{sec:method:iso}

Given the isophotal threshold of \oiiil{} at 3 \uintensity{} rest-frame, we define the emission-line region area as the area enclosed by the isophote, and the radius as the distance from the galactic centroid to the most distant point within the isophote. 
However, as the isophotes are affected by noise and occasionally include other nearby sources on the image (galaxies or stars), we impose two additional constraints on the selection of the isophotes. 

First, to exclude nearby sources, such as companion galaxies and foreground stars, we define a central region with a radius of 5\arcsec{}. Any isophotal contour that does not overlap with this central region is excluded. 
If the signal coming from companion galaxies or tidal tails is ionized by the AGN such as in \citet{Lintott2009} and \citet{Villar-Martin2010}, it perhaps should be included. 
These ionized regions tend to be closer to the AGN so would not be excluded. 
Our visual inspection finds that 19 objects ($6\%$) of the objects have companions or foreground stars within a radius of 5\arcsec{}, but the number is small enough that it does not impact the results significantly. 

Second, positive noise fluctuations could be interpreted as signal and thus introduce a positive bias to the area measurement. This could be potentially problematic as the bias is dependent on the noise level and thus on the redshift and the AGN luminosity.  To mitigate this effect, we exclude isolated isophote regions that have a connected area smaller than 5 pixels (0.14 arcsec$^2$). Real signals have a finite point-spread function that is typically larger than 0.4 arcsec$^2$ and thus would not be excluded with this area limit. 

With these definitions, Fig. \ref{fig:map_example} shows an example of the selected isophote contour in thick white lines. 
The \oiii{} region is considered unresolved with respect to the area (radius) measurement if the area (radius) is less than the PSF FWHM squared (PSF FWHM). 

There are a number of sources of uncertainty in the isophotal measurements. 
We estimate the uncertainties that are due to noise and point spread function using a set of simulations as described in Appendix \ref{sec:append:sim}. 
The simulations are based on area measurements. 
We find little systematic bias but random scatter due to both the noise and the finite size of the PSF. We conservatively assign the uncertainties due to the image noise to be proportional to the square root of the area ($A$), and the uncertainty due to the PSF to be 
\begin{equation}
\sigma_{\log{A}} = \log(1+\mathrm{PSF_{FWHM}}/\sqrt{A}).
\label{eq:err:psf}
\end{equation} 
The final error in the area is taken as the quadratic sum of these two error terms in log space. The median error in area is 0.3 dex.  The errors in the log radii are taken as half of the errors in log area. 

These errors do not account for other sources of uncertainties such as residuals in the continuum subtraction, emission lines from star formation, image saturation of the galaxy nucleus in low redshift systems, and contamination from foreground stars or companion galaxies. 
Some of these uncertainties only affect a small subset of the sample. 
For area measurements, these unaccounted uncertainties should be subdominant compared to the assigned error for most galaxies. But as the radius measurements are dominated by the most distant points and thus more susceptible to, for example, companion galaxies and noise, the assigned errors may be underestimated.

\section{Results} \label{sec:res}
\begin{figure*}
    \centering
    \hbox{
    \includegraphics[width=3.5in]{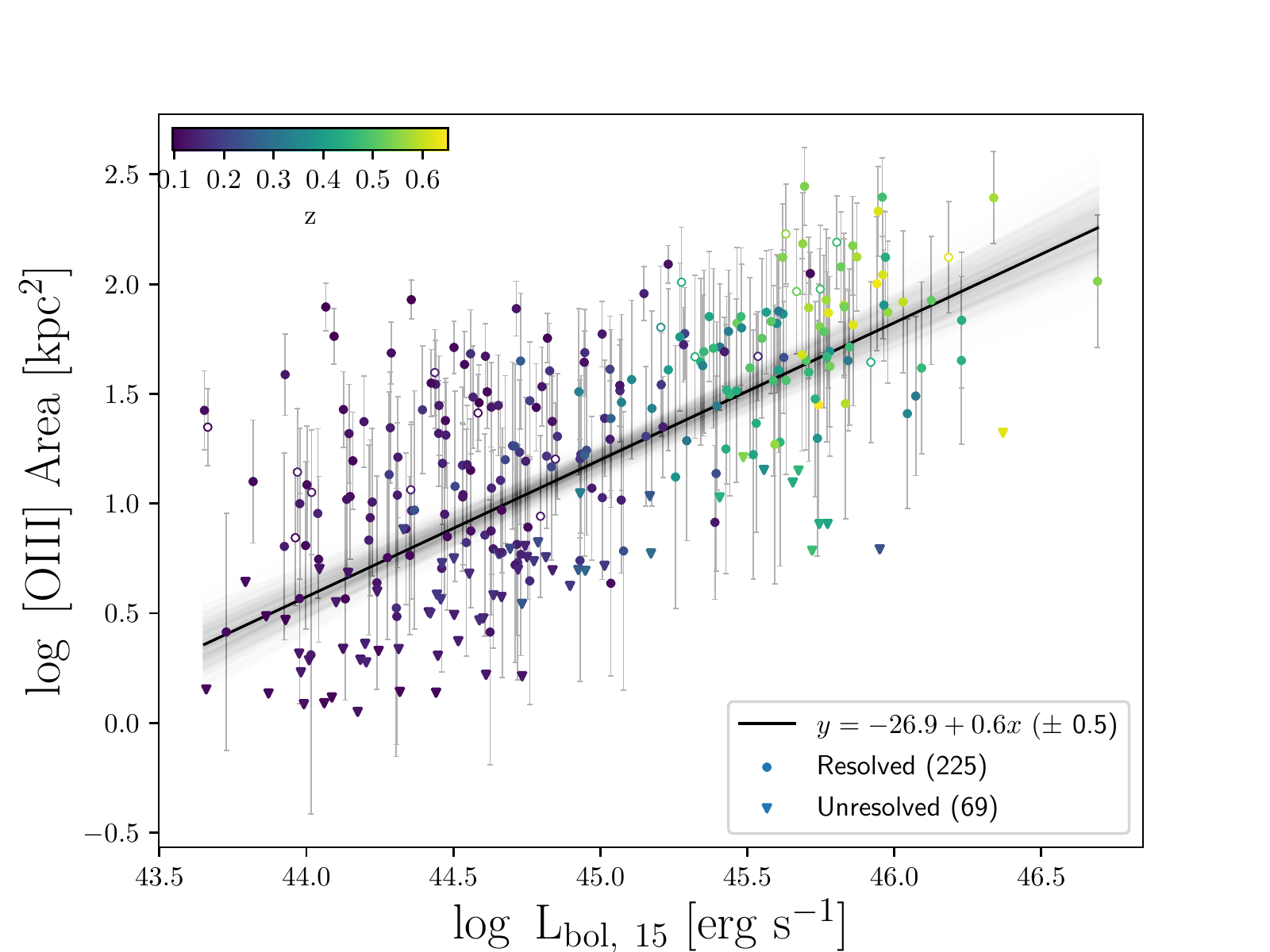}
    \includegraphics[width=3.5in]{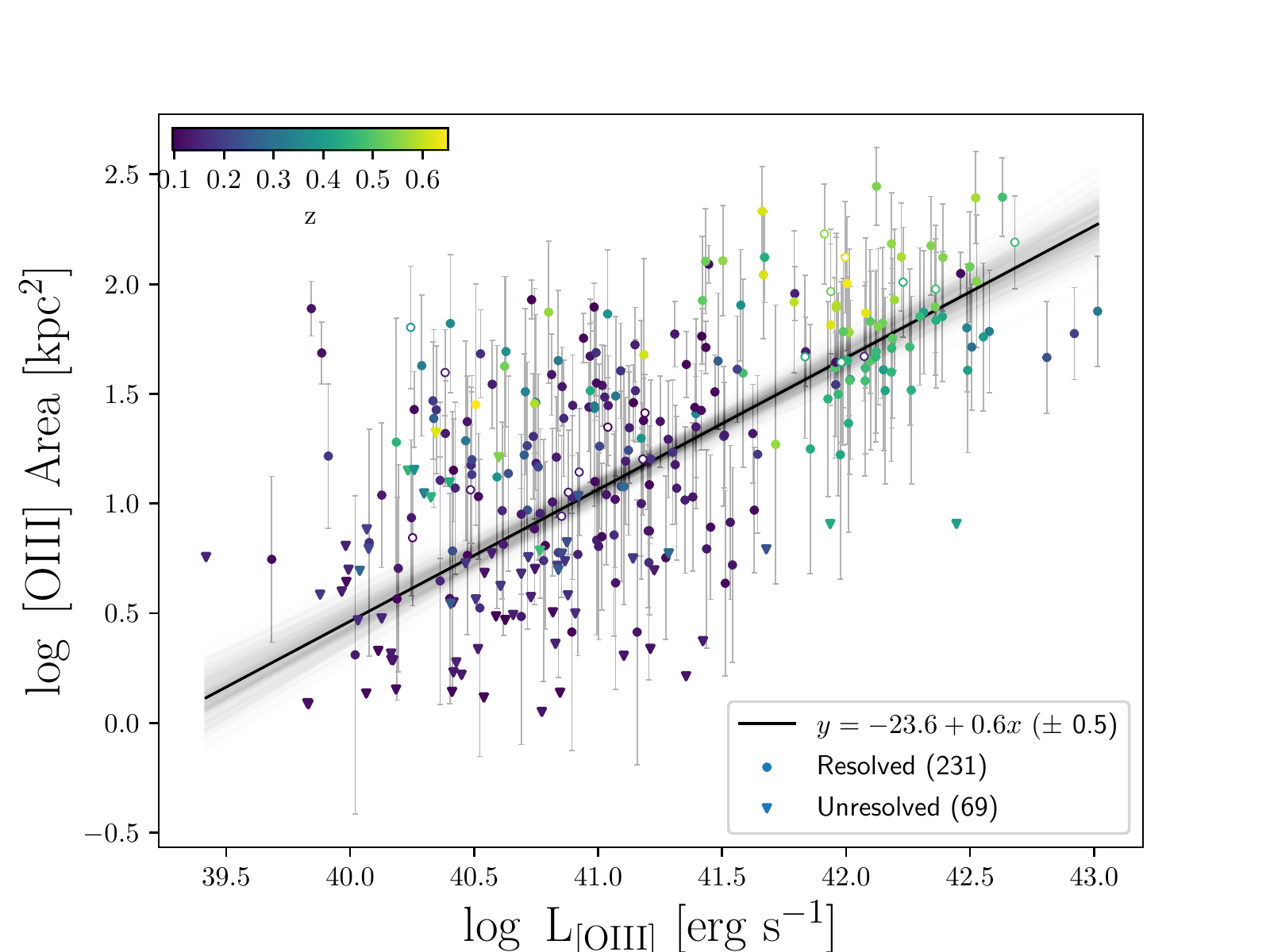}
    }
    \hbox{
    \includegraphics[width=3.5in]{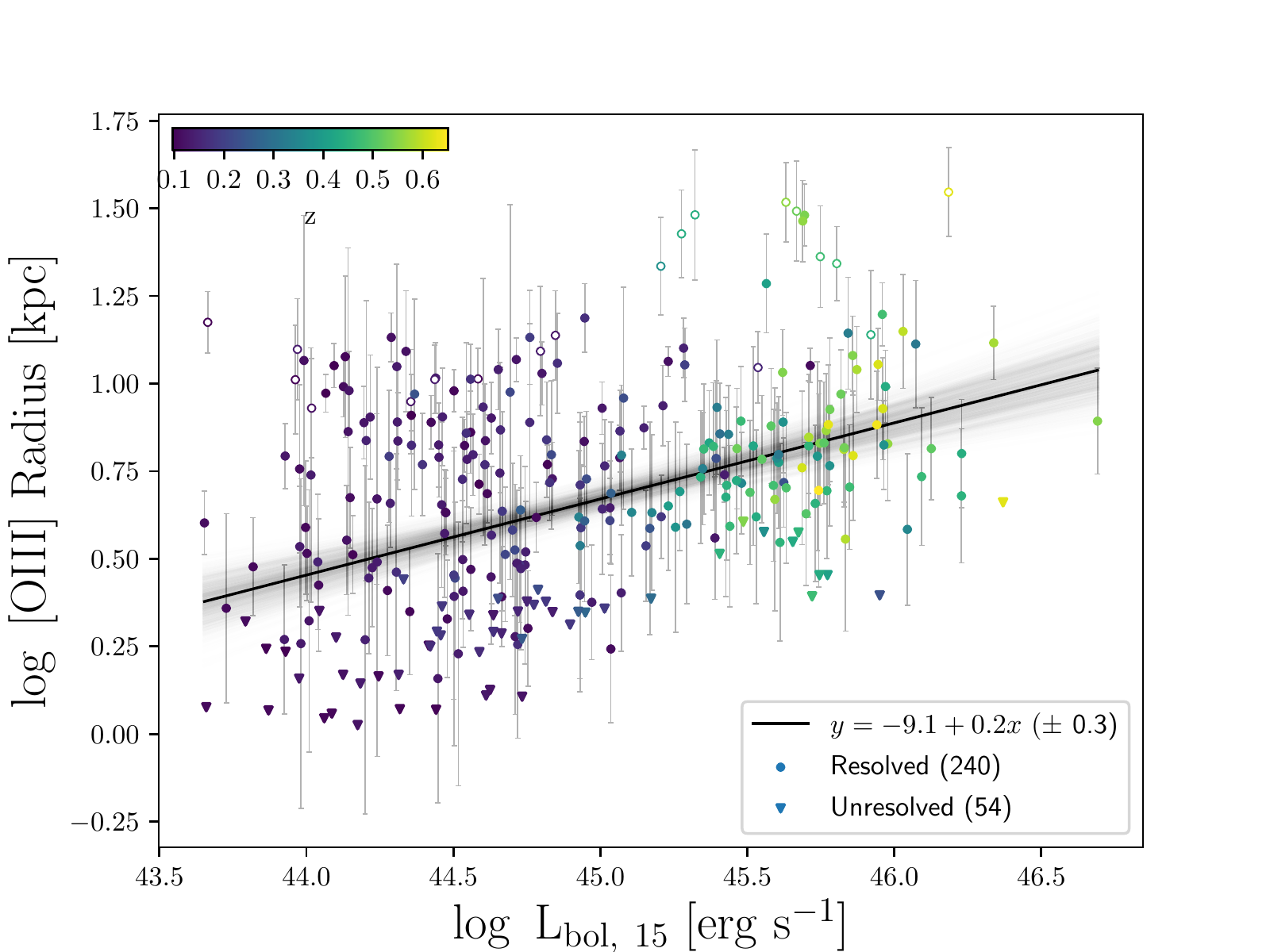}
    \includegraphics[width=3.5in]{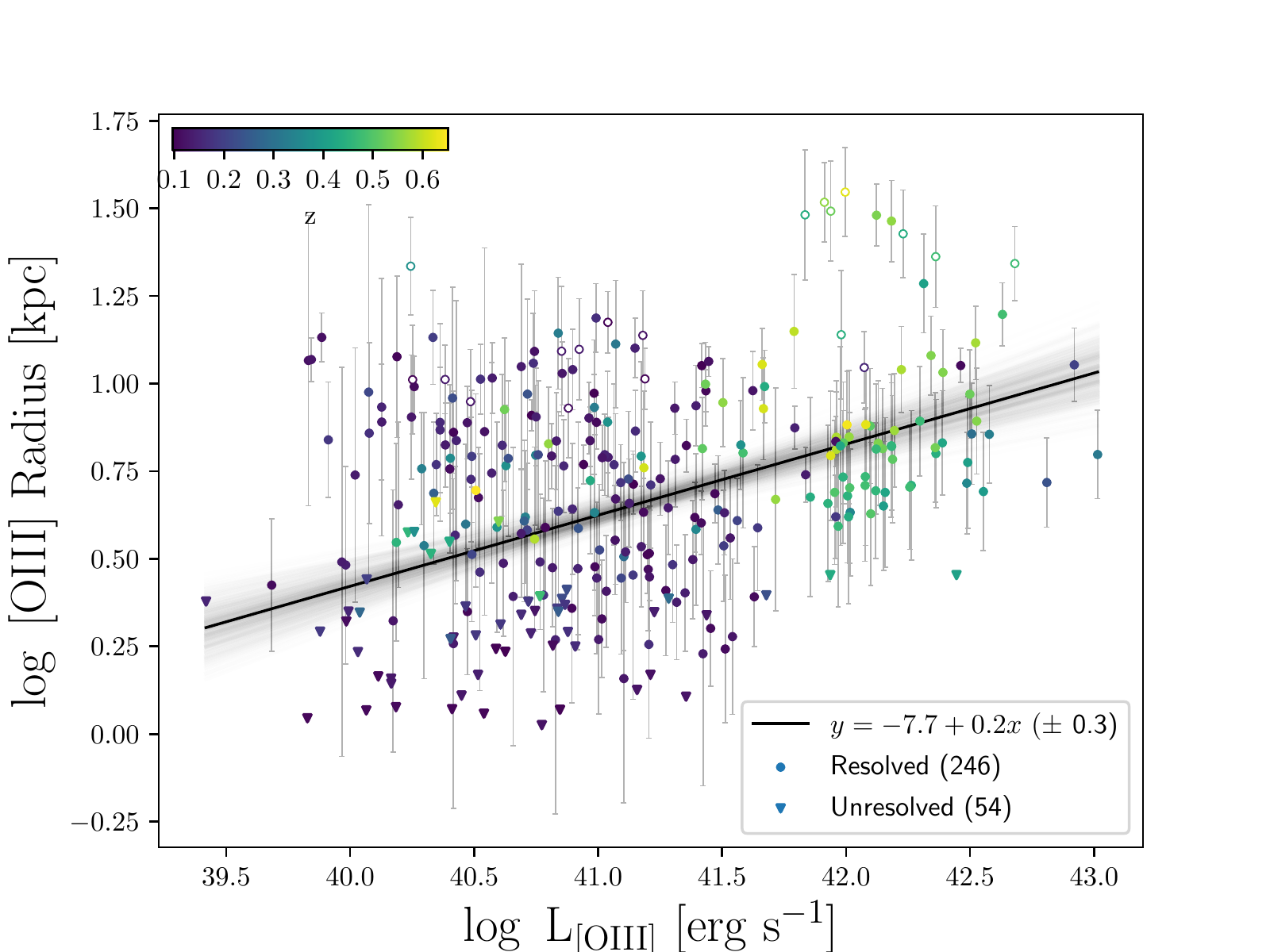}
    }
    \hbox{
    \includegraphics[width=3.5in]{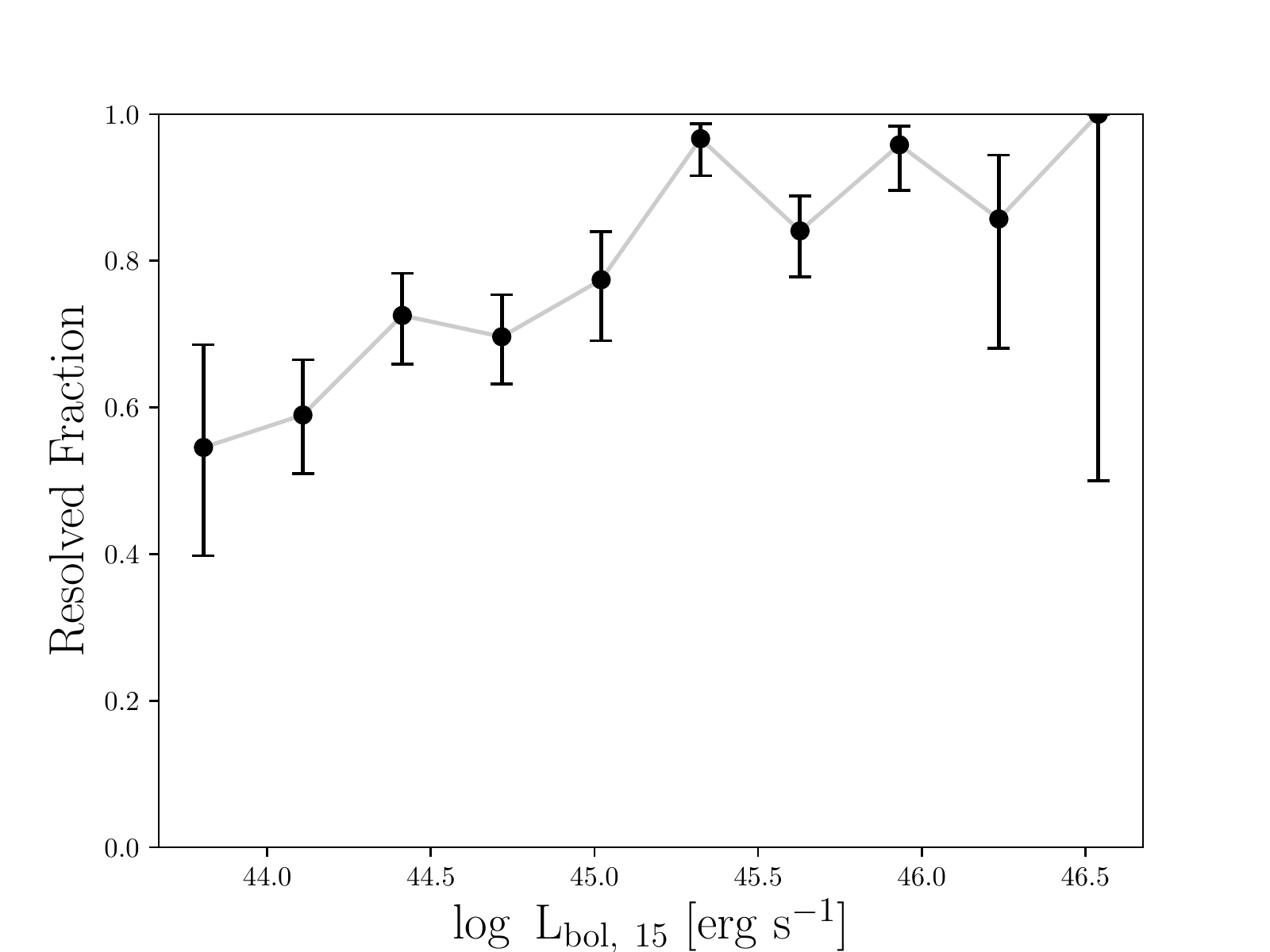}
    \includegraphics[width=3.5in]{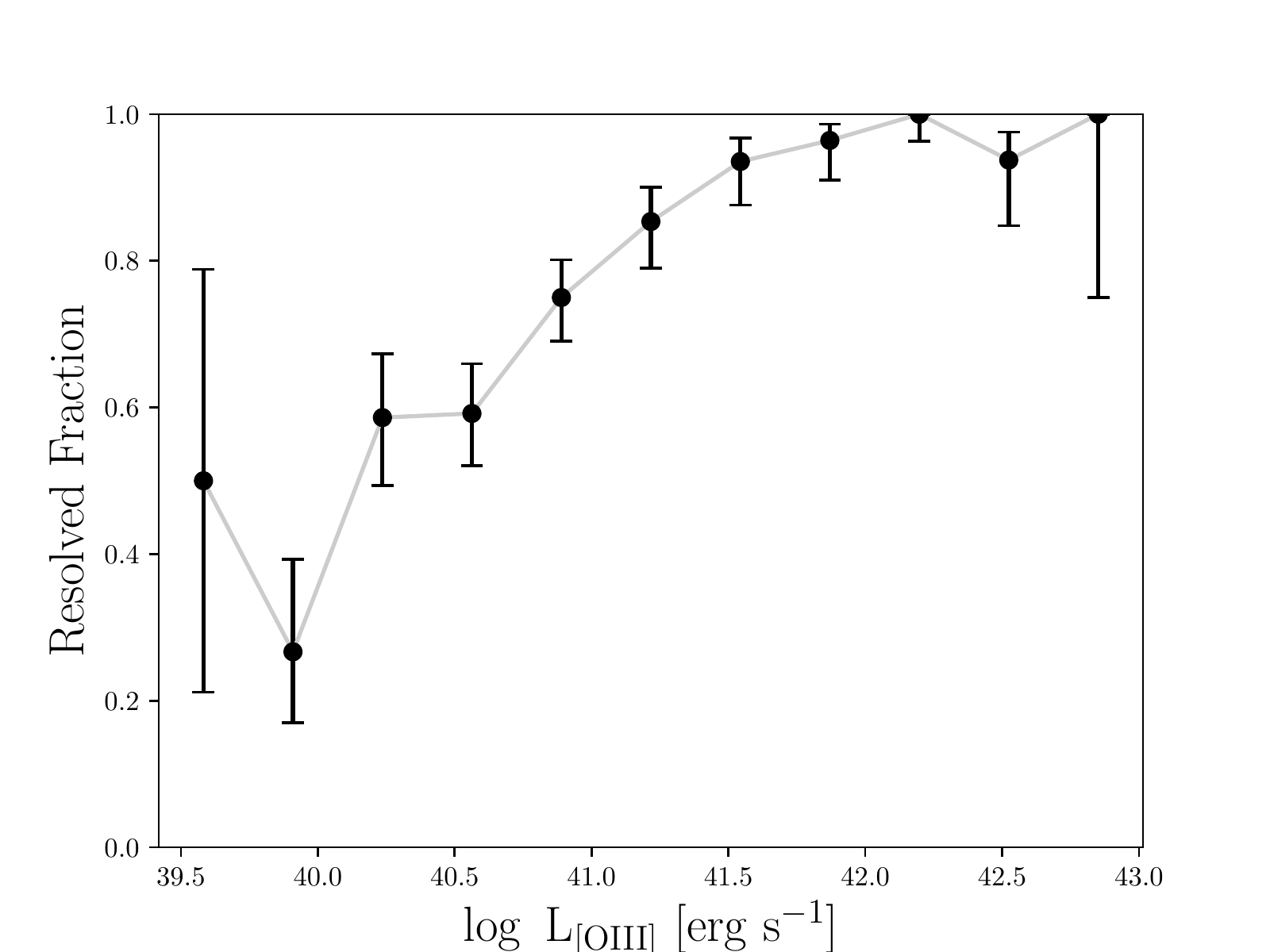}
    }
    \caption{The size -- luminosity relations of AGN \oiiil{} emission-line regions, as described in Sec. \ref{sec:res:sizelum}. 
    The top four panels are isophotal area ({\it top}) or radius ({\it middle}) as a function of AGN bolometric luminosity inferred from rest-frame 15 \micron{} luminosity ({\it left}) or \oiiil{} luminosity ({\it right}).  
	Circles are resolved objects and triangles, which have no error bars, are unresolved or undetected. 
    Empty symbols are systems with companion galaxies or foreground stars such that the area and radius may be overestimated. 
    The color coding indicates redshift. 
    The black line is the best-fit power-law with its grey shallow showing its uncertainty (posterior distribution). 
	{\it Bottom:} 
	The fraction of resolved objects, i.e., with area larger than PSF FWHM squared, as a function of bolometric luminosity inferred from 15 \micron{} luminosity ({\it left}) or \oiiil{} luminosity ({\it right}). The error bars show the 68$\%$ confidence Wilson score interval \citep{Wilson1927}. 
    }
    \label{fig:sizelum}
\end{figure*}

\begin{figure*}
    \centering
    \includegraphics[width=3.5in]{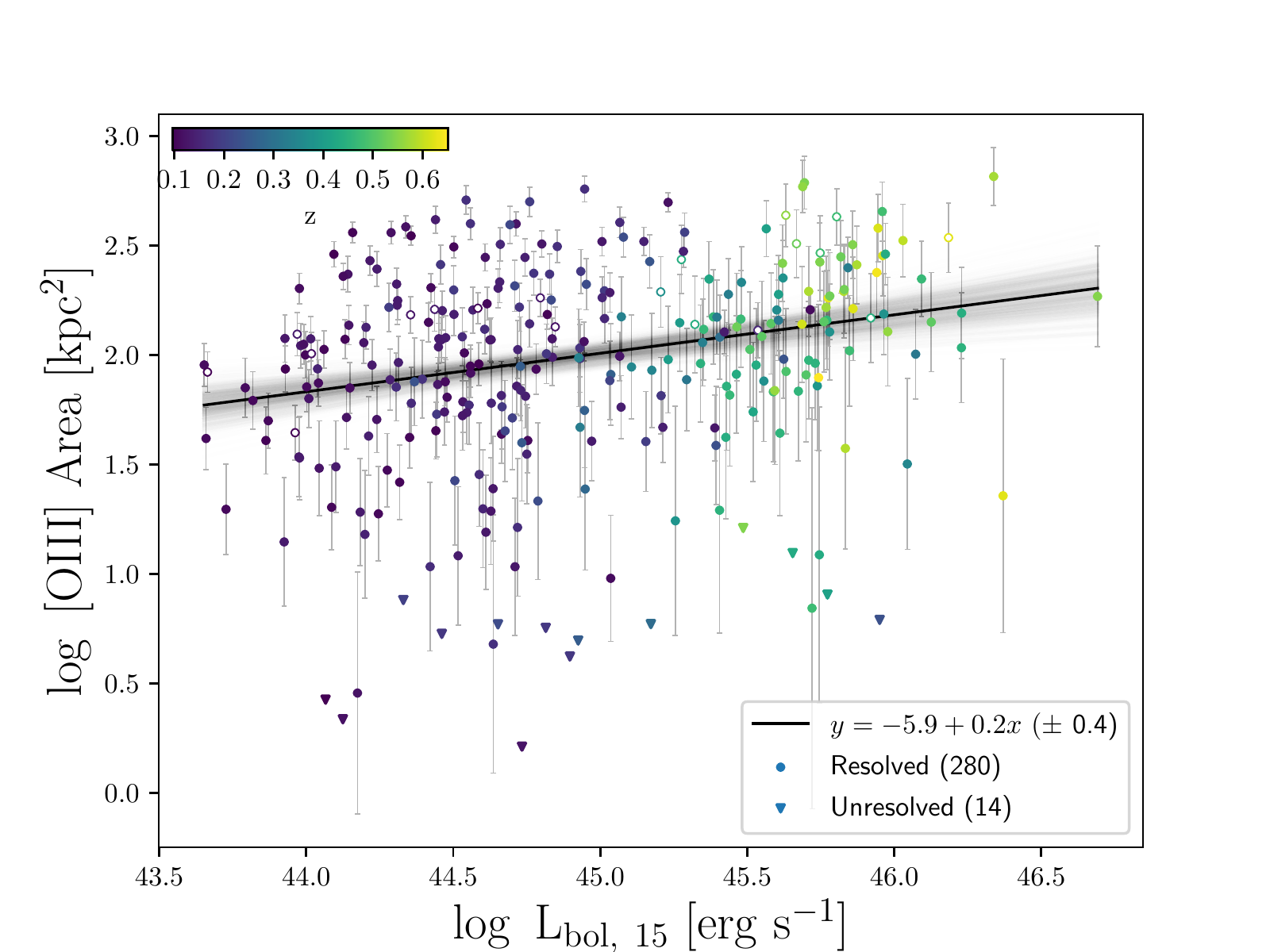}
    \caption{A comparison of size -- luminosity relations similar to the {\it upper left} panel of Fig. \ref{fig:sizelum} but with \oiii{} area measured at a lower isophote level of 1 \uintensity{}. The correlation is much weaker, suggesting that the signal at this isophote level is not dominated by AGN powered emission lines, see discussion in Sec. \ref{sec:method:isolevel}. 
    }
    \label{fig:sizelum_lowcut}
\end{figure*}

\begin{table*}
\begin{center}
\caption{Regression of Size -- Luminosity Relations}
\begin{tabular}{ccccccc}
\hline
\hline
Relation & Intercept $\alpha$  & Slope $\beta$ & Intrinsic Scatter $\epsilon$ & Pearson's $r$ & $p$-value \\
\hline
Area $- \lbolfifteen{}$    & $-26.91^{+2.52}_{-2.47}$ & $0.62^{+0.05}_{-0.06}$ & $0.51^{+0.03}_{-0.03}$ & 0.68 & $2.15\times10^{-31}$ \\
Area $- \loiii{}$          & $-23.55^{+1.94}_{-1.95}$ & $0.60^{+0.05}_{-0.05}$ & $0.49^{+0.03}_{-0.03}$ & 0.59 & $6.25\times10^{-23}$ \\
Radius $ - \lbolfifteen{}$ & $ -9.10^{+1.53}_{-1.52}$ & $0.22^{+0.03}_{-0.03}$ & $0.33^{+0.02}_{-0.02}$ & 0.32 & $5.69\times10^{-7} $ \\
Radius $ - \loiii{}$       & $ -7.69^{+1.22}_{-1.22}$ & $0.20^{+0.03}_{-0.03}$ & $0.33^{+0.02}_{-0.02}$ & 0.19 & $2.39\times10^{-3} $ \\
\hline
Area $- \lbolfifteen{}$  (w/o c.)  & $-27.61^{+2.70}_{-2.68}$ & $0.64^{+0.06}_{-0.06}$ & $0.53^{+0.03}_{-0.03}$ & 0.67 & $7.03\times10^{-28}$ \\
\hline
\end{tabular} \\
\textit{Note.}
The linear regression results and correlation coefficients of the size -- luminosity relations as described in Sec. \ref{sec:res:sizelum}. 
The first three columns are the intercept $\alpha$, slope $\beta$, and intrinsic scatter $\epsilon$ of the relation, and the numbers quoted are the posterior mean and the 68\% credible interval. The last two columns are the Pearson's $r$ correlation coefficient and its $p$-value calculated from only the resolved objects. In the first four rows, objects with companion galaxies or foreground stars (empty symbols in Fig. \ref{fig:sizelum}) are included. 
As a comparison, the last row shows the Area -- \lbolfifteen{} relation excluding these contaminated objects. The parameters are not significantly different. 
\label{tab:sizelum}
\end{center}
\end{table*}

\subsection{Size -- Luminosity Relation} \label{sec:res:sizelum}
More luminous AGN are expected to create larger narrow emission-line regions. In this section, we quantify this relation with our \oiii{} emission line map measurements. 
Fig. \ref{fig:sizelum} shows the relations between the sizes of the \oiiil{} emission-line regions, as measured by isophotal area and radius (Sec. \ref{sec:method:iso}), and the luminosities of the AGN. 
Two indicators of the AGN bolometric luminosities are used -- the rest-frame mid-IR 15 \micron{} luminosities from WISE (Sec. \ref{sec:data:wise}), and the \oiiil{} luminosities. 
Among the 300 galaxies in our primary sample, 6 do not have WISE detection in every band, resulting in 294 galaxies in the size - \lbolfifteen{} relations and 300 in the size - \loiii{} relations. 

We do not measure the \oiiil{} luminosity from our reconstructed \oiii{} map because the systemic uncertainty that may arise from saturation of the galactic nucleus, contaminations from star-forming regions, companion galaxies, and stars, can dominate over the random uncertainty and is hard to correct for.
Instead, the \oiiil{} luminosities are measured from the continuum subtracted SDSS and BOSS spectrum by integrating the line flux within a velocity range of $\pm 1400$ \kms{}. 
The SDSS and BOSS fibers are 3\arcsec{} and 2\arcsec{} in diameter, respectively, which is often smaller than the extended emission-line regions in our sample. 
Although the SDSS and BOSS spectrophotometry has an aperture correction applied such that the fluxes of point sources are correct, additional flux may be missed due to the extended sizes our objects. 
Therefore, the \loiii{} measurement is not homogeneous among the sample and aperture correction is needed for the BOSS sample, by an amount which will depend on the source radial profile. 
We here adopt a statistical approach using the \loiii{} -- \lbolfifteen{} relation to calibrate the amount of aperture correction required, as described in Appendix \ref{sec:append:aptcorr}. 
We apply an aperture correction of 0.7 dex to the \loiii{} of the \citet{Yuan2016} sample taken with BOSS spectrograph. After this correction, the \loiii{} and \lbolfifteen{} relation is linear for our primary sample, agreeing with what is seen in the \citet{Mullaney2013} parent sample (Appendix \ref{sec:append:aptcorr}). 
No extinction correction is applied to \oiiil{} because such a correction increases the scatter in the \loiii{} -- \lbolfifteen{} relation. 
However, due to the inhomogeneous nature of the \loiii{} measurements of our sample, we will concentrate on results based on \lbolfifteen{} for the conclusions of this paper. 

First, we find that the fraction of spatially resolved objects increases with the AGN luminosity, despite the fact that more luminous AGN are systematically at higher redshifts and are more affected by the noise and the PSF. 
Second, the measured areas and radii of the resolved objects increase with the luminosities in a statistically significant way. 
The Pearson's-$r$ correlation coefficient gives small $p$-values $\lesssim 3\times10^{-3}$, see Tab. \ref{tab:sizelum}. 
The correlations are not just driven by a small number of high luminosity objects. 
Correlations persist even after excluding the 30\% of the most luminous objects. 
Both of these findings are consistent with the expectation that more luminous AGN create larger and more prominent emission-line regions. 
This result also validates our assumption that our reconstructed \oiii{} maps trace AGN emission-line regions. 

To quantify the relations in Fig. \ref{fig:sizelum}, we adopt the Bayesian linear regression approach developed by \citet{Kelly2007}, which uses a Markov chain Monte Carlo sampling method and accounts for heteroscedastic errors, intrinsic scatter, and censored data. 
The model is a single power law plus Gaussian intrinsic scatter in log scale, 
\begin{equation}
\log(y) = \alpha + \beta \times \log(x) + N(\sigma = \epsilon{}), 
\label{eq:kelly}
\end{equation}
where the intercept $\alpha$, slope $\beta$, and the 1-$\sigma$ width of the Gaussian intrinsic scatter $\epsilon$ are free parameters. The dependent variable $y$ represents the \oiiil{} region area or radius, and the independent variable $x$ is the AGN luminosity -- \lbolfifteen{} or \loiii{}. 
We assume that the measurement errors in the isophotal area and radius are Gaussian in log scale with amplitudes quantified in Sec. \ref{sec:method:iso}, but that there are no uncertainties in the AGN luminosities \lbolfifteen{} or \loiii{}. 
Unresolved and undetected emission-line regions are treated as non-detections in area and radius with upper limits set to the size of the PSF. 

The resulting best-fit (posterior) relationships are plotted as black (grey) lines in Fig. \ref{fig:sizelum} and tabulated in Table \ref{tab:sizelum}. 
The posterior mean and the statisical uncertainty (68\% credible interval) of the slope is $0.62^{+0.05}_{-0.06}$ and $0.60^{+0.05}_{-0.05}$ for the Area - \lbolfifteen{} and the Area - \loiii{} relations, respectively. 
The slope of the Radius - \lbolfifteen{} and Radius - \loiii{} relations are $0.22^{+0.03}_{-0.03}$ and $0.20^{+0.03}_{-0.03}$, respectively. 
These quoted errors only represent the statistical uncertainties due to the PSF and noise, and could be smaller than the systemic errors, as we now discuss. 

First, companion galaxies or foreground stars could contribute to additional uncertainties, especially in the radius -- luminosity relations. 
The radius -- luminosity relations have a larger scatter and a larger number of outliers than the area -- luminosity relations. 
A closer inspection finds many of the outliers have companion galaxies or foreground stars which lead to over-estimation of the radius. 
Indeed, as radius only depends on the most distant point from the galaxy centroid, it is more susceptible to uncertainties on the outskirts. 
We visually inspect our sample and find that out of the 300 galaxies, 19 (6\%) still have measurements affected by contaminants that are close enough to the AGN such that they are not excluded by the 5\arcsec{} distance constraint (see Sec. \ref{sec:method:iso}). Marked as empty symbols in Fig. \ref{fig:sizelum}, these systems are preferentially outliers in the radius -- luminosity relations, although this is a small fraction and excluding them does not change the fits significantly, see the bottom row of Tab. \ref{tab:sizelum}. 
Some of these systems could also be companion galaxies ionized by the AGN so may carry astrophysical significance. 
However, this highlights the fact that the radius measurements, in the context of broadband imaging observations, are more susceptible to contaminants than the area measurements. 
We therefore adopt area as the primary quantity for our discussions of the size -- luminosity relation in Sec. \ref{sec:discussion}. 

Emission lines from star formation regions and uncertainties from stellar continuum subtraction can contribute additional uncertainty. 
As discussed, we adopt a higher isophotal threshold to mitigate some of these issues.  
As a comparison, we show in Fig. \ref{fig:sizelum_lowcut} that the area -- luminosity relation with a lower isophote of 1 \uintensity{}, a value used by previous spectroscopic studies \citep{Liu2013b,Hainline2013,Sun2017}. The correlation becomes weak, suggesting that the signal at this isophote level is not dominated by AGN powered emission lines. 
Such effects may also be present in the current relations with higher isophotes, but should be limited to the systems with the highest star formation rates. 
The stellar continuum subtraction is subject to a 10\% uncertainty because of the uncertainty in determining the continuum color, see \ref{sec:method:conti}. Galaxies with strong continuum color variation may have stronger continuum subtraction residuals but should only constitute a small fraction of the sample. 

The fitted relations do not take the errors in the luminosities into account. Unlike the dependent variable, measurement errors in the independent variable bias the slope to lower values.  The uncertainties in \lbolfifteen{} are dominated by the WISE W4 photometry, which has a typical uncertainty of 0.1 dex for the objects in our sample. The bolometric correction may introduce additional scatter. 
We find that a conservative error of 0.2 dex in \lbolfifteen{} would increase the slope of the Area - \lbolfifteen{} relation from 0.62 to 0.69. 
So the errors in the luminosities contribute to a systemic error of $<$0.1 in the slope. 

The intrinsic scattering determined from the MCMC fit of the area -- luminosity relation is 0.5 dex in area. The fit only takes into account observational uncertainties that arise from image noise and PSF, so the systemic uncertainties we have been discussing should account for some of the 0.5 dex of scatter. 
Having said this, we expect that true intrinsic scatter does exist, due to variations in the extent and density profile of the interstellar medium, AGN variability, viewing angles among the sources, and so on. 

We do not find significant differences in the slope when we use \lbolfifteen{} and \loiii{} as the AGN luminosity indicator. 
But this could be an result of the aperture correction applied to the \oiiil{} measurements, which is calibrated to the \loiii{} -- \lbolfifteen{} relation. 
In addition, as each of the four parent samples of type 2 AGN have different selection criteria, selection effects on \loiii{} are inhomogeneous and hard to quantify. 
These effects would introduce additional systematics to the size -- \loiii{} relations over the \lbolfifteen{} based variations. 
Indeed, we find that the correlations in the size -- \lbolfifteen{} relation are stronger and more significant, see Tab. \ref{tab:sizelum}. 
Thus, we take the slope of the Area -- \lbolfifteen{} relation as representative for the size -- luminosity relation. 

In summary, our best estimates of the size -- luminosity relation comes from the Area -- \lbolfifteen{} relation. In addition to random uncertainties due to image noise and the finite PSF, the slope of this relation is also subject to a number of systemic uncertainties arising from, for example, companion galaxies, foreground stars, star formation, and uncertainties in the AGN luminosities, which altogether may account for a uncertainty of $\pm 0.1$ in the slope.  
Our estimate for the slope of the Area -- \lbolfifteen{} relation is thus $0.62^{+0.05}_{-0.06}\pm0.10$, where the errors represent statistical and systemic uncertainties, respectively. 
The presence of a strong correlation suggests that our reconstructed \oiii{} map indeed traces extended emission-line regions. 
The implications of this result are discussed in Sec. \ref{sec:discussion}.

\subsection{Extended Emission-Line Region Candidates} \label{sec:res:candidates}

\begin{figure*}
    \centering
    \vbox{
    \hbox{
    \includegraphics[width=3.in]{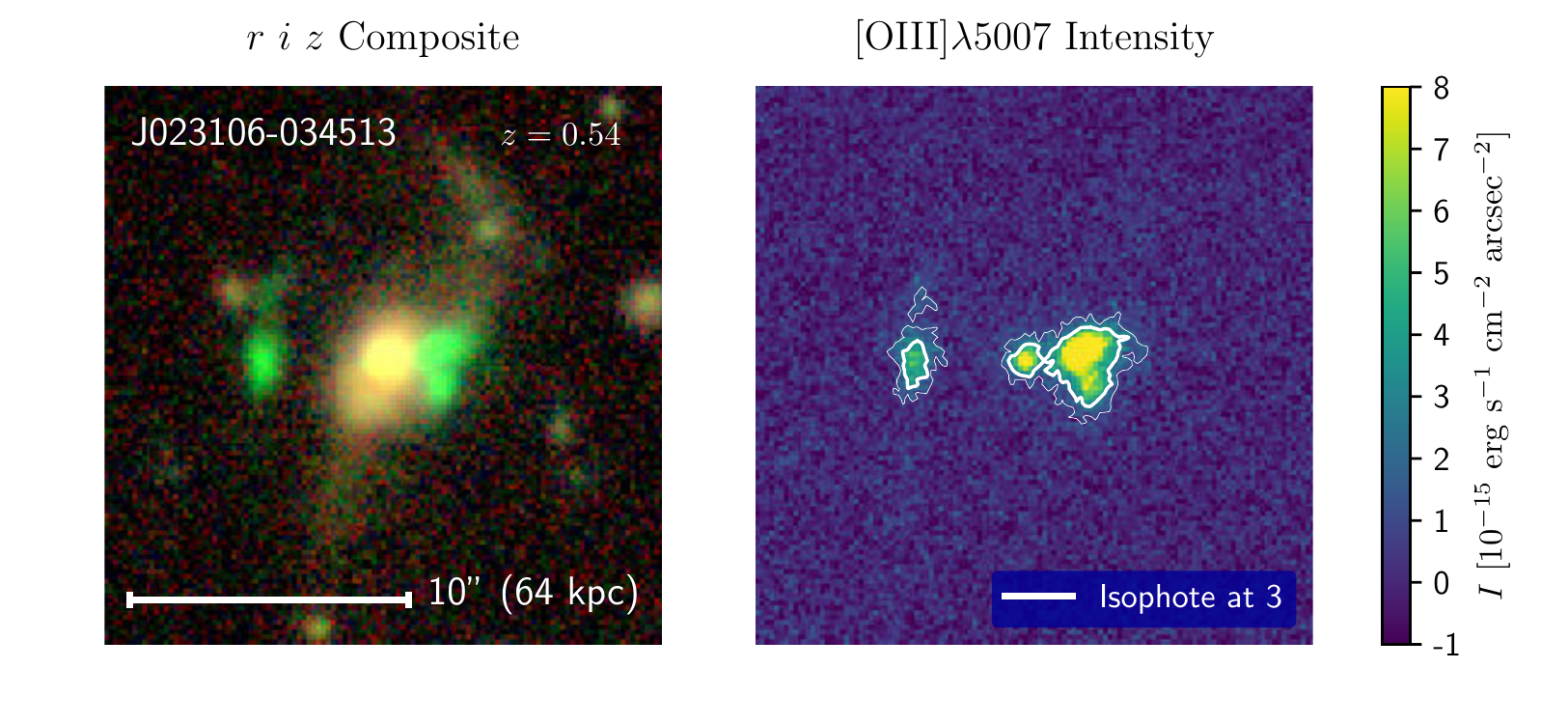}
    \includegraphics[width=3.75in]{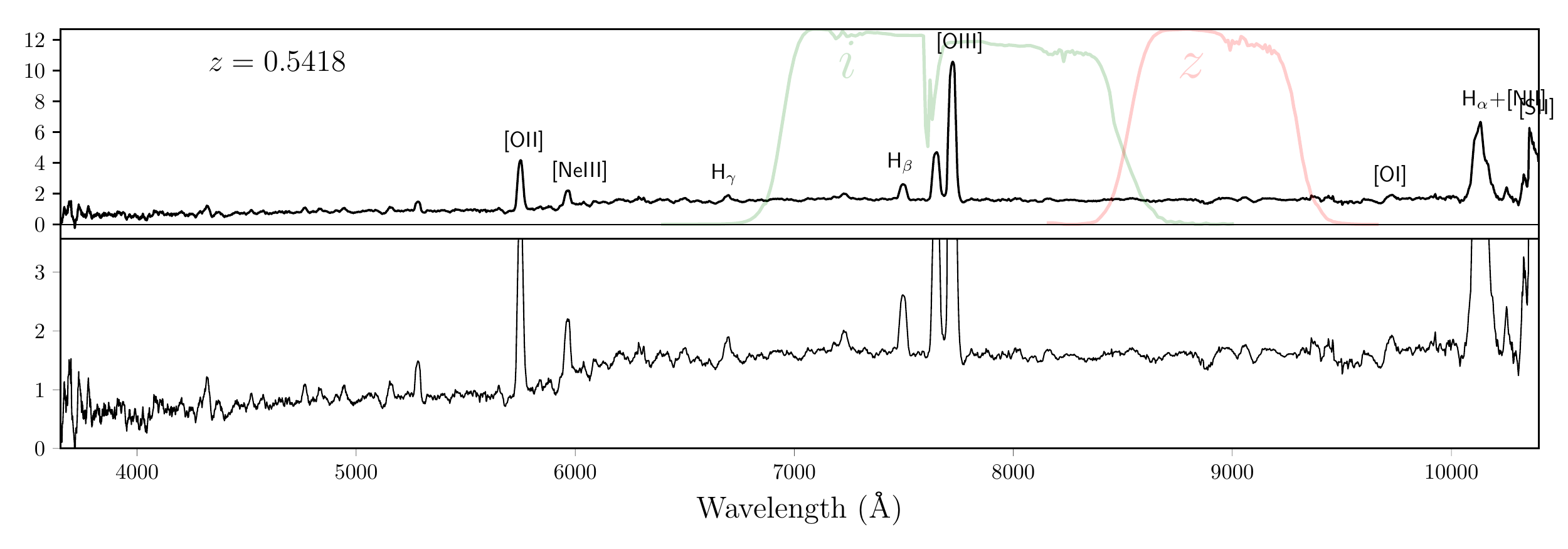}
    }
    \hbox{
    \includegraphics[width=3.in]{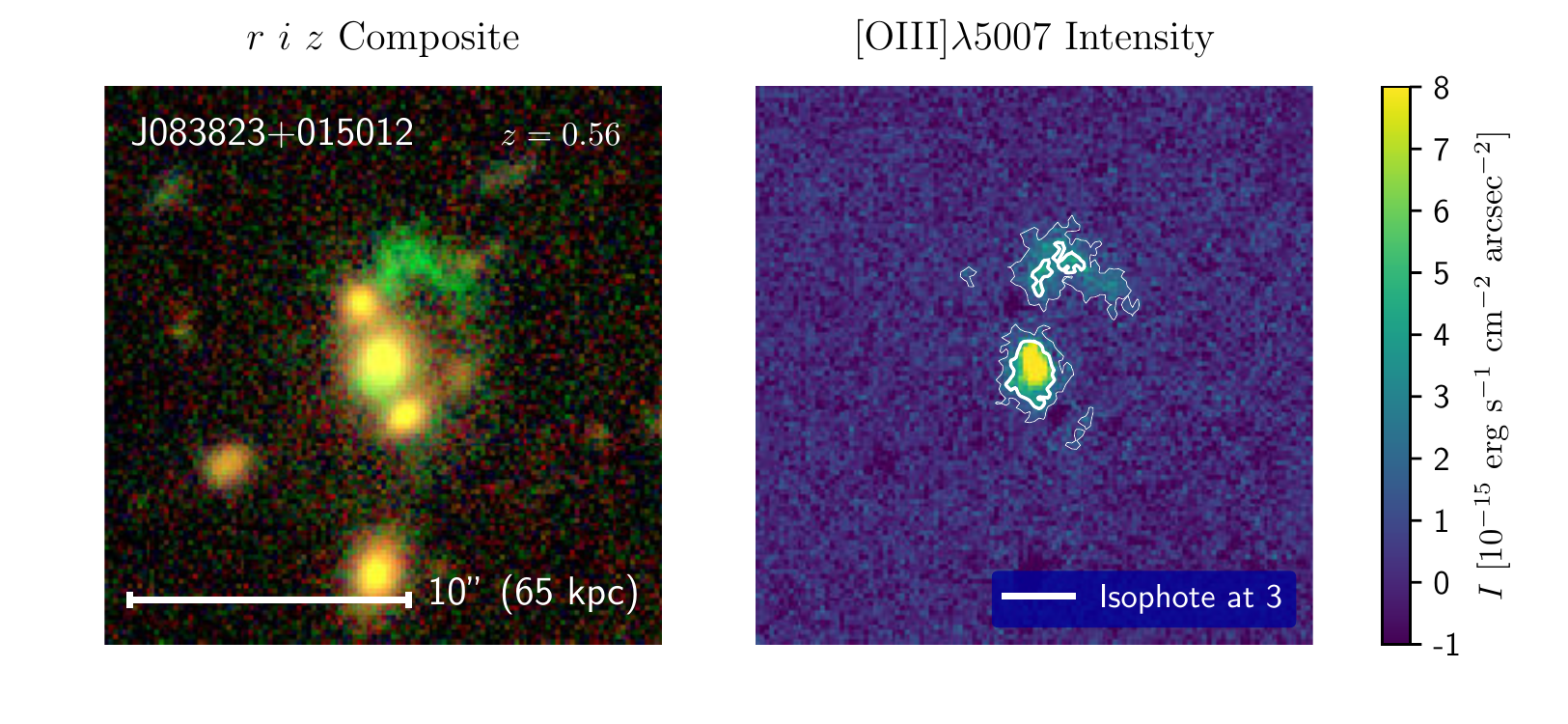}
    \includegraphics[width=3.75in]{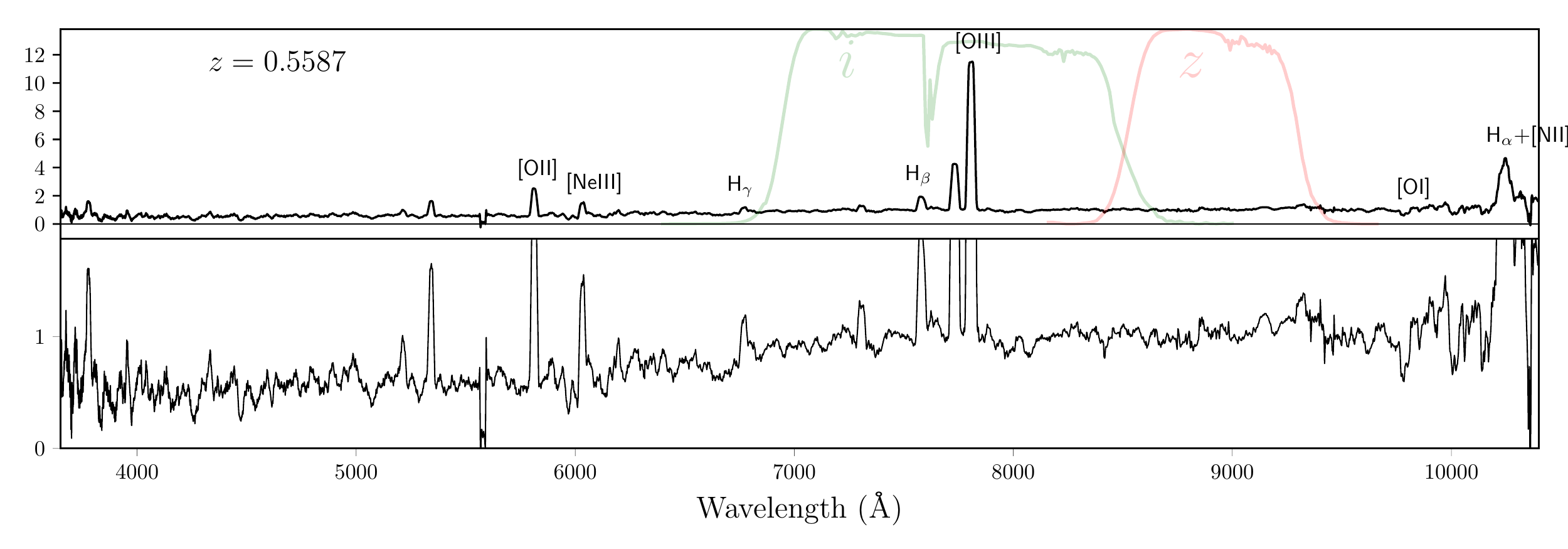}
    }
    \hbox{
    \includegraphics[width=3.in]{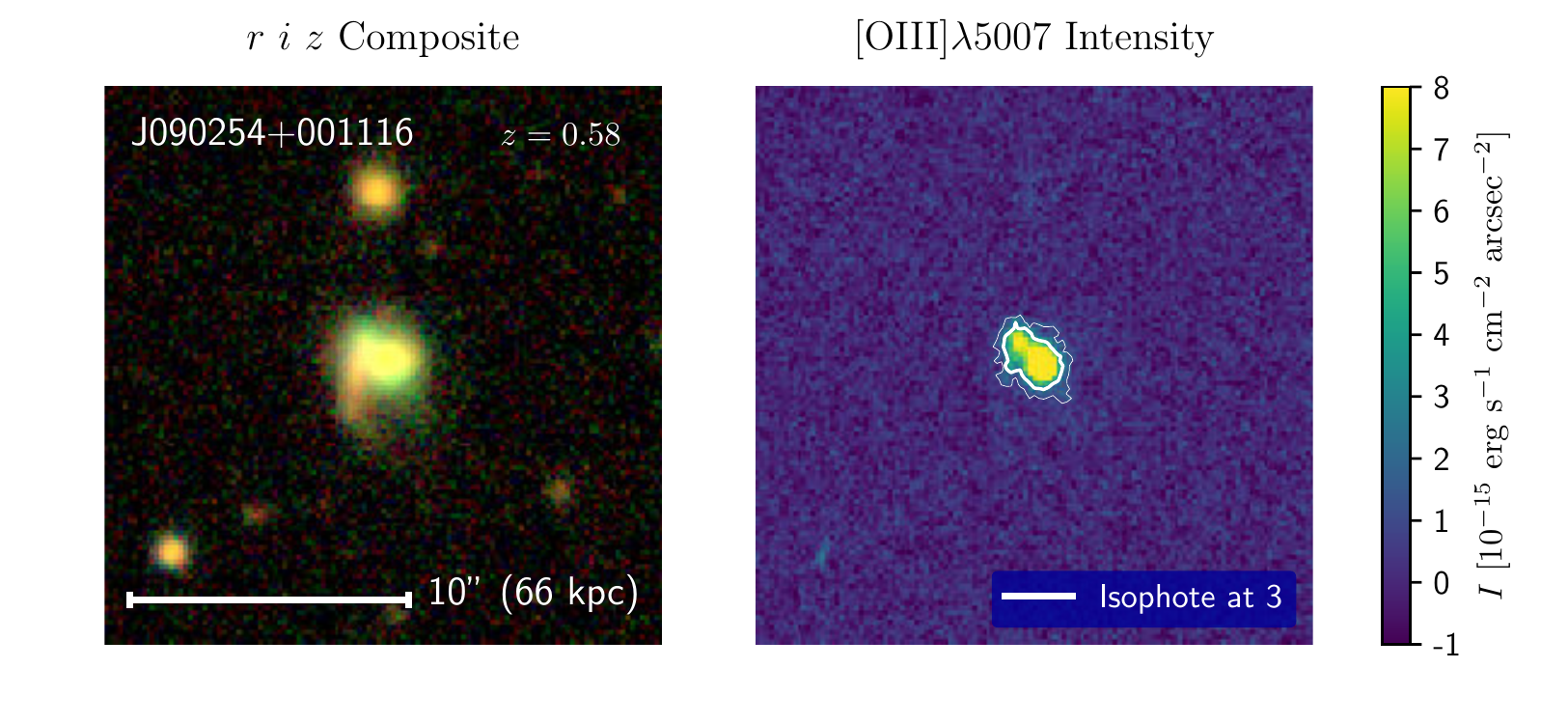}
    \includegraphics[width=3.75in]{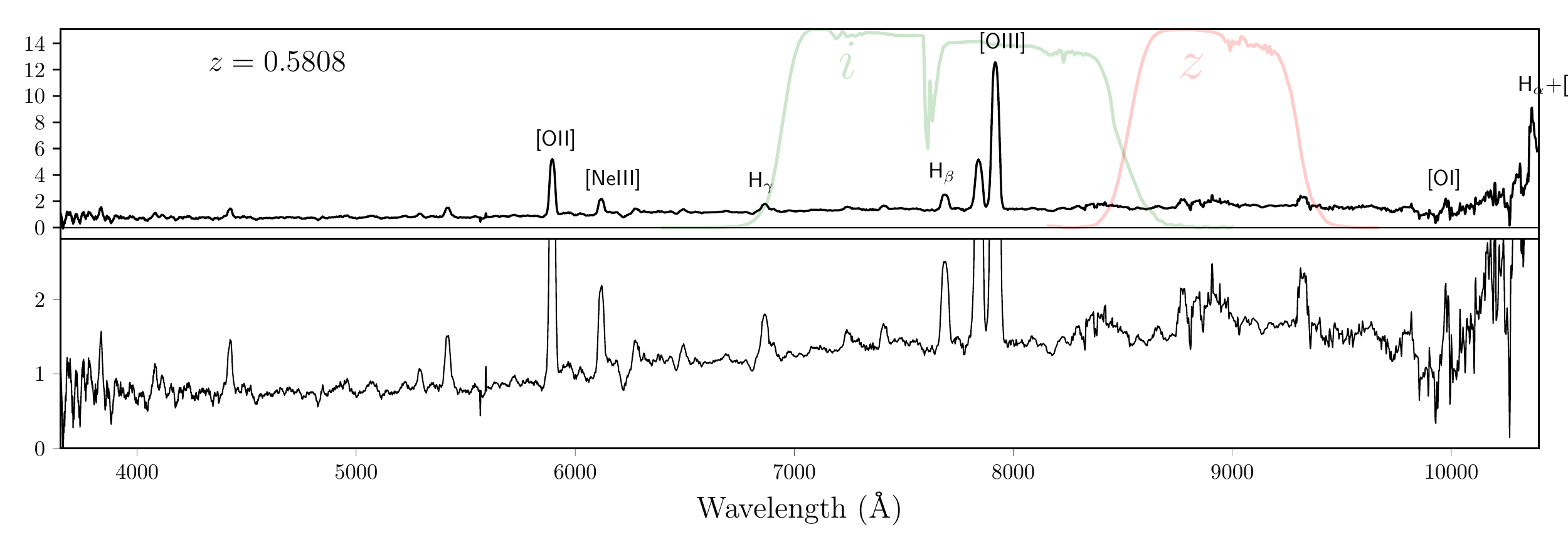}
    }
    \hbox{
    \includegraphics[width=3.in]{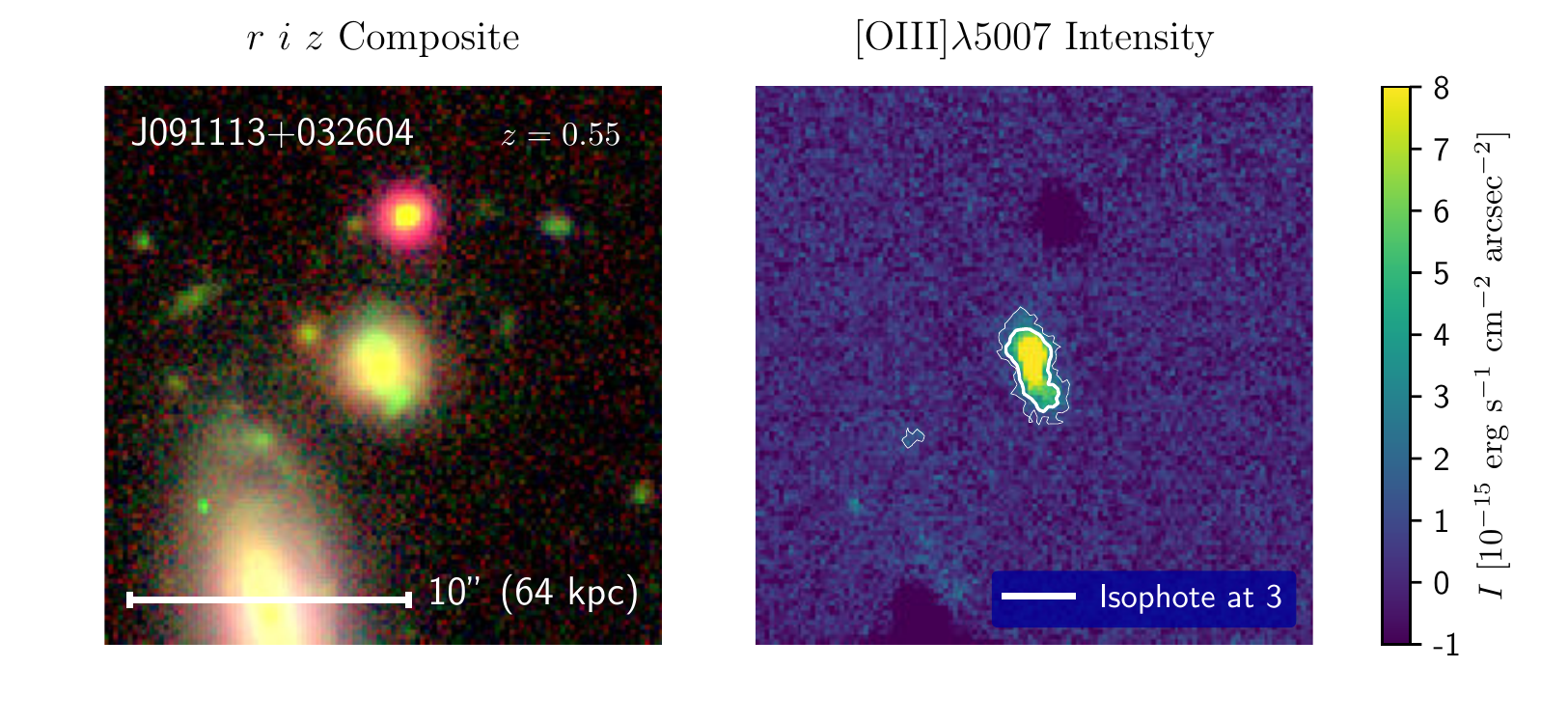}
    \includegraphics[width=3.75in]{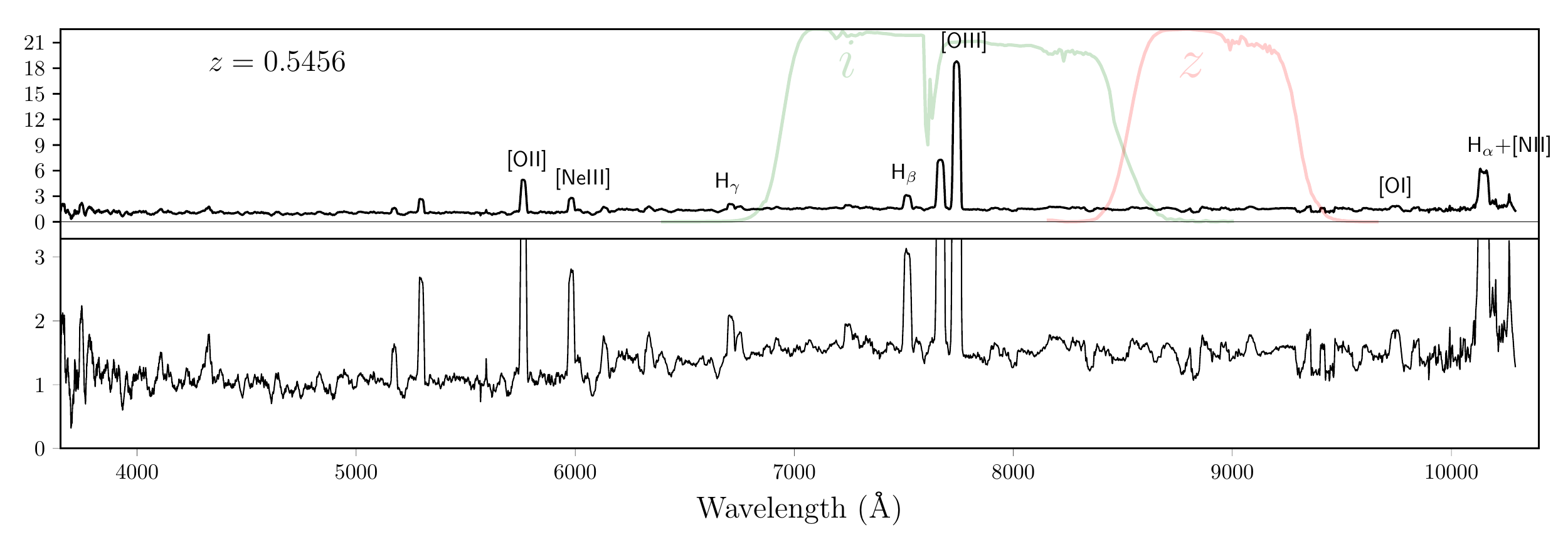}
    }
    \hbox{
    \includegraphics[width=3.in]{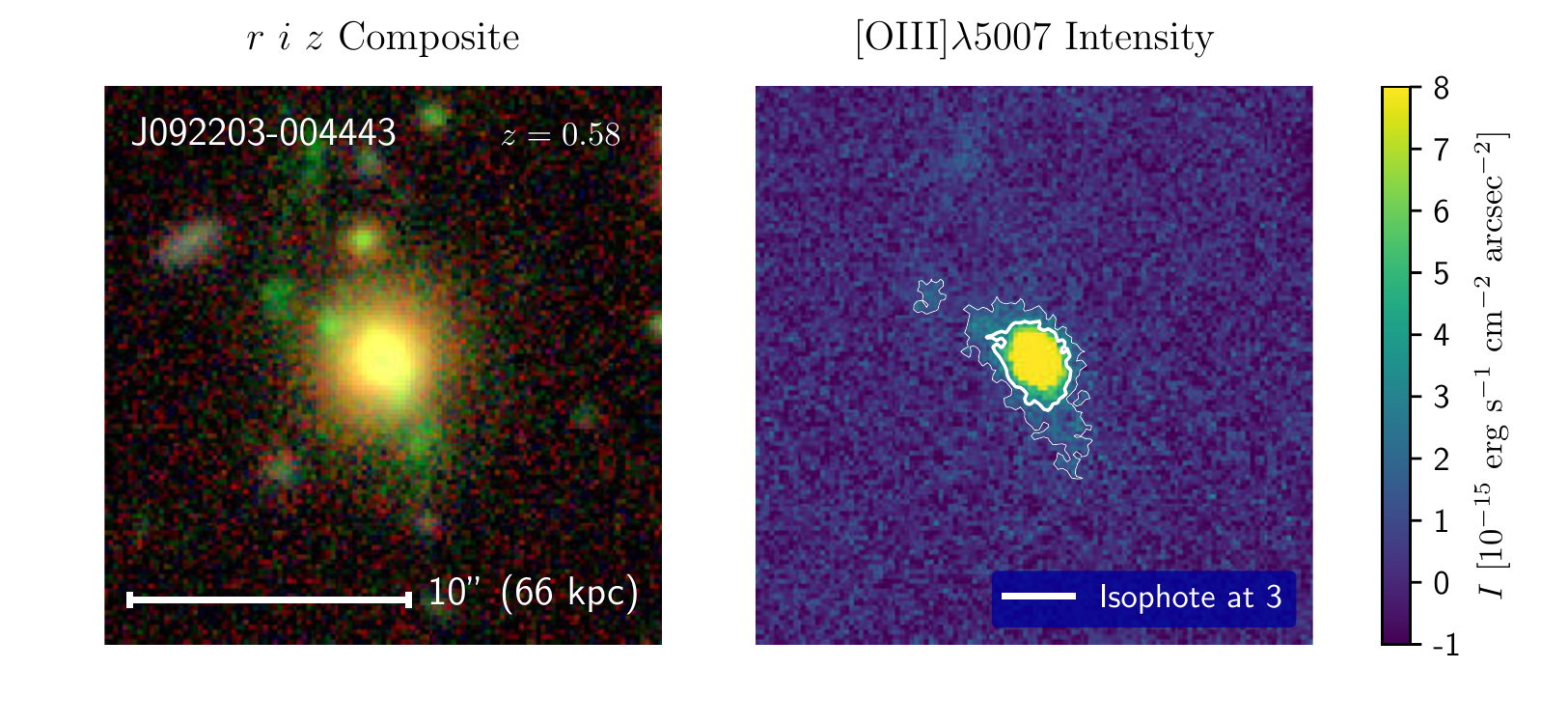}
    \includegraphics[width=3.75in]{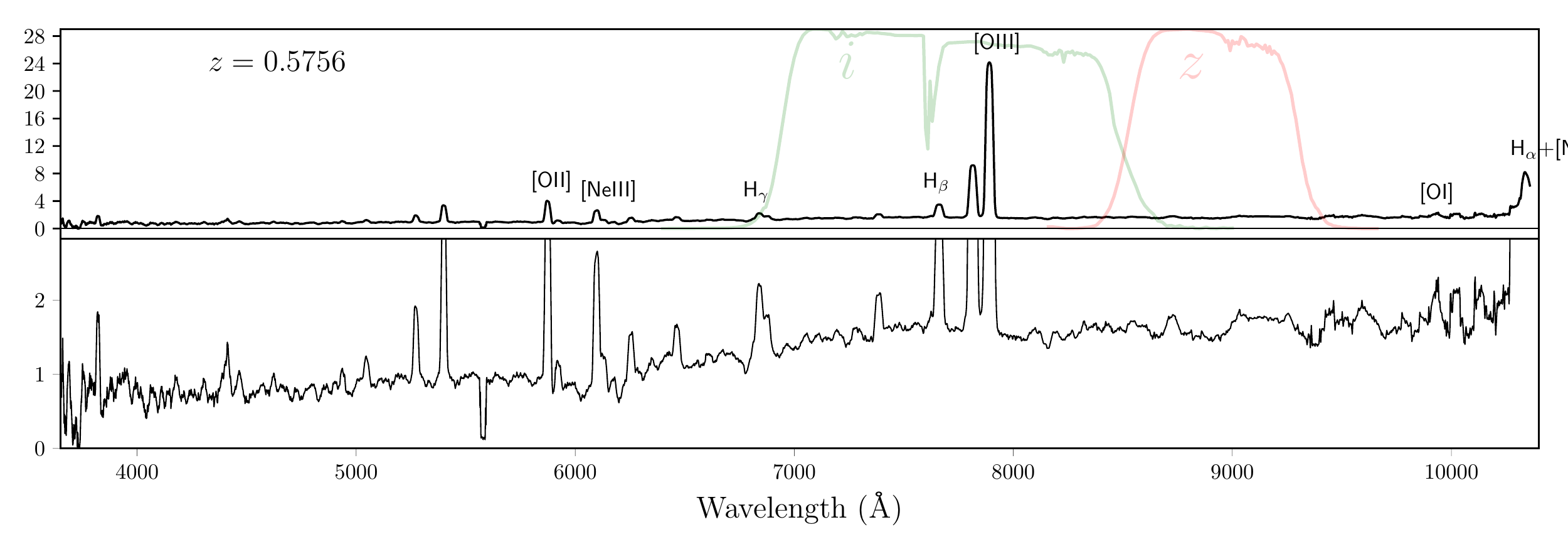}
    }
    \hbox{
    \includegraphics[width=3.in]{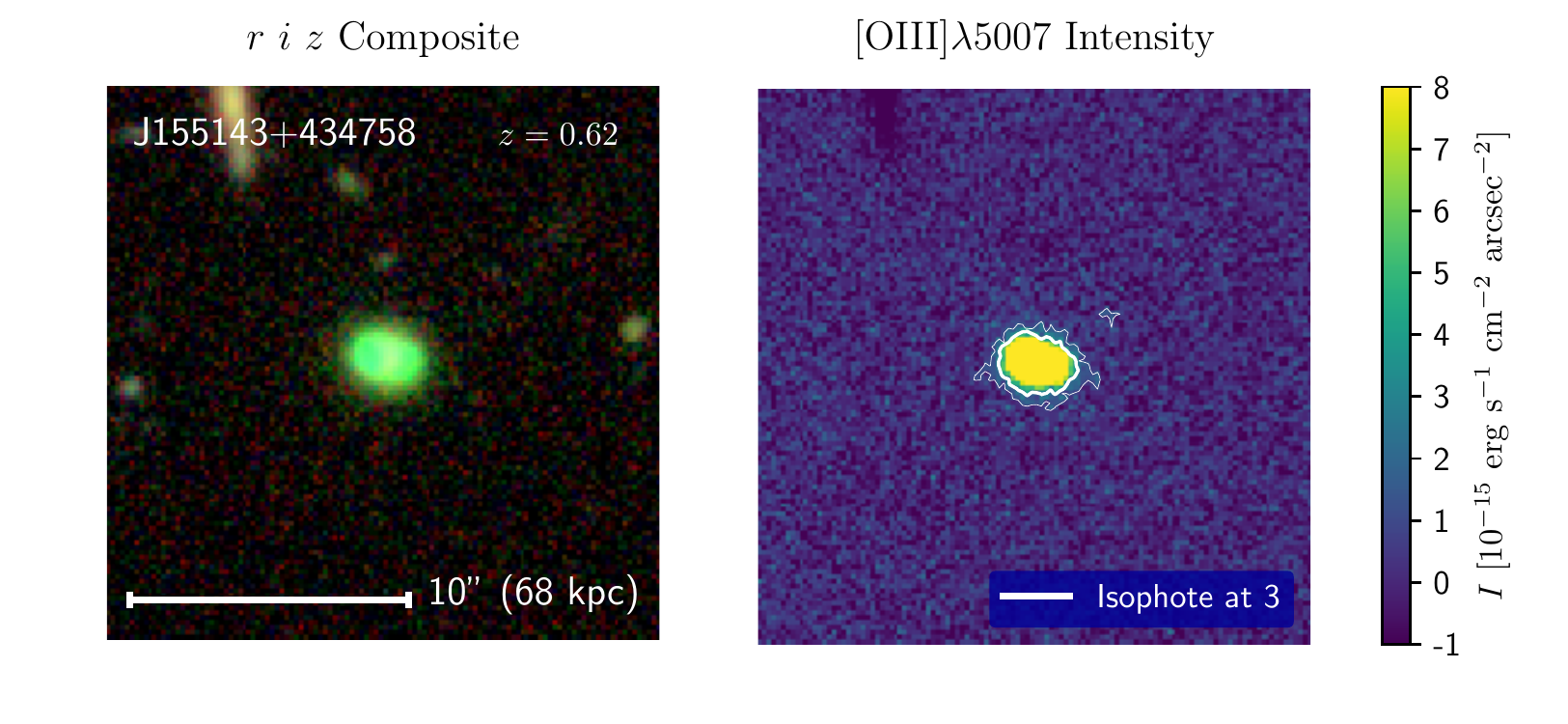}
    \includegraphics[width=3.75in]{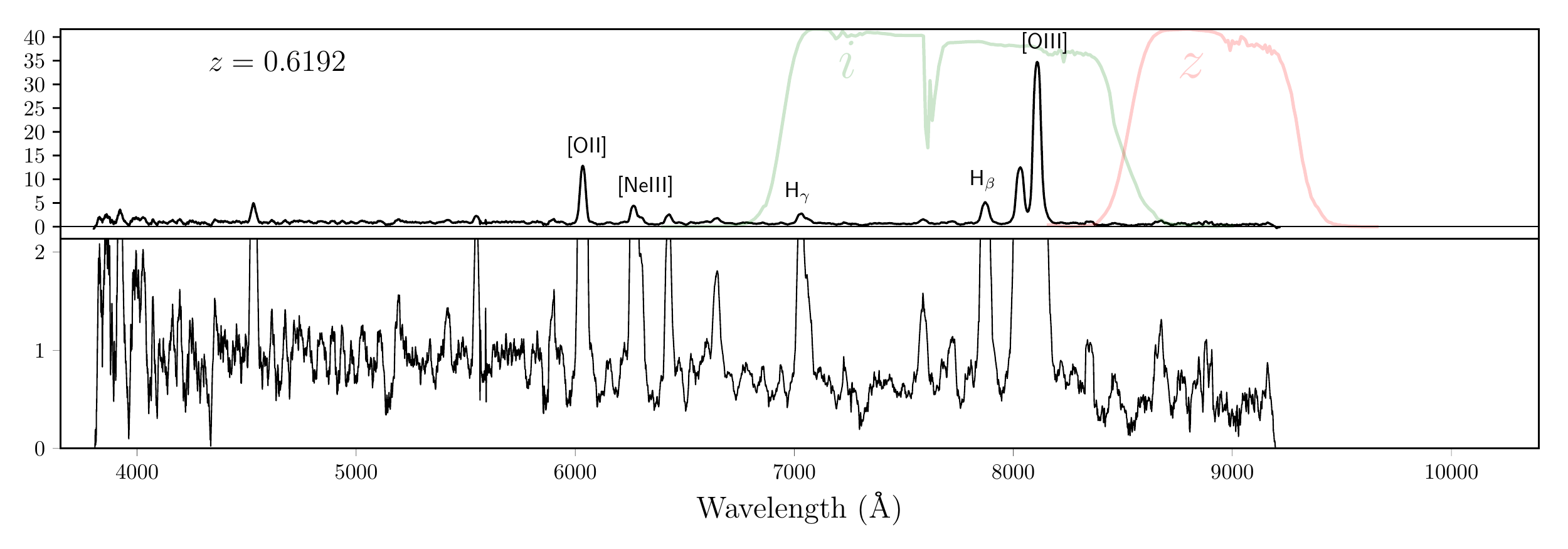}
    }
    }
    \caption{The top 10 extended emission-line region candidates with largest isophotal area.  
    {\it Left}: the $r$-, $i$-, $z$-band color composite HSC image and the reconstructed \oiii{} map. The symbols are as described in Fig. \ref{fig:map_example}. {\it Right}: The SDSS spectrum. The HSC filters of the line-band and the continuum-band are superposed. The bottom panel is zoomed in to show the continuum. The unit on the $y$-axis is $10^{-15}~\mathrm{erg~s^{-1}~cm^{-2}~\AA^{-1}}$. 
    }
    \label{fig:map_extcandi}
\end{figure*}

We present the ten largest extended emission-line candidates in area in Figure \ref{fig:map_extcandi} and Table \ref{tab:extcandi}. They have areas larger than 120 kpc$^{2}$ and radii larger than 8 kpc, that is, larger than the typical stellar size of galaxies. 
Targets affected by companion galaxies or foreground stars from visual inspection are not included here as their areas are overestimated. 
Because the sizes of emission-line regions correlate strongly with the AGN luminosity, these candidates are predominantly at higher luminosities ($\log$\lbolfifteen{}$ > 45.6 $) and higher redshifts ($z>$0.4). 
Their \oiii{} regions are well resolved, showing morphologies that are quite distinct from those of their host galaxies. This finding suggests that we are not mapping the stellar continuum but emission-line regions. 

Most of them have elongated shapes possibly due to the bi-polar cones of AGN ionization \citep{Mulchaey1996a,Zakamska2005,Obied2016}. Some exhibit extended filamentary shapes that may be associated with tidal features or outflows, e.g., J083823+015012 and J092203+004443. 
Indeed, some of the most extended \oiii{} regions on tens of kpc scale discovered in the past are associated with companion galaxies or tidal features ionized by luminous AGN, e.g., J1255$-$0339 \citep{Sun2017}, and J0123+0044 \citep{Villar-Martin2010}. 
There are also extended emission-line regions that are found to be in fact strong galactic outflows, which are valuable for studies of AGN feedback, e.g., J1356+1026 \citep{Greene2012,Sun2014}. 
Follow-up spectroscopic study, such as with long-slit or integral field unit (IFU) spectroscopy, is required to confirm the nature of these extended emission-line region candidates. For example, with kinematic measurements, one can distinguish between a photo-ionized medium that has virialized kinematics and extended outflows with much larger dispersion (e.g., $>600$ \kms{}). 

Our success in mapping emission-line regions using broad-band photometric data alone demonstrates that our continuum-subtraction technique is useful for searching large numbers of galaxies for extended emission-line regions. It also allows morphology studies of emission-line regions with statistical samples for future investigations. 
As the current methodology relies on spectroscopic preselection of type 2 AGN, AGN light echoes such as Hanny's Voorwerp \citep{Lintott2009}, where the AGN has faded, would not enter our selection. With improved methodology, it may be possible to search for these light echoes in a similar manner, as discused in Sec. \ref{sec:discussion:broadband}. 
The entire primary sample is presented in Appendix \ref{sec:append:wholesample}.

\renewcommand{\thefigure}{\arabic{figure} (Cont.)}
\addtocounter{figure}{-1}

\begin{figure*}
    \centering
    \vbox{
    \hbox{
    \includegraphics[width=3.in]{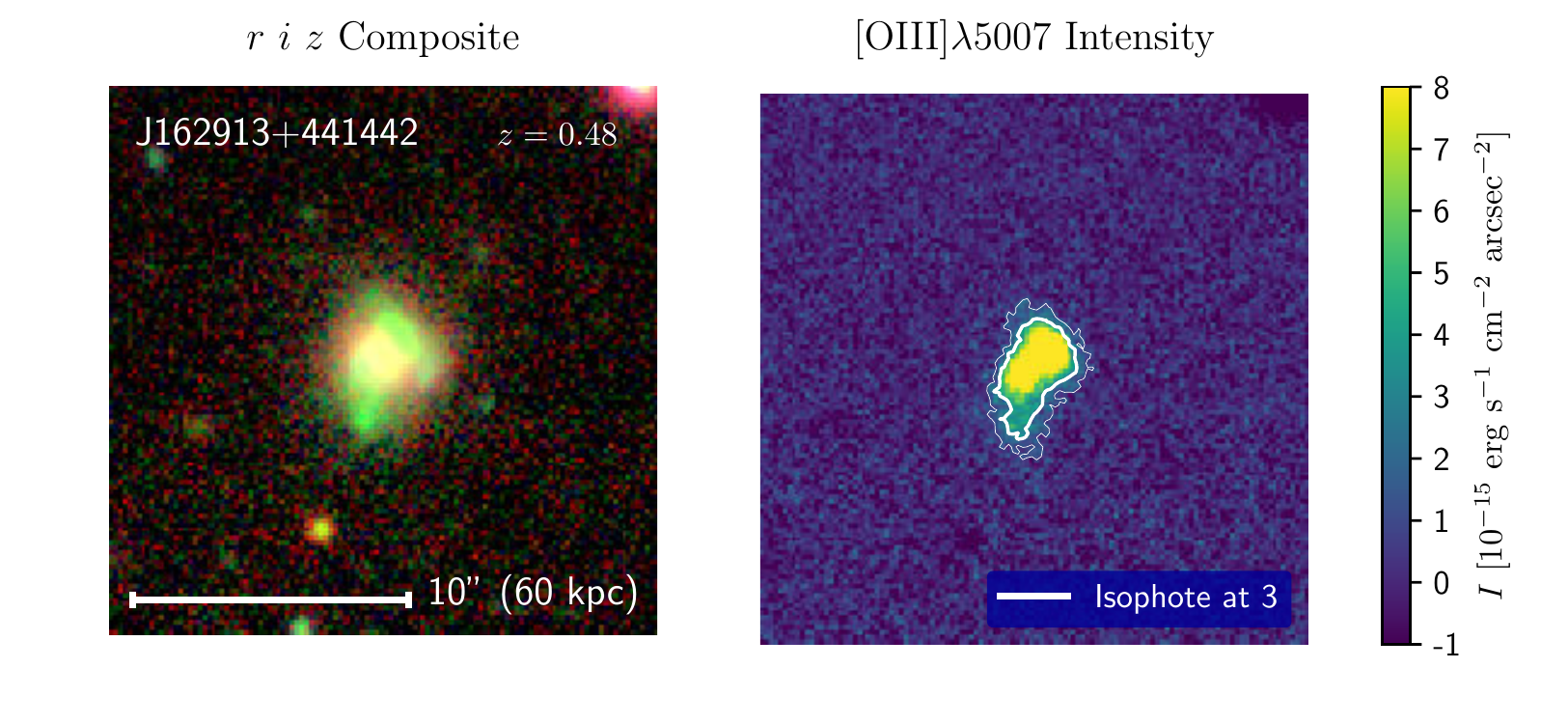}
    \includegraphics[width=3.75in]{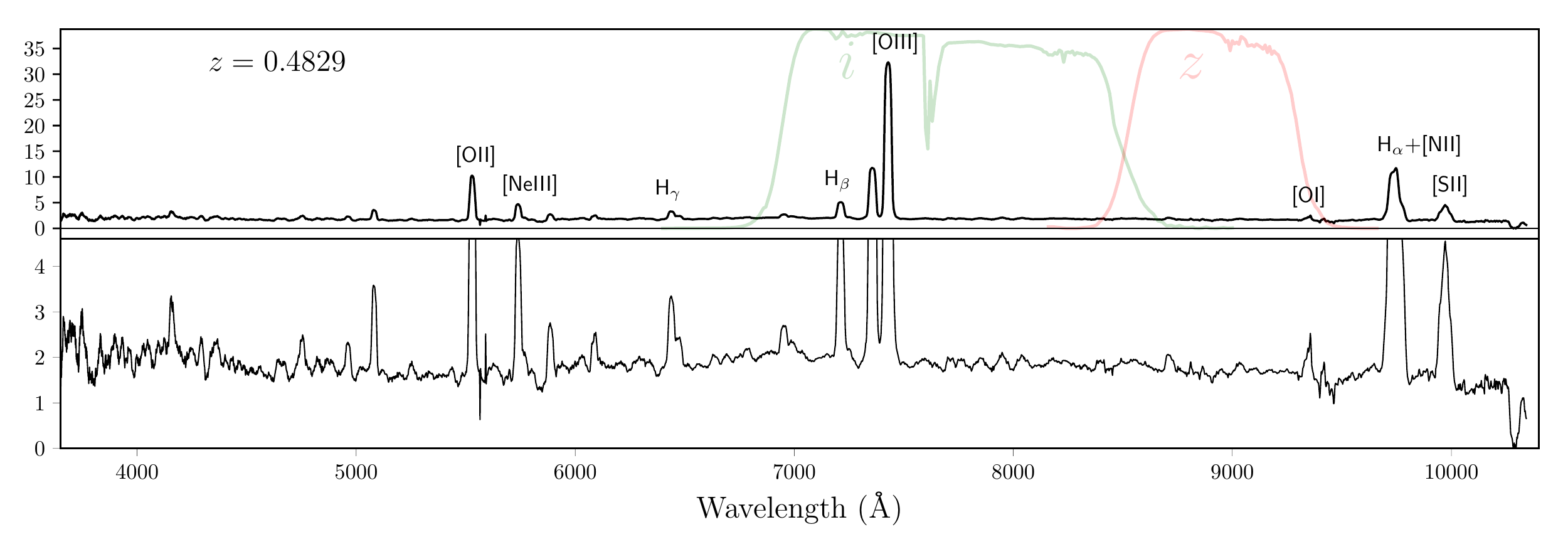}
    }
    \hbox{
    \includegraphics[width=3.in]{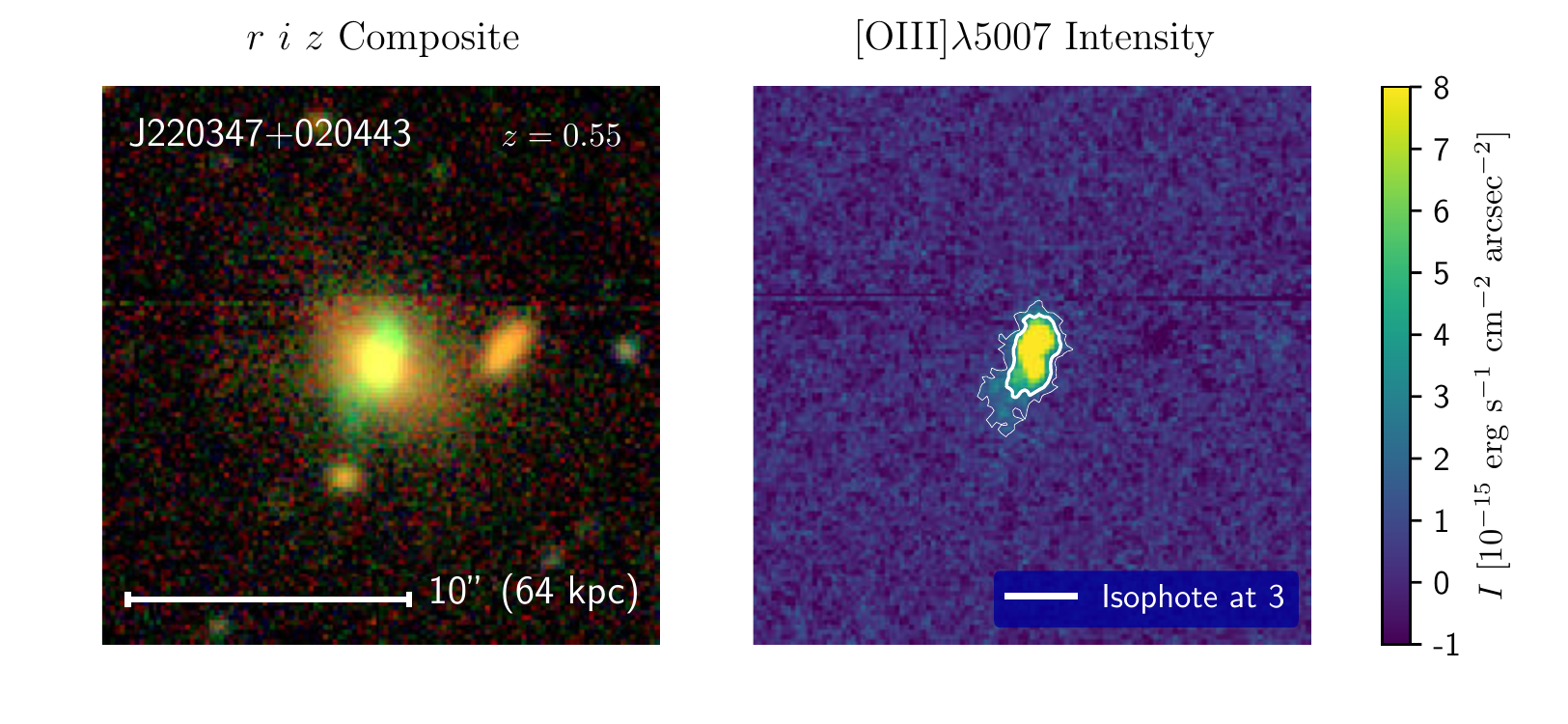}
    \includegraphics[width=3.75in]{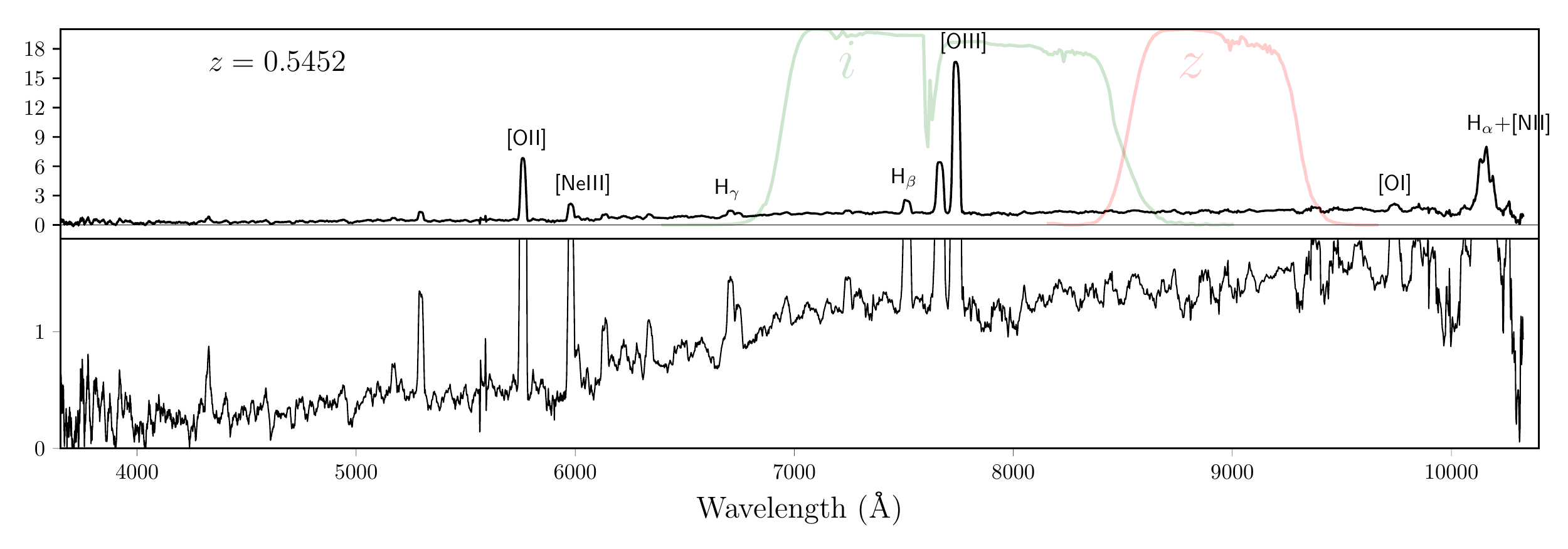}
    }
    \hbox{
    \includegraphics[width=3.in]{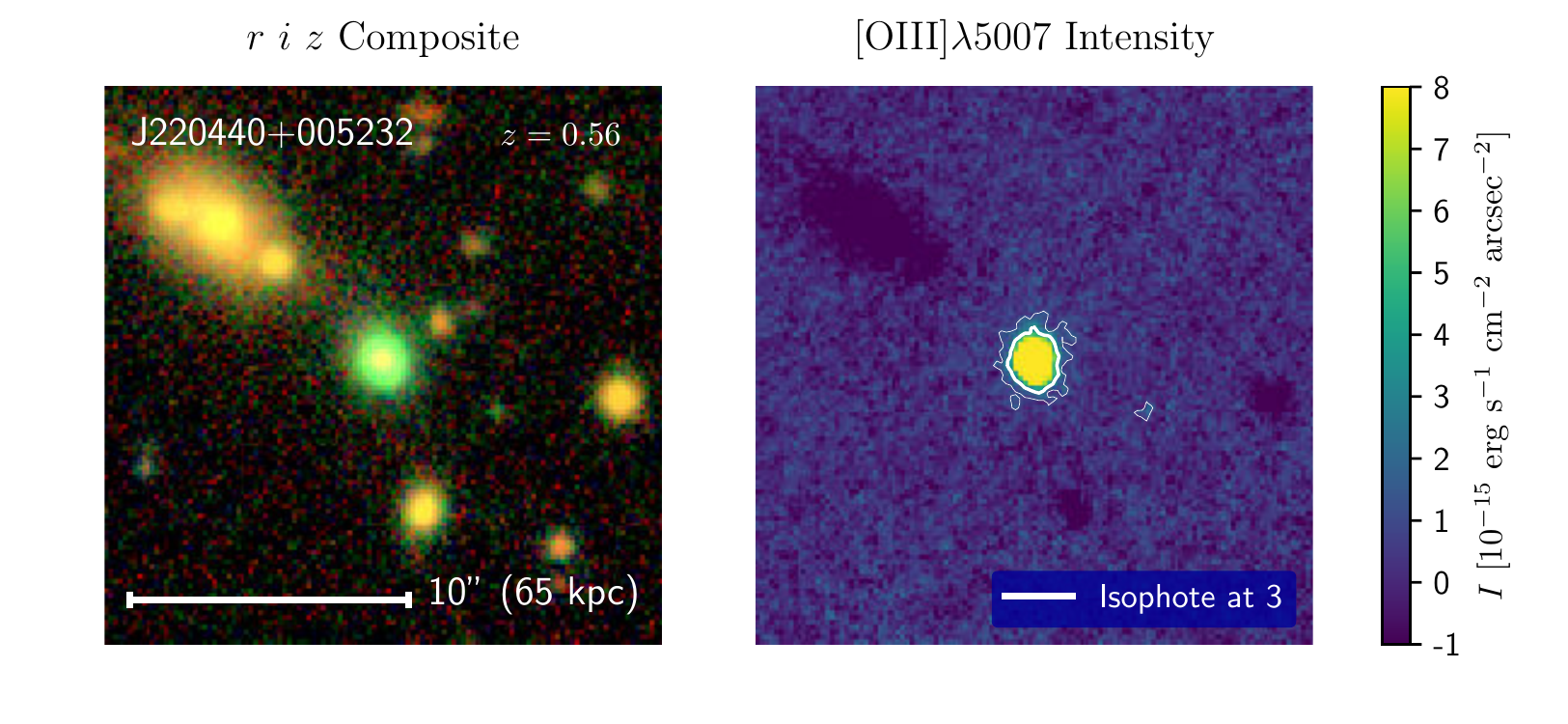}
    \includegraphics[width=3.75in]{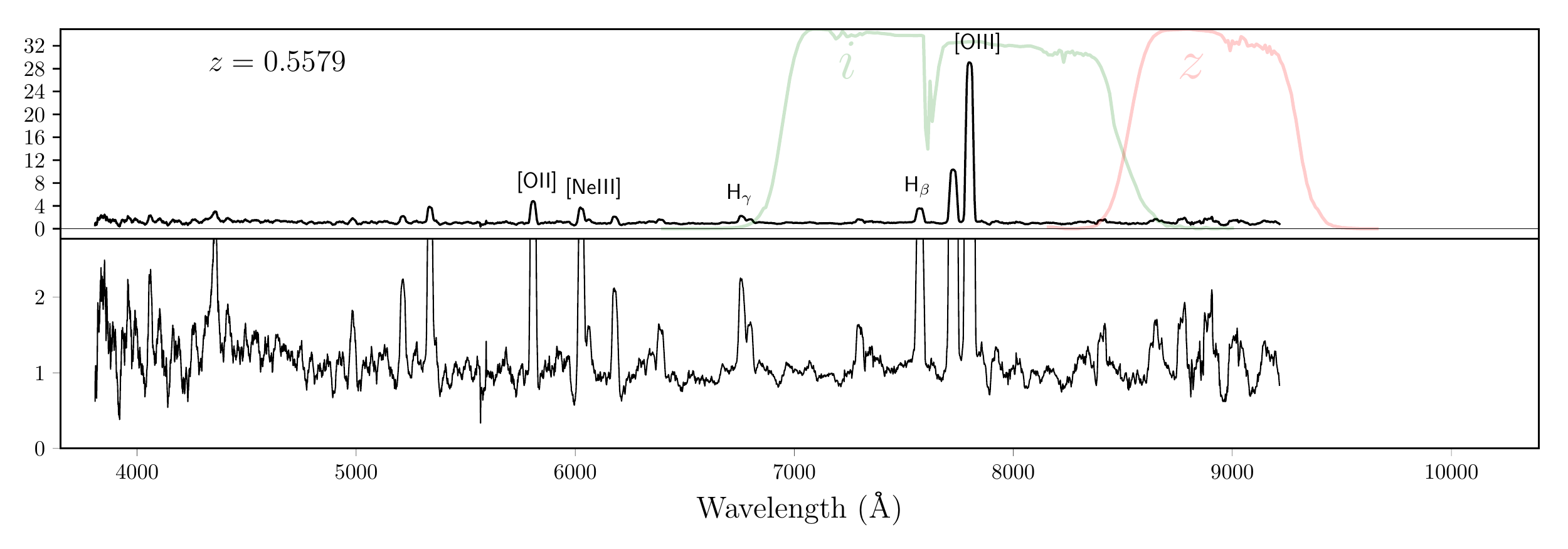}
    }
    \hbox{
    \includegraphics[width=3.in]{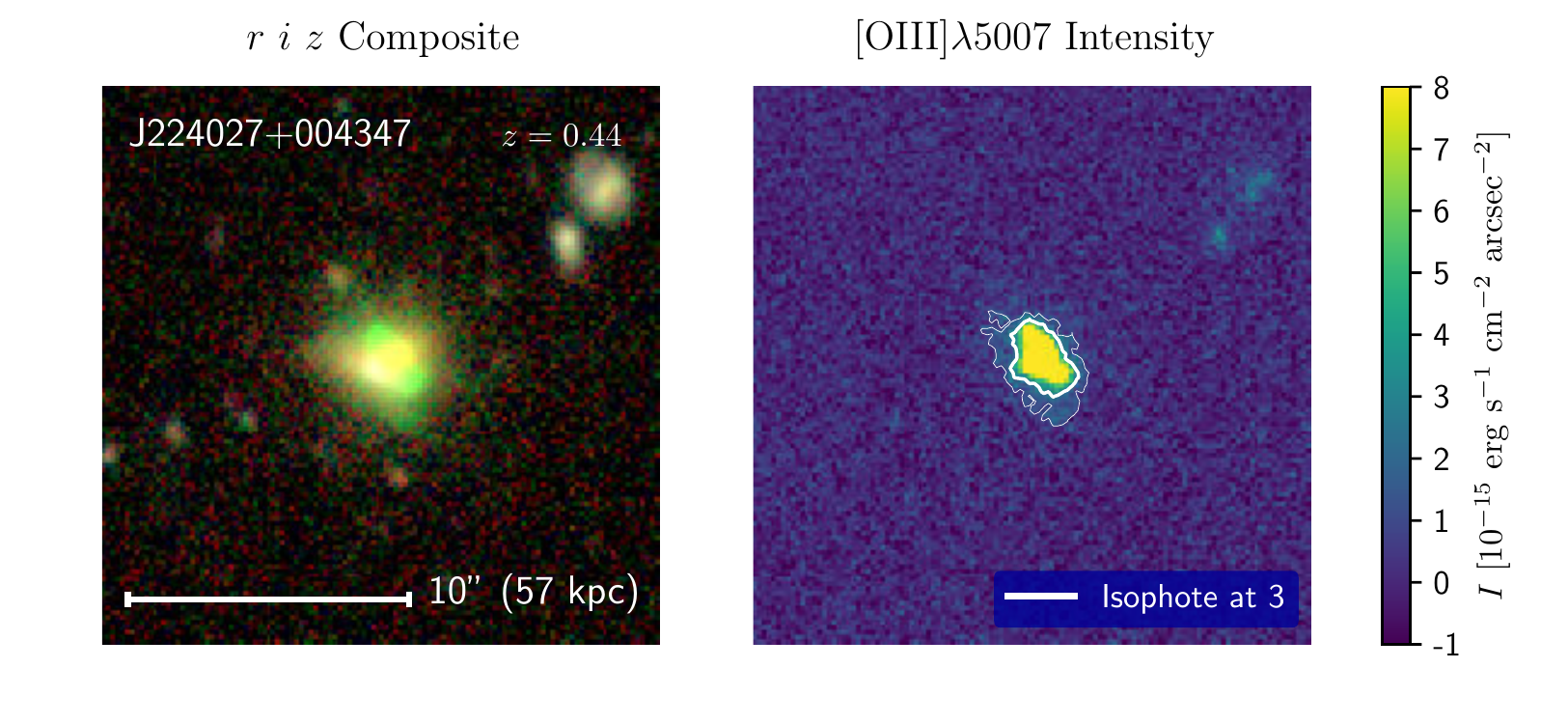}
    \includegraphics[width=3.75in]{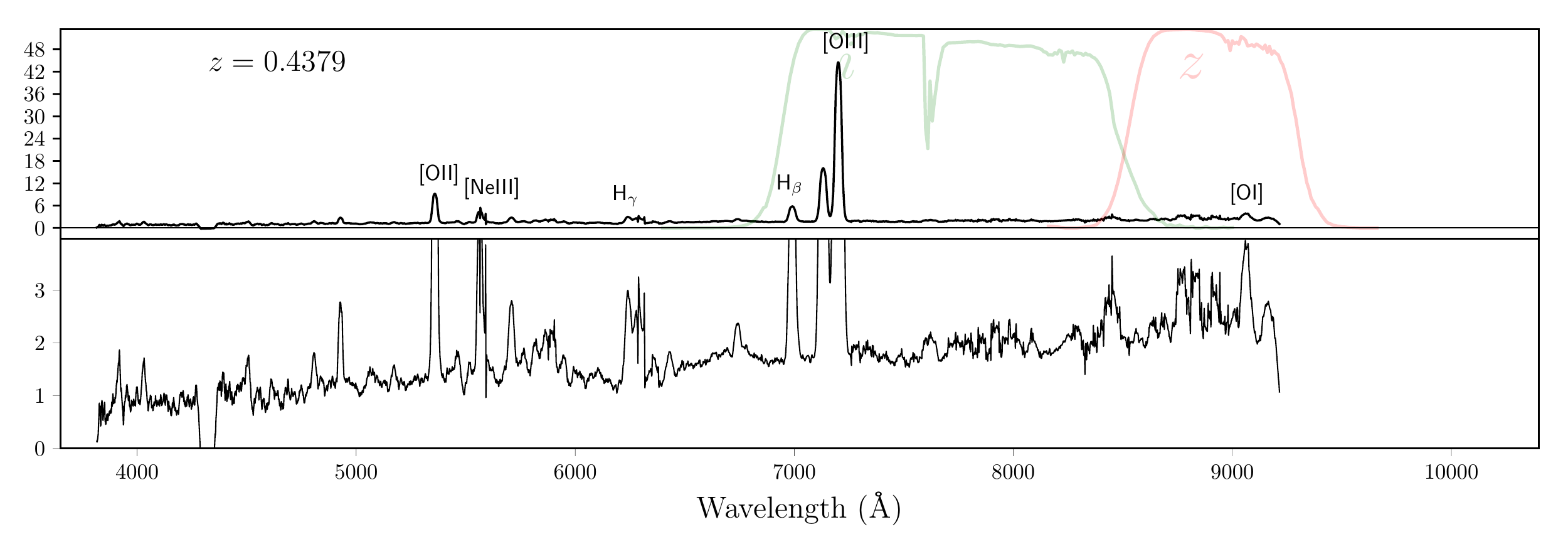}
    }
    }
    \caption{
    }
\end{figure*}

\renewcommand{\thefigure}{\arabic{figure}}

\begin{table*}
\begin{center}
\caption{Top 10 Candidates of Extended Emission-line Regions}
\begin{tabular}{cccccccccccc}
\hline
\hline
Name & R.A.  & Dec.  & $z$ & Bands & Source & zmag & ymag & \lbolfifteen{} & \loiii{} & Area & Radius\\
 & (J2000)  & (J2000)  &  &  &  & (mag) & (mag) & (\ergs{}) & (\ergs{}) & (kpc$^2$) & (kpc) \\
(1) & (2) & (3) & (4) & (5) & (6) & (7) & (8) & (9) & (10) & (11) & (12) \\
\hline
J023106$-$034513 & 02:31:06.5 &  $-$03:45:13.6 & 0.542 & i, z & Y & $-$23.37 & $-$23.74  & 45.69   & 42.12 & 278 & 30 \\
J083823+015012 & 08:38:23.1 &  +01:50:12.4 & 0.559 & i, z & Y & $-$22.74 & $-$22.97  & 45.69   & 42.18 & 152 & 29 \\
J090254+001116 & 09:02:54.2 &  +00:11:16.6 & 0.581 & i, z & Y & $-$23.15 & $-$23.38  & 45.87   & 42.22 & 132 & 10 \\
J091113+032604 & 09:11:13.9 &  +03:26:04.7 & 0.546 & i, z & Y & $-$22.99 & $-$23.21  & 45.62   & 42.39 & 132 & 10 \\
J092203$-$004443 & 09:22:03.2 &  $-$00:44:43.4 & 0.576 & i, z & Y & $-$23.44 & $-$23.70  & 46.34   & 42.52 & 247 & 13 \\
J155143+434758 & 15:51:43.9 &  +43:47:58.1 & 0.619 & i, z & R & $-$22.01 & $-$22.26  & 45.95   & 41.66 & 214 & 11 \\
J162913+441442 & 16:29:13.2 &  +44:14:42.8 & 0.483 & i, z & Y & $-$23.10 & $-$23.69  & 45.96   & 42.63 & 248 & 15 \\
J220347+020443 & 22:03:47.4 &  +02:04:43.8 & 0.545 & i, z & Y & $-$20.25 & \nodata & 45.86   & 42.34 & 149 & 12 \\
J220440+005232 & 22:04:40.7 &  +00:52:32.2 & 0.558 & i, z & R & $-$21.97 & $-$22.44  & \nodata & 41.50 & 127 & 8 \\
J224027+004347 & 22:40:27.0 &  +00:43:47.4 & 0.438 & i, z & R & $-$22.49 & $-$23.18  & 45.97   & 41.67 & 132 & 9 \\
\hline
\end{tabular}
\textit{Note.}
The top 10 extended emission-line region candidates with largest isophotal area. Column 1-4: the name, coordinates, and redshift of the target. Column 5: the line-band and continuum-band used for constructing the \oiii{} map. 
Column 6: the source of the object: Z, R, M, and Y stand for \citet{Zakamska2003,Reyes2008,Mullaney2013}; and \citet{Yuan2016}, respectively. 
Column 7-8: absolute HSC $z$- and $y$-band Kron AB magnitudes as a proxy for the stellar mass. K-correction is not applied.  
Column 9: the log AGN bolometric luminosity in units of \ergs{} inferred from intrapolated WISE rest-frame 15 \micron{} luminosity with a bolometric correction factor of 9. Column 10: the log \oiiil{} luminosity in units of \ergs{} measured from the SDSS/BOSS spectrum, as described in Sec. \ref{sec:res:sizelum}. Column 11 \& 12: the area and radius of the \oiiil{} emission-line region measured with a isophote level of rest-frame 3 \uintensity{}, as described in Sec. \ref{sec:method:iso}. Zero values in Col. 11 and 12 denote undetected or unresolved objects. 
\label{tab:extcandi}
\end{center}
\end{table*}

\section{Discussion} \label{sec:discussion}
\subsection{Size luminosity relation} \label{sec:discussion:sizelum}

In this section, we put our results of the size -- luminosity relations in the context of previous spectroscopic and narrow-line imaging studies. 
A correlation between the \oiii{} size and luminosity was quantified with studies of type 1 and 2 AGN using Hubble Space Telescope (HST) narrow band imaging \citep{Bennert2002, Schmitt2003}, but these two works found different slopes, $\beta = 0.5$ and 0.3, respectively, of the size -- luminosity relation. 
This discrepancy may be due to the relatively shallow HST images and, most importantly, the fact that these sizes were not defined with a fixed surface brightness threshold and thus depend on the sensitivity of the observations. 

With deeper ground-based integral field unit (IFU) spectroscopy, \citet{Husemann2013} recovered extended emission-line region in type 1 AGN that are a factor of a few larger than measured by \citet{Bennert2002} with sizes of $\sim$ 10 kpc. 
In a small number of instances, particularly in disky galaxies, they found that the extended emission-line region could be dominated by star formation, in line with the finding in Seyfert galaxies by \citet{Bennert2006a,Bennert2006}. 
Interestingly, \citet{Husemann2013} further found that the size of extended emission-line region is not significantly correlated with the total \oiii{} luminosity (but is correlated with the AGN continuum), although they adopted a radius that encloses 95\% of the \oiii{} flux, $r_{95}$, as the size, which tends to be smaller when the nucleus is brighter in \oiii{}, which might weaken the correlation. A strong correlation indeed exists in their data if isophotal size were used. 

\citet{Greene2011} focus on type 2 luminous AGN and find a shallower slope of $0.22\pm0.04$ with long-slit spectroscopy. Compared to type 1 AGN, type 2 AGN have the benefit that the broad \hbeta{} line and AGN continuum are obscured, making it easier to measure the \oiii{} lines robustly. 
\citet{Liu2013b} perform a similar IFU survey to \citet{Husemann2013} but on luminous type 2 luminous AGN. 
Most importantly, \citet{Liu2013b} use a cosmological-dimming-corrected constant surface brightness cut, such that the size measurement is independent of both the depth of the observation and the redshift of the object. 
Combined with previous spectroscopic studies of lower luminosity type 2 AGN \citep{Fraquelli2003,Bennert2006,Humphrey2010,Greene2011}, for which \citet{Liu2013b} determine the new isophotal radius from the data, they find a tight correlation between the \oiii{} radius and luminosity over 5 decades in luminosity. 
The slope of the relation ($\beta=$0.250$\pm$0.018) is shallower than previously found. 
However, this slope is dominated by lower luminosity AGN \citep{Fraquelli2003} that have isophotal radii estimated by extrapolation of the \oiii{} radial profile instead of directly measured. Therefore, more uniform observations are needed to confirm this slope. 
With more long-slit observations and careful PSF deconvolution, \citet{Hainline2013} and \citet{Hainline2014a} suggested that there is an upper limit on the size of the extended emission-line regions at about 10 kpc and a bolometric luminosity about $10^{46}$ \ergs{}. This relation is further confirmed by long-slit observations by \citet{Sun2017}, but the faint end slope remains ill-constrained. 

In this paper, the new broadband imaging technique and the Subaru Hyper Suprime-Cam survey allow us to measure the \oiii{} isophotal size uniformly over three decades in AGN luminosity. 
The imaging technique, which does not require observationally expensive follow-up spectroscopy, allows us to probe an unprecedented sample size and therefore offers new statistical power to constrain the size -- luminosity relation. 
With a higher isophotal cut, our measured radii are in general a factor of a few smaller than constrained by \citet{Liu2013b,Hainline2013}; and \citet{Sun2017}, but the correlation persists and the slope is broadly in agreement with \citet{Schmitt2003,Greene2011} and \citet{Liu2013b}. 
As discussed in Sec. \ref{sec:res:sizelum}, our best estimate is the Area -- \lbolfifteen{} relation that gives a slope of $0.62^{+0.05}_{-0.06}\pm0.10$ (statistical and systemic uncertainty). 
Given that area is proportional to radius squared, this would imply a slope of $\sim$ 0.31 in the radius -- luminosity relation. 
Such a slope is within the range of the spectroscopic studies mentioned above. 

There are not enough data points above \lbol{} $= 10^{46}$ \ergs{} in our sample to verify the flattening of the size -- luminosity relation found by \citet{Hainline2013,Hainline2014b}; and \citet{Sun2017}. There are also few outliers that are significantly above or below the relation, given the large size uncertainties. Larger samples are needed to constrain these two properties of the size -- luminosity relation. 

Broadband imaging is a complementary technique to spatially resolved spectroscopy to quantify extended emission-line gas. 
While our method boasts a significantly expanded sample size, it comes with limitations.
Emission lines from star formation or shocks cannot be isolated, which is especially an issue for lower luminosity AGN. 
Variations in continuum color, either due to dust obscuration, different stellar populations, companion galaxies, or foreground stars, leave imprints in the emission line maps. 
The depth of imaging that is required to detect diffuse line emission sometimes results in a saturated AGN nucleus in nearby objects. 
In this paper, we mitigate these issues with a more conservative isophotal threshold (Sec. \ref{sec:method:isolevel}) and with the adoption of the area as the primary size measurement. 
There are also other challenges associated with a large flux-limited sample. As luminous systems are found predominantly at higher redshifts, there might be redshift dependent biases in the size -- luminosity relation that are yet to be discovered. 
The consistency of our measured slope of the size -- luminosity relation with previous studies strongly supports our methodology. 
However, our results, in particular, the faint end slope, should be further tested by additional spectroscopic studies.

\subsection{The nature of extended emission-line regions} \label{sec:discussion:nature}

To interpret the slope of the size -- luminosity relation, it is important to consider not only the density profile but also the density inhomogeneity of the line emitting gas. 
Unlike Str{\"o}mgren's sphere around HII regions ionized by O stars, the radial profiles of \oiiil{} intensity of extended emission-line regions do not have sharp cut-offs, but follow a power-law decline with a range of power-law slopes \citep{Liu2013b}. 
The gas in narrow-line regions is likely clumpy, as suggested by the relatively high electron density \citep[few hundred cm$^{-3}$, e.g., from \oii{}$\lambda 3726\AA$/$\lambda 3729 \AA$ diagnostics, ][]{Greene2011} and low volume-filling factor observed \citep[$\sim10^{-6}$ from Balmer line luminosities,][]{Nesvadba2006}. 
The \oiii{} emission could be dominated by these dense clouds because of the $n^2$ steep dependence in the emissivity. 
In addition, in order to have \oiii{} ions the gas cannot be under-ionized or over-ionized. 
The size -- luminosity relation should then depend on the properties of the clouds and their radial distribution. 

In a recent theoretical work, \citet{Dempsey2018} consider a model where sparsely distributed dense clouds are ionized and pressure-confined by the AGN radiation pressure. 
They find that at the high-luminosity end the model sizes are in quantitative agreement with the observed ones as long as the narrow-line region has any ionization-bounded clouds, i.e., clouds with high enough column density that ionizing photons cannot penetrate. 
However, the predicted slopes of the size-[OIII] luminosity relationships ($\beta \sim$0.5) are steeper than the observed radius-luminosity relation ($\beta \sim$0.22 -- 0.33). 
It is not clear if this implies that the assumed gas geometry and distribution in \citet{Dempsey2018} are wrong, or, alternatively, that the measured slopes are biased by various measurement errors.
Contamination of star formation emission could lead to overestimation of the size especially in less luminous AGN. 
Unaccounted errors in the luminosities could flatten the observed slope. 
Further investigations on both the theoretical and observational sides are needed to resolve this discrepancy.

\subsection{On Broadband Imaging Technique and Future Surveys}\label{sec:discussion:broadband}

As the Subaru HSC survey continues to expand in area, we can expect a factor of 8 increase in the sample size when the survey is completed (1400 deg$^2$ in the wide layer). In addition, the deep and ultra-deep layer of the HSC survey could allow mapping objects at higher redshifts than explored in this paper. 

The method used in this paper can also be applied to other broadband imaging datasets, for example, the Dark Energy Survey. 
The depth of the survey is critical for mapping extended emission lines. 
The signal-to-noise ratio reached with shallower surveys, such as the SDSS, is lower by a factor of $\gtrsim$ 15 than HSC. This limits possible redshift ranges to $\lesssim$ 0.2, where high luminosity AGN is rare. 
Even where detection is possible, the noise will become a dominant issue and algorithms for image denoising, e.g., wavelet denoising, may be required \citep[see,][]{Sun2016}. 
The advantage of SDSS is that, by construction, all objects in the spectroscopic type 2 AGN samples \citep[e.g.,][]{Zakamska2003,Reyes2008,Mullaney2013,Yuan2016} have corresponding images. This increases the sample size and allows searches for rare but prominent emission-line regions. 

The Large Synoptic Survey Telescope (LSST) will be the leading imaging survey in the coming decade. 
As it is deeper and much wider (20,000 deg$^2$) than HSC, the increased sample size will offer much improved statistical power, but it will also come with challenges associated with the large data volume. 
Furthermore, as the spectroscopic coverage of LSST will be limited, it is important to develop imaging techniques for the emission-line regions in the absence of spectroscopic inputs. For example, one would need to rely on the phototmetric data to infer the redshifts and the galaxy continuum color of the system. 
Variations of image deblending techniques that take advantage of expected properties of spatial structure may also be useful when spectral information is limited \citep[e.g.,][]{Joseph2016,Melchior2018}. 
These types of methodologies can bypass the spectroscopic selection that the current method relies on and could extend the search to AGN light echoes in which the emission-line region exists but the AGN has faded. 

The technique explored in this paper could also help with studies of other emission-line systems. Type 1 AGN are not included in this paper because they would require additional treatment to remove the glaring nuclear point source and to infer the galaxy color when the continuum is dominated by the AGN. But the detection of extended emission-line region in these systems is in principle possible once these effects are taken into account. In addition to \oiiil{}, Lyman-$\alpha$ of high-redshift objects is another emission line that is strong enough to be imaged with broadband images. The large cosmological volume offered by broadband surveys would make it complementary to narrowband searches of Lyman-$\alpha$ emitters as it is optimal for the searches of the most prominent, extended, and rarest systems \citep[e.g.,][]{Hennawi2015}.

\section{Summary} \label{sec:summary}

Narrow-line regions are one of the defining features of active galactic nuclei. 
Extended emission-line regions on scales of kpc to tens of kpc, probe the effect of AGN radiation on their host galaxies. They are useful for studying various aspects of AGN physics, including photo-ionization \citep[e.g.,][]{Liu2013b}, outflows and feedback \citep[e.g.,][]{Sun2017}, the obscuring geometry and the unification model \citep[e.g.,][]{Mulchaey1996a,He2018}, and AGN variability \citep[e.g.,][]{Keel2017}. 
In this paper, we develop a new broadband imaging method to detect and resolve these narrow-line regions with the Subaru Hyper Suprime-Cam survey. 
Leveraging on the data volume and image quality offered by this survey, our method probe an sample size that has not been possible for traditionally targeted spectroscopy or narrow-band imaging observations. 

Isolating the strong \oiii{}$\lambda 4959, 5007$ emissions requires a careful subtraction of the host galaxy continuum from the broadband image, which is based on an image in a line-free band, and spectrophotometry from the SDSS. 
Applying this method to the Subaru HSC data, we image the narrow-line regions of 300 SDSS selected type 2 AGN with redshifts between 0.10 and 0.69, among which 231 are detected and spatially resolved. 

We revisit the size -- luminosity relation with uniform isophotal measurements across three decades in AGN luminosity. 
In order to mitigate a number of contaminants that become important in low luminosity AGN, including star formation and continuum subtraction residuals, we adopt a higher and therefore more robust surface brightness limit than in previous works, and use area as the primary size measurement. 
The area correlates strongly with the AGN luminosity. We infer a power-law index of the area -- luminosity relation of $0.62^{+0.05}_{-0.06}\pm0.10$ (statistical and systemic errors) with luminosities inferred from the interpolated rest-frame 15 \micron \emph{WISE} luminosity. 
Given the expectation that area should be about radius squared, this is consistent with previous spectroscopic works that find a radius -- luminosity relation of $R \propto L^{0.22-0.33}$ \citep{Schmitt2003b,Greene2011,Liu2013b}. 

We show 10 extended emission-line region candidates, which are larger than 100 kpc$^2$ in area. Their \oiii{} morphologies are distinct from their host galaxies, often exhibiting bi-polar or irregular shapes, possibly associated with ionization cones, outflows, or jets. 
More follow-up observations are needed to determine the nature of these extended emission-line regions. 

As the Subaru HSC survey continues to expand in area, we can expect a factor of several increase in sample size. This technique is also useful for the next generation of imaging surveys, e.g., LSST. 
This larger sample size will enable statistical studies of emission-line regions and allow the discovery of rare and prominent candidates for studies of AGN physics.

\section*{Acknowledgements}

A.-L. Sun would like to thank James E. Gunn, Robert Lupton, Jim Bosch, Peter Melchior, James H. H. Chan, Bau-Ching Hsieh, and Masami Ouchi for enlightening discussions. 
The Hyper Suprime-Cam (HSC) collaboration includes the astronomical communities of Japan and Taiwan, and Princeton University.  The HSC instrumentation and software were developed by the National Astronomical Observatory of Japan (NAOJ), the Kavli Institute for the Physics and Mathematics of the Universe (Kavli IPMU), the University of Tokyo, the High Energy Accelerator Research Organization (KEK), the Academia Sinica Institute for Astronomy and Astrophysics in Taiwan (ASIAA), and Princeton University.  Funding was contributed by the FIRST program from Japanese Cabinet Office, the Ministry of Education, Culture, Sports, Science and Technology (MEXT), the Japan Society for the Promotion of Science (JSPS),  Japan Science and Technology Agency  (JST),  the Toray Science  Foundation, NAOJ, Kavli IPMU, KEK, ASIAA,  and Princeton University.
This paper makes use of software developed for the Large Synoptic Survey Telescope. We thank the LSST Project for making their code available as free software at http://dm.lsst.org.
This paper is based on data collected at the Subaru Telescope and retrieved from the HSC data archive system, which is operated by the Subaru Telescope and Astronomy Data Center at National Astronomical Observatory of Japan.
The Pan-STARRS1 Surveys (PS1) have been made possible through contributions of the Institute for Astronomy, the University of Hawaii, the Pan-STARRS Project Office, the Max-Planck Society and its participating institutes, the Max Planck Institute for Astronomy, Heidelberg and the Max Planck Institute for Extraterrestrial Physics, Garching, The Johns Hopkins University, Durham University, the University of Edinburgh, Queen's University Belfast, the Harvard-Smithsonian Center for Astrophysics, the Las Cumbres Observatory Global Telescope Network Incorporated, the National Central University of Taiwan, the Space Telescope Science Institute, the National Aeronautics and Space Administration under Grant No. NNX08AR22G issued through the Planetary Science Division of the NASA Science Mission Directorate, the National Science Foundation under Grant No. AST-1238877, the University of Maryland, and Eotvos Lorand University (ELTE).
Funding for SDSS-III has been provided by the Alfred P. Sloan Foundation, the Participating Institutions, the National Science Foundation, and the U.S. Department of Energy Office of Science. The SDSS-III web site is http://www.sdss3.org/.
SDSS-III is managed by the Astrophysical Research Consortium for the Participating Institutions of the SDSS-III Collaboration including the University of Arizona, the Brazilian Participation Group, Brookhaven National Laboratory, Carnegie Mellon University, University of Florida, the French Participation Group, the German Participation Group, Harvard University, the Instituto de Astrofisica de Canarias, the Michigan State/Notre Dame/JINA Participation Group, Johns Hopkins University, Lawrence Berkeley National Laboratory, Max Planck Institute for Astrophysics, Max Planck Institute for Extraterrestrial Physics, New Mexico State University, New York University, Ohio State University, Pennsylvania State University, University of Portsmouth, Princeton University, the Spanish Participation Group, University of Tokyo, University of Utah, Vanderbilt University, University of Virginia, University of Washington, and Yale University.
This publication makes use of data products from the Wide-field Infrared Survey Explorer, which is a joint project of the University of California, Los Angeles, and the Jet Propulsion Laboratory/California Institute of Technology, funded by the National Aeronautics and Space Administration. 
This paper makes use of packages available in Python's open scientific ecosystem, including NumPy \citep{VanderWalt2011}, SciPy\footnote{http://www.scipy.org/}, Matplotlib \citep{Hunter2007}, IPython \citep{Perez2007}, scikit-learn \citep{Pedregosa2011}, and scikit-image \citep{VanderWalt2014}. 
This research made use of Astropy, a community-developed core Python package for Astronomy \citep{TheAstropyCollaboration2013,TheAstropyCollaboration2018}.
This research made use of HumVI\footnote{https://github.com/drphilmarshall/HumVI} \citep{Marshall2016}, a open source software to generate color composite images.








\appendix

\section{Simulations to Quantify Isophote Area Uncertainties}
\label{sec:append:sim}

\subsection{Uncertainties from Image Noise}
\label{sec:append:sim:noise}

Noise in the \oiii{} image could introduce systematic and random uncertainties in the measurement of the isophotal area. To quantify these effects, we performed a set of Monte Carlo simulations inserting fake noise to our images. 
There are 69 galaxies in our sample that have low noise in the \oiiil{} image ($<0.2$ \uintensity{} per pixel). We use their \oiiil{} image and insert Gaussian white noise with an amplitude of 1 \uintensity{}, which is the maximum noise level among our sample, see Fig. \ref{fig:qc}. It is assumed that there is no correlated noise among pixels.  We run 100 noise realizations for each image and measure their isophote area in the same way it is measured on the original image. The bias and random uncertainties are estimated from the mean and the standard deviation of the distribution of the noisy isophotal areas. 

\begin{figure*}
    \centering
    \vbox{
    \includegraphics[width=3.in]{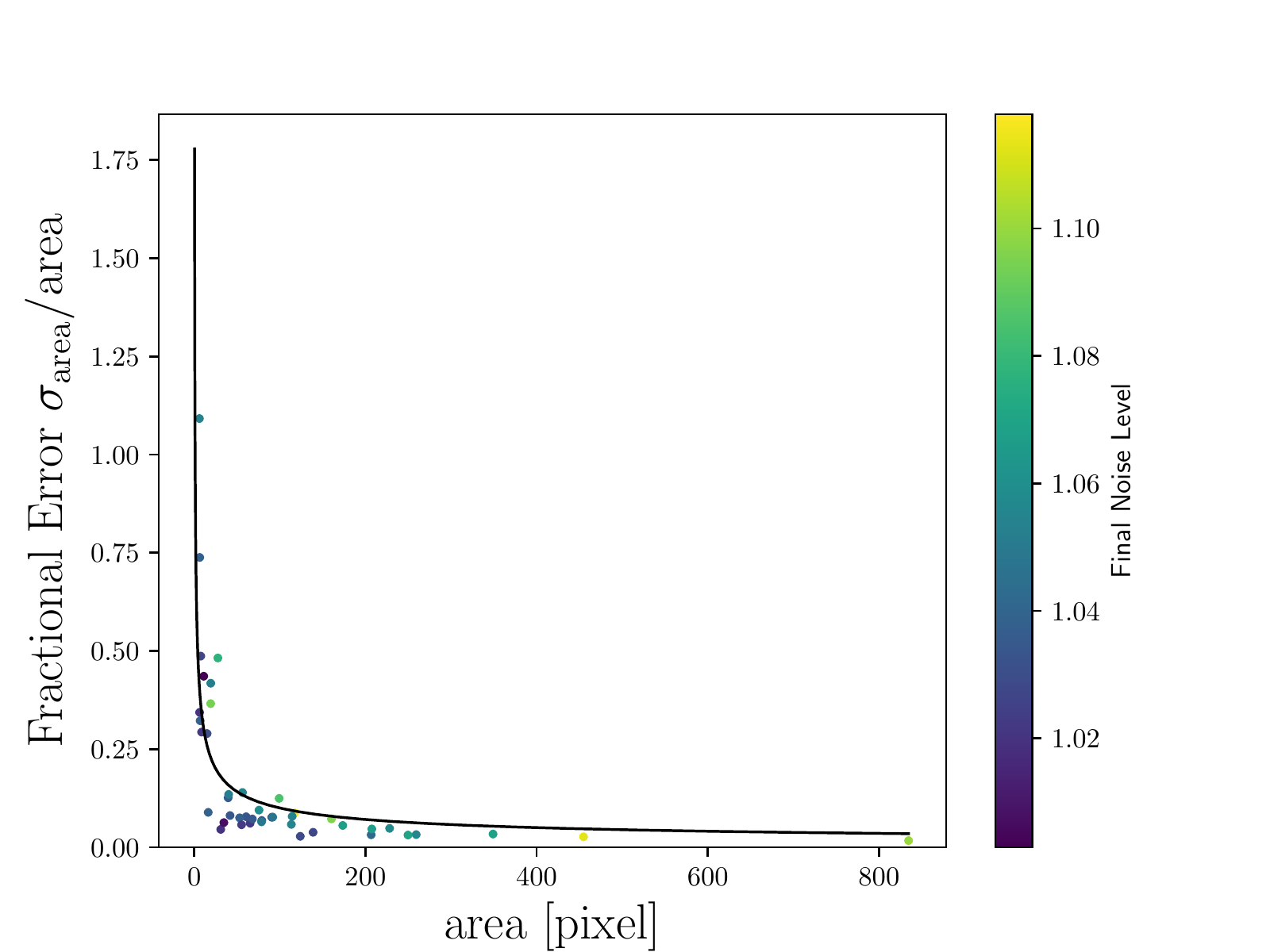}
    }
    \caption{The random uncertainty in isophotal area due to image noise estimated from simulations, see Appendix \ref{sec:append:sim:noise}. 
    Plotted is the fractional random uncertainty versus the original area in pixels ($n$). The black line shows the function $1/\sqrt{n}$. 
    Despite large scatter, the simulated data in general follows or falls below this relation. 
    }
    \label{fig:append:simnoise}
\end{figure*}

The biases are very small, with an average among galaxies of -0.018 arcsec$^2$, 
about 1\% of the median measured isophotal area. Therefore, we conclude that the bias in isophotal area introduced by image noise is negligible. 
We found that the random uncertainty in area increases with the area, see Fig. \ref{fig:append:simnoise}. Although the scatter is large, the uncertainty roughly follows the function form of 
\begin{equation}
\sigma_{n}=\sqrt{n}, 
\end{equation}
where $n$ is the isophotal area in units of pixels and $\sigma_{n}$ is its random uncertainty in the same units. 
This $\sqrt{n}$ dependence is possibly due to the fact that the noise only affects the area measurement at its isophote boundaries, which has a length proportional to the square root of the area. 

As the noise inserted in the simulation is higher than the typical noise level in our images, the real uncertainty should be smaller than or comparable to the one described by this functional form. We therefore adopt $\sigma_{n}=\sqrt{n}$ as the random uncertainty in the isophotal area measurement due to noise in the image. 

\subsection{Uncertainties from Point Spread Function}
\label{sec:append:sim:psf}

The point-spread function (PSF) defines the smallest structure that can be resolved in an image. To estimate how the PSF affects the isophotal area measurement, we simulate its effect using our real \oiii{} images with small PSFs (PSF FWHM $< 0.6$\arcsec{}). These images are convolved with Moffat PSF kernels with a range of widths and a Moffat power index of 2.5, which is typical for the PSF of our images, to create a convolved image with a final PSF FWHM of 1.0\arcsec{}. 
We measure the isophotal area of this convolved final image in the same way as we do for the science data. 

\begin{figure*}
    \centering
    \vbox{
    \includegraphics[width=3.in]{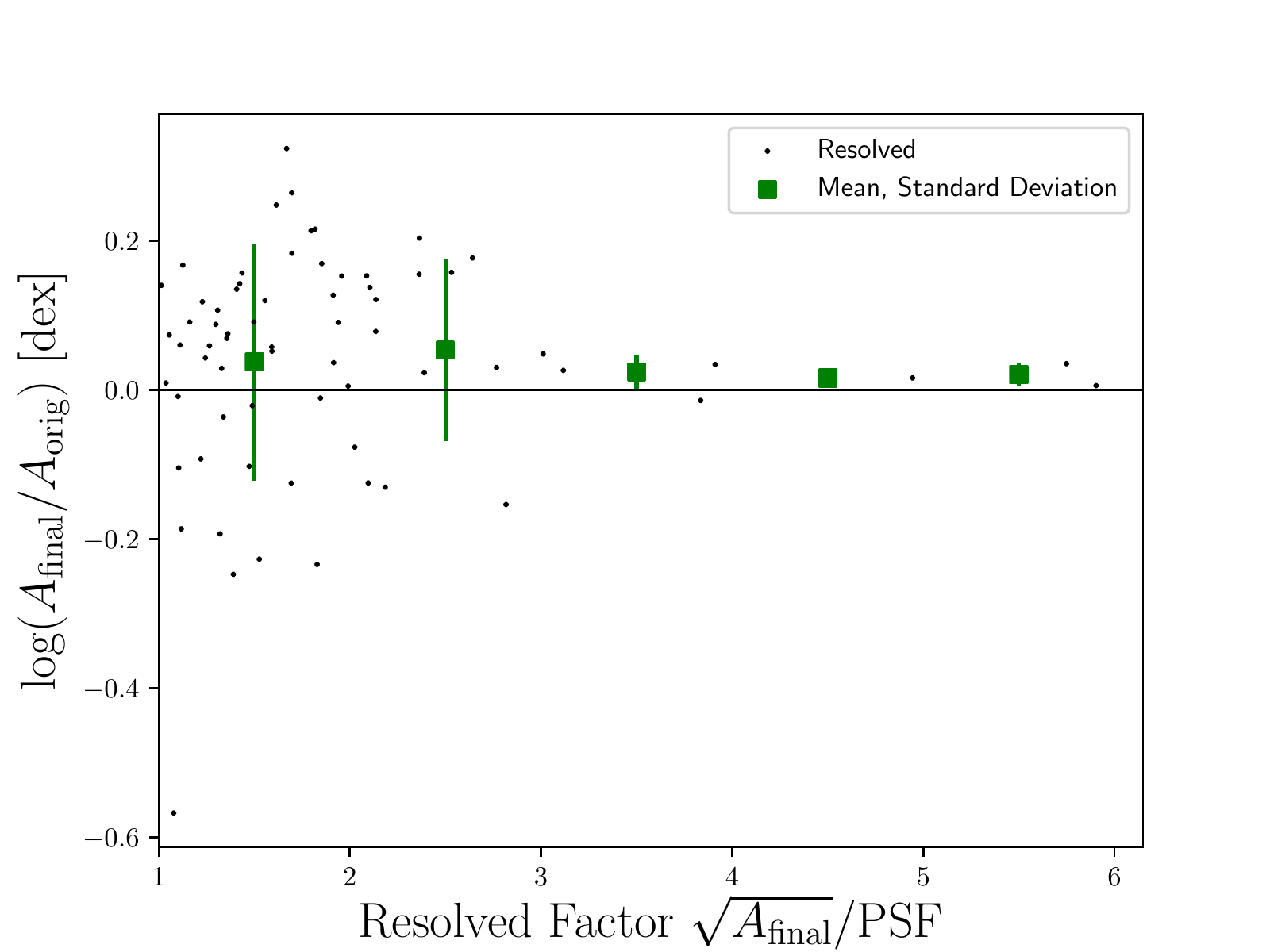}
    \includegraphics[width=3.in]{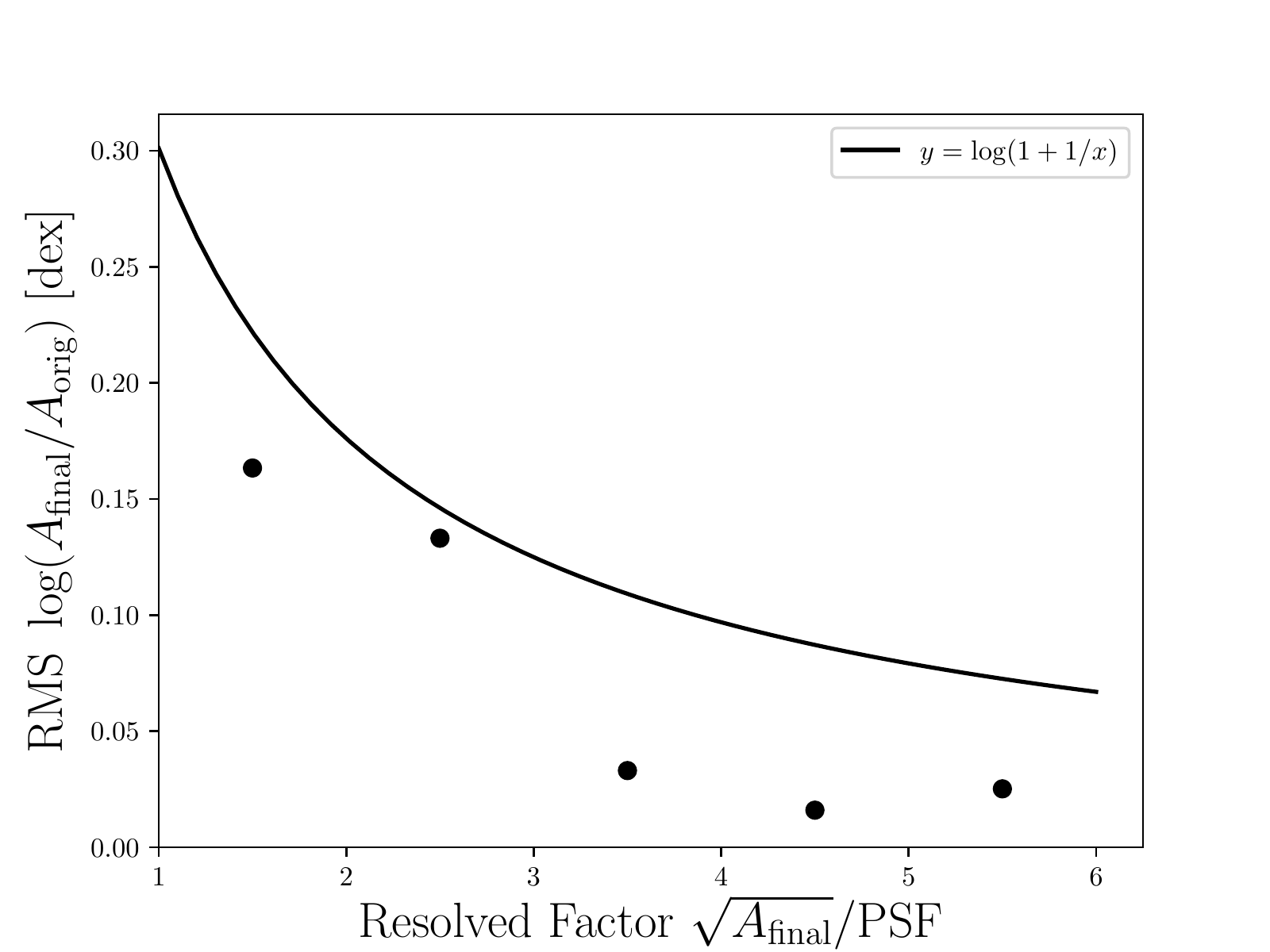}
    }
    \caption{
    The uncertainty in isophotal area due to PSF for resolved objects. {\it Left}: The ratio between the PSF convolved area and the origial area versus the resolved factor. The log difference between the two areas is smaller when the object is more resolved.  {\it Right}: The RMS log error introduced by PSF as a function of the resolved factor. The function $y=\log{1+1/x}$ (black line) is used to assign uncertainties to area measurements. 
    }
    \label{fig:append:simpsf}
\end{figure*}

The final image is either resolved ($\sqrt{A} > PSF$) or unresolved ($\sqrt{A} < PSF$) with respect to its new PSF. For the resolved ones, the comparison between the convolved and the original area is shown in Fig. \ref{fig:append:simpsf}. There is little systematic deviation, but the fractional random uncertainty is larger for objects that are less well-resolved (left panel). 

If we assume that the difference in the area that is caused by the PSF is equal to $\sqrt{A_{orig}} * PSF$, which is about the area of a perimeter ring with a thickness of the PSF FWHM, then the error would follow the functional form  $y=\log{1+1/x}$.
This is plotted as the black line on the right panel of Fig. \ref{fig:append:simpsf}, where $x$ is the resolved factor $\sqrt{A}/PSF$ and $y$ is the error in log scale. 
This functional forms an upper envelope encompassing the simulated RMS error. 
We thus adopt the function in Eq. \ref{eq:err:psf} as the uncertainty in area introduced by the PSF. 

For the unresolved objects, which have a convolved area smaller than the square of the PSF, we compare the original area to the size of the PSF. We find that for most of the cases the original area is indeed smaller than the PSF. This is consistent with the square of PSF being the upper limit of the area, with an uncertainty of about 0.2 dex. 

\section{Aperture Correction on \loiii{}}
\label{sec:append:aptcorr}

\begin{figure*}
    \centering
    \includegraphics[width=3.in]{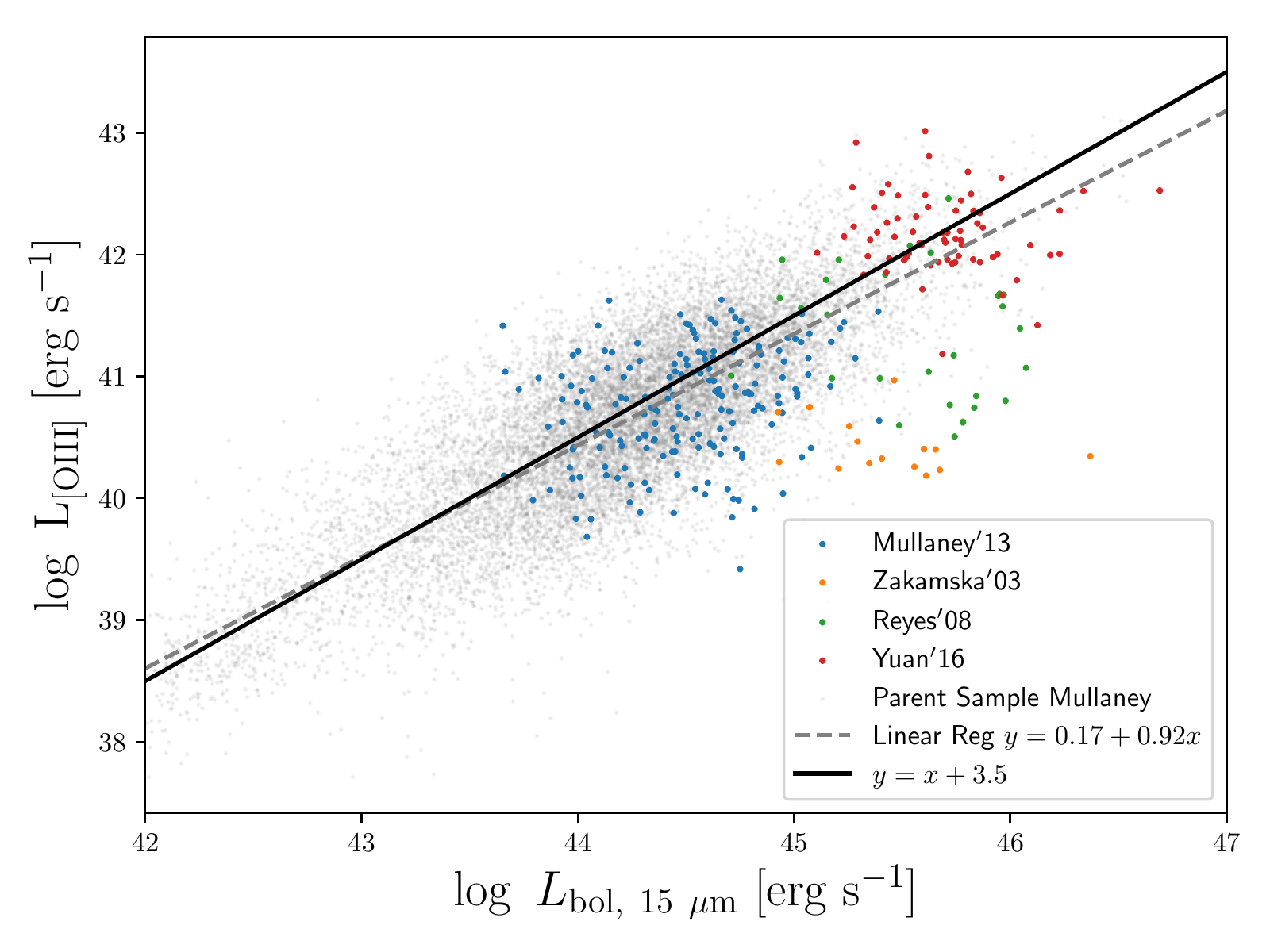}
    \caption{
    The relation between \loiii{} and \lbolfifteen{}.  The colored dots are the primary sample with colors indicating the source of the target. An aperture correction of 0.7 dex is applied to the \citet{Yuan2016} sample which is taken with the BOSS spectrograph. As a comparison, the background grey dots show the type 2 AGN from the \citet{Mullaney2013}, where the \loiii{} is measured homogeneously with the SDSS spectrograph. They exihibit an approximately linear relation with a power-law slope of 0.92 (grey dashed line). The black line shows a linear relation between \loiii{} and \lbolfifteen{}. 
    }
    \label{fig:append:loiiilbol}
\end{figure*}

As discussed in Sec. \ref{sec:res:sizelum}, the \loiii{} is not measured homogeneously among the sample. The \citet{Yuan2016} sample uses BOSS spectrum which has a smaller fiber aperture (2\arcsec{}) than the SDSS spectrum (3\arcsec{}) used in the other three samples. 
As the extended emission-line regions are often larger than the spectroscopic aperture, an aperture correction is needed. For simplicity, we only correct for the inhomogeneities that may arise from the smaller spectroscopy aperture in the BOSS sample. 
We use the \loiii{} -- \lbolfifteen{} relation to calibrate the \loiii{} relation and determine the amount of correction needed. As shown in Fig. \ref{fig:append:loiiilbol}, from the parent sample of \citet{Mullaney2013}, \loiii{} is approximately linearly proportional to \lbolfifteen{}, with a power-law slope of 0.92. 
The \loiii{}/\lbolfifteen{} is 0.7 dex lower in the \citet{Yuan2016} sample than in the \citet{Mullaney2013} sample.  We therefore apply an aperture correction of 0.7 dex to the \loiii{} of the \citet{Yuan2016} sample such that the \loiii{} in our primary sample is about linearly proportional to \lbolfifteen{}. The \citet{Zakamska2003} and \citet{Reyes2008} samples also 
fall below the \loiii{} -- \lbolfifteen{} relation of the \citet{Mullaney2013} sample, but it is unclear why that is the case given that they are all measured with the SDSS spectrograph. \loiii{} is not used for the main results of this paper due to its inhomogeneous nature and uncertainties associated with aperture correction. 

\section{The primary sample}
\label{sec:append:wholesample}

\begin{figure*}
    \centering
    \vbox{
    \hbox{
    \includegraphics[width=3.in]{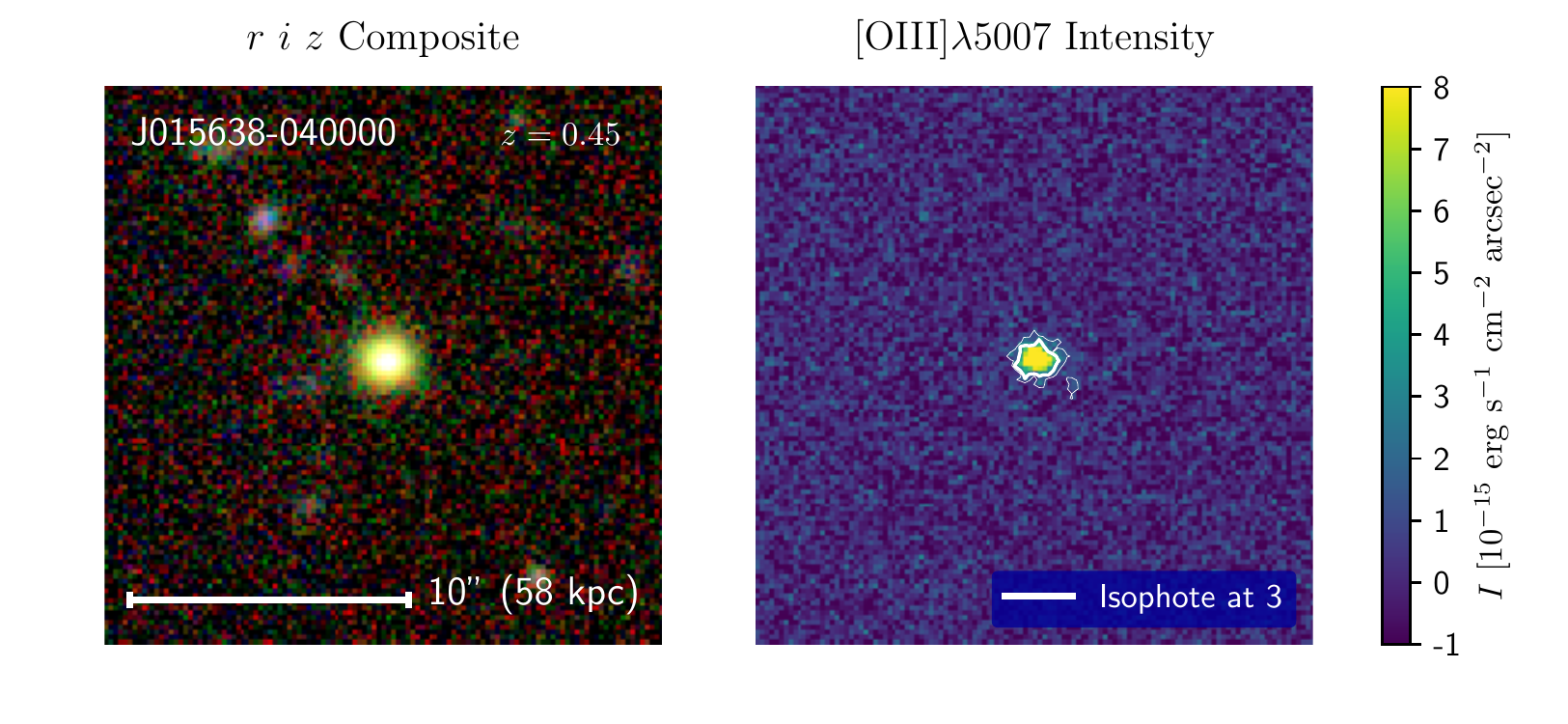}
    \includegraphics[width=3.75in]{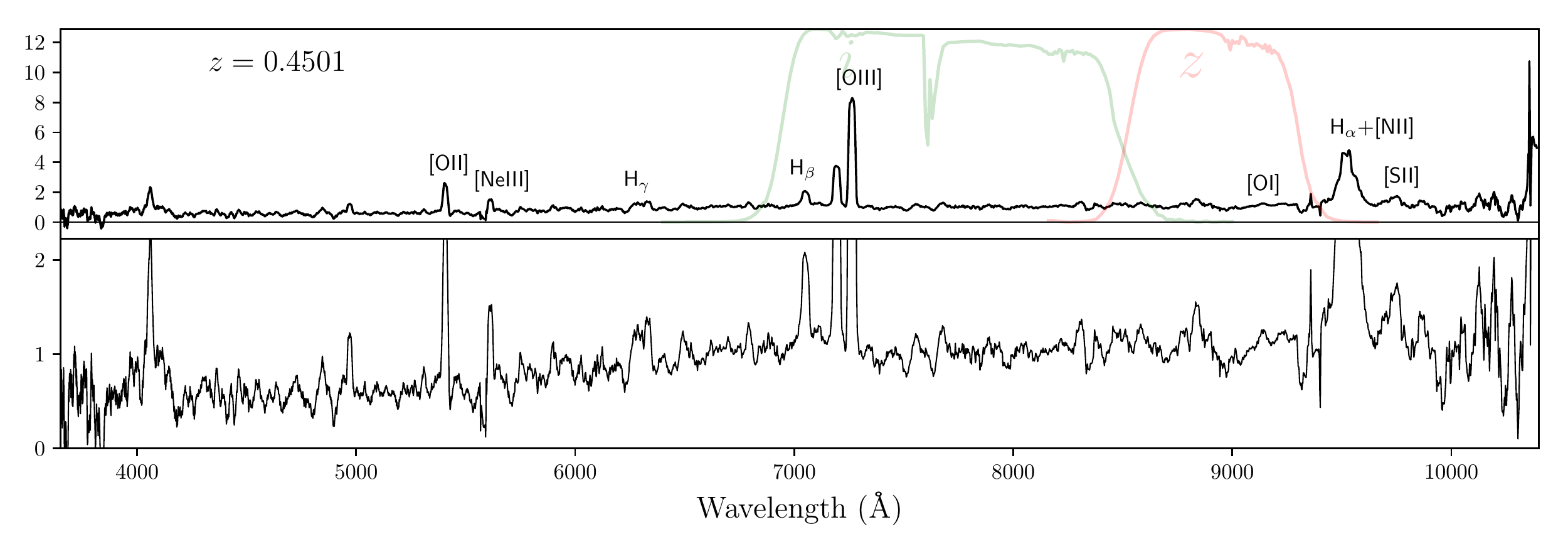}
    }
    \hbox{
    \includegraphics[width=3.in]{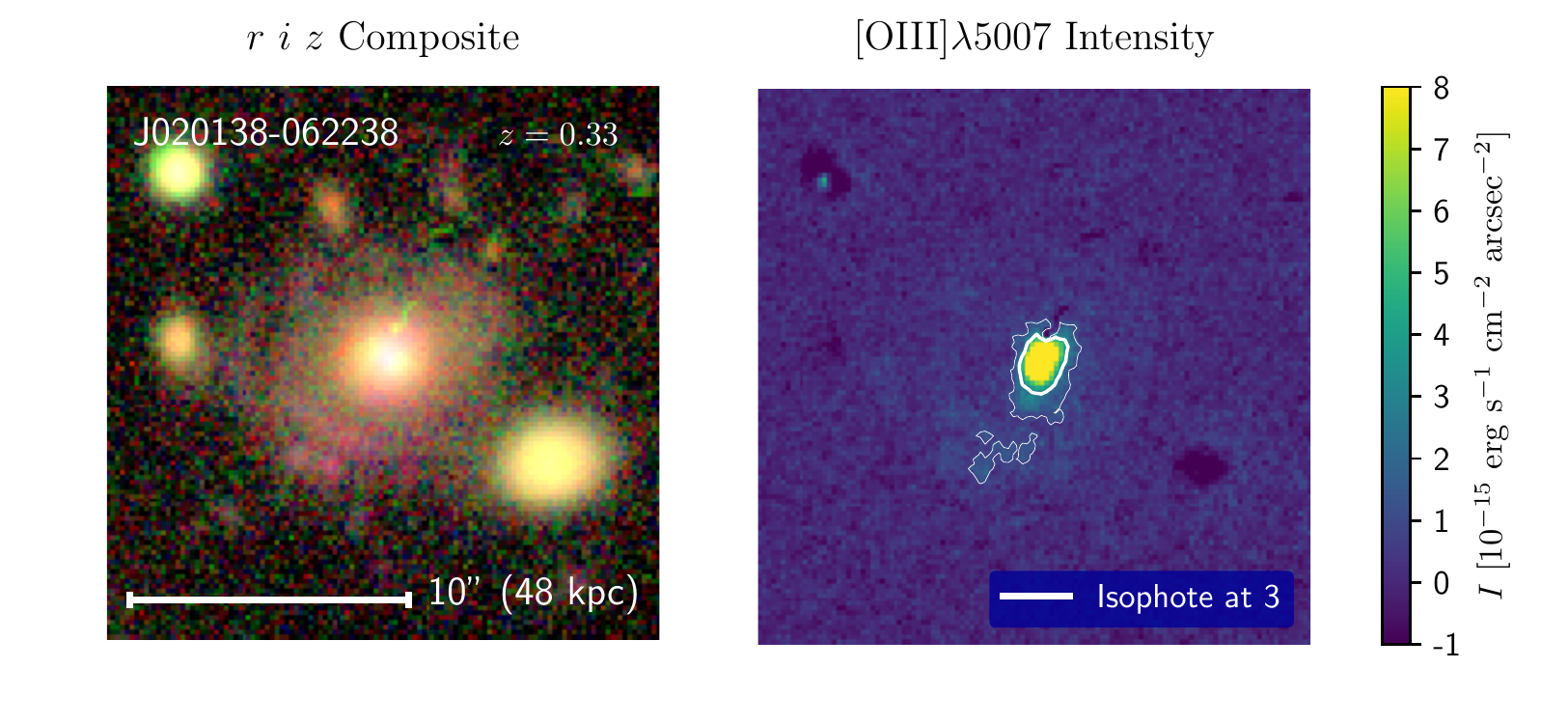}
    \includegraphics[width=3.75in]{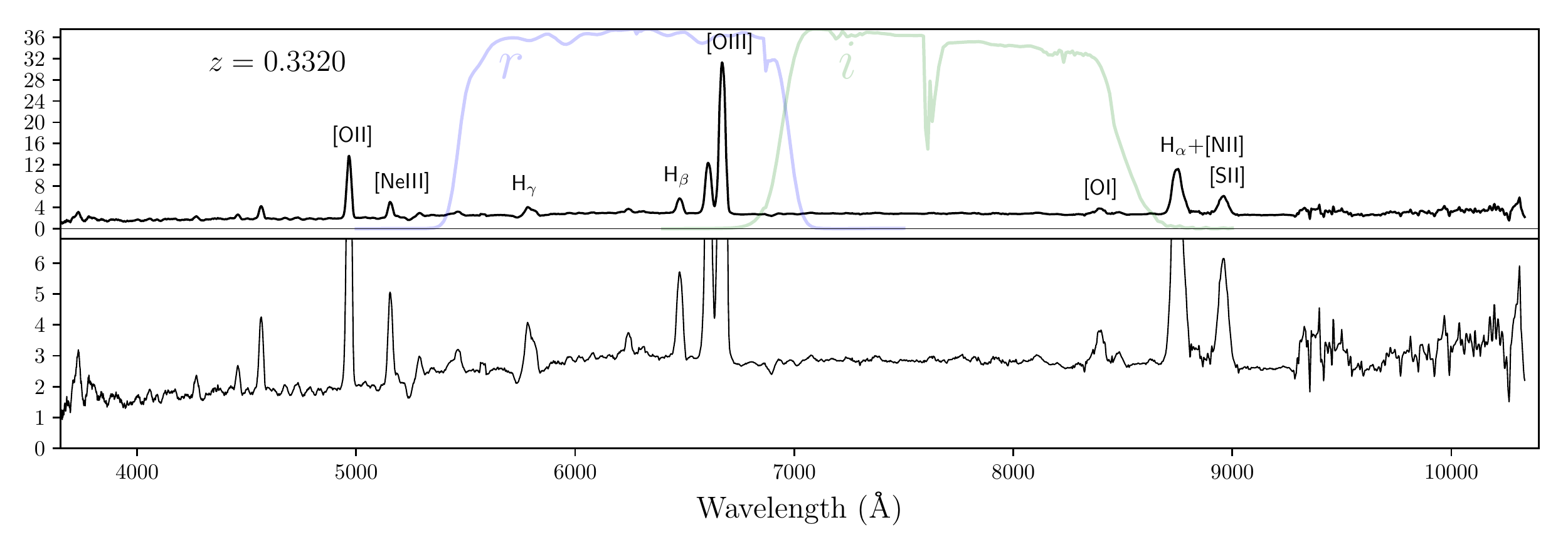}
    }
    \hbox{
    \includegraphics[width=3.in]{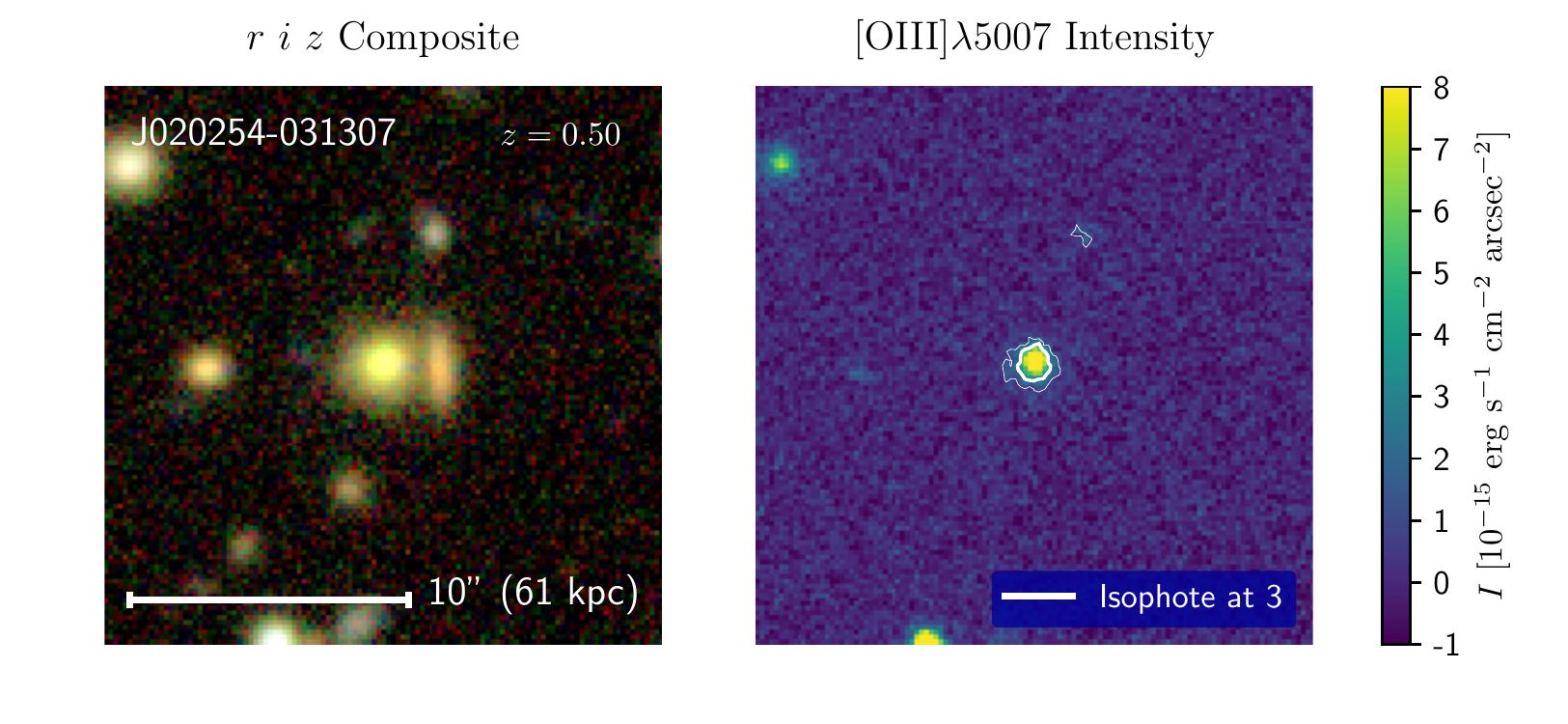}
    \includegraphics[width=3.75in]{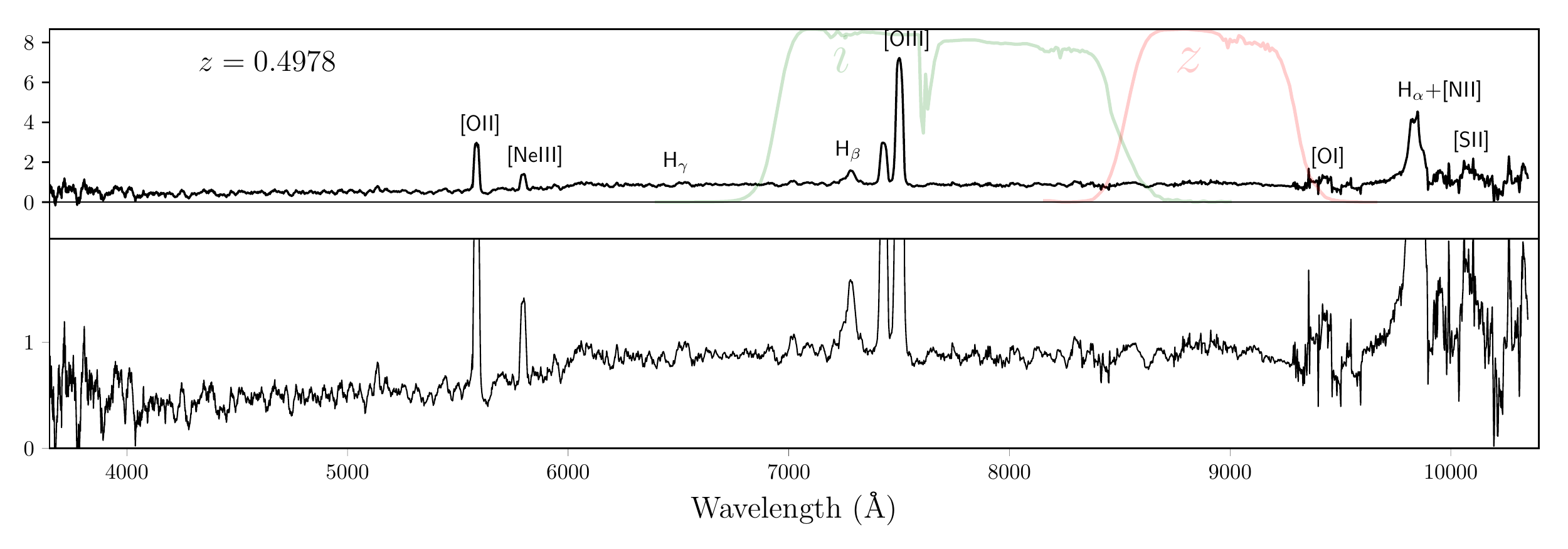}
    }
    \hbox{
    \includegraphics[width=3.in]{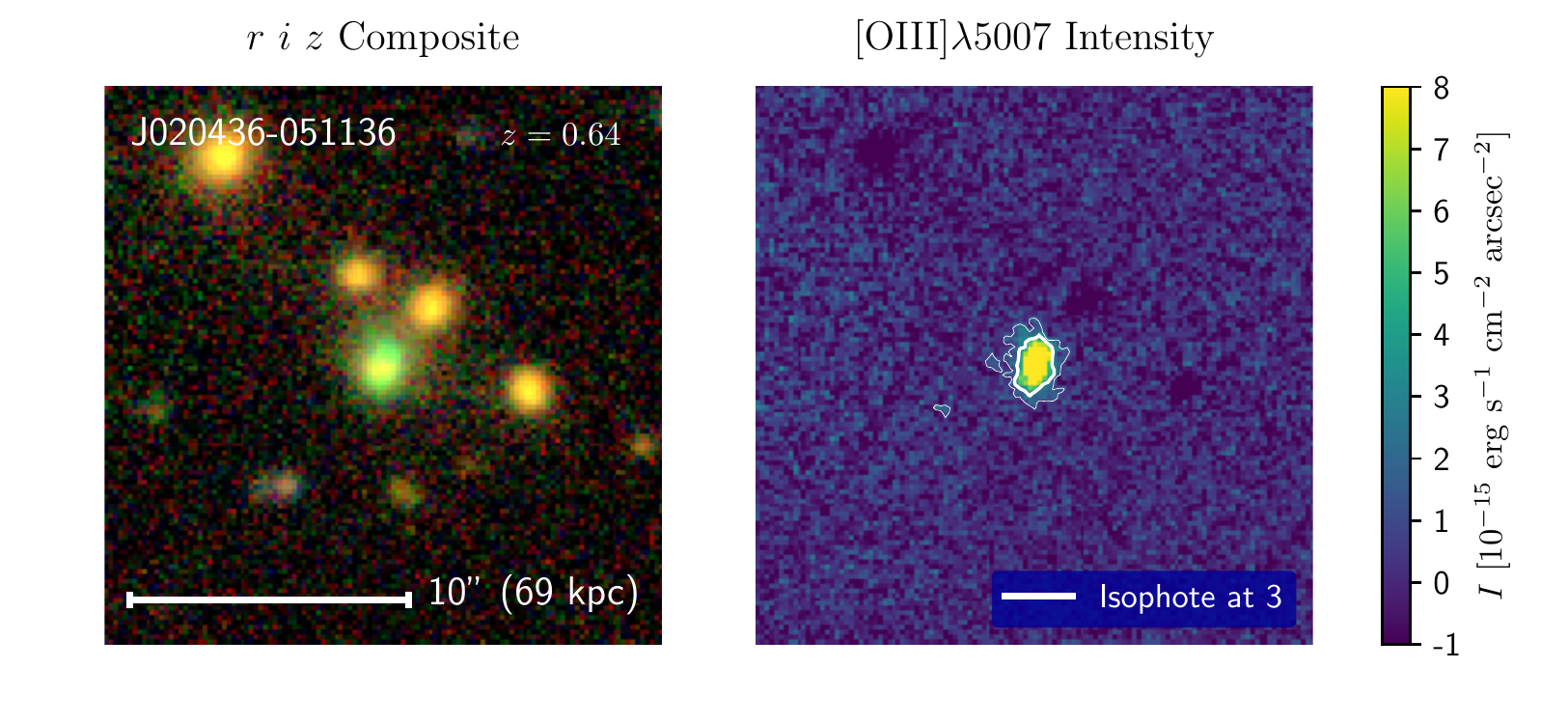}
    \includegraphics[width=3.75in]{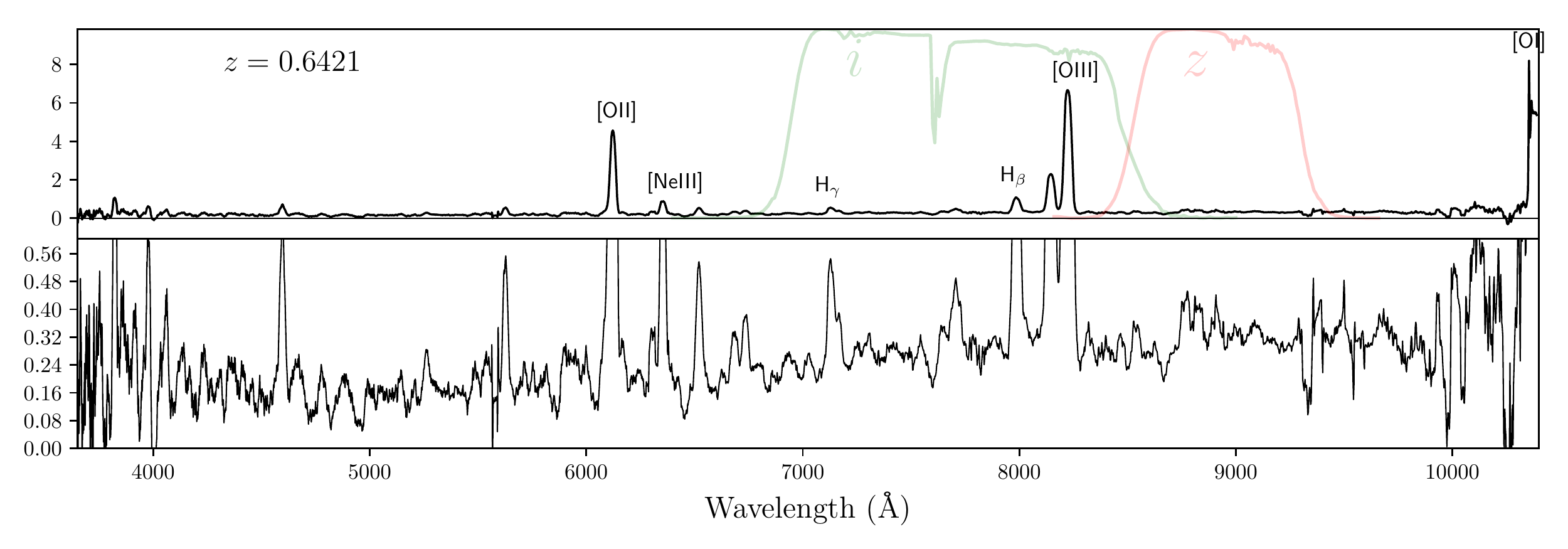}
    }
    \hbox{
    \includegraphics[width=3.in]{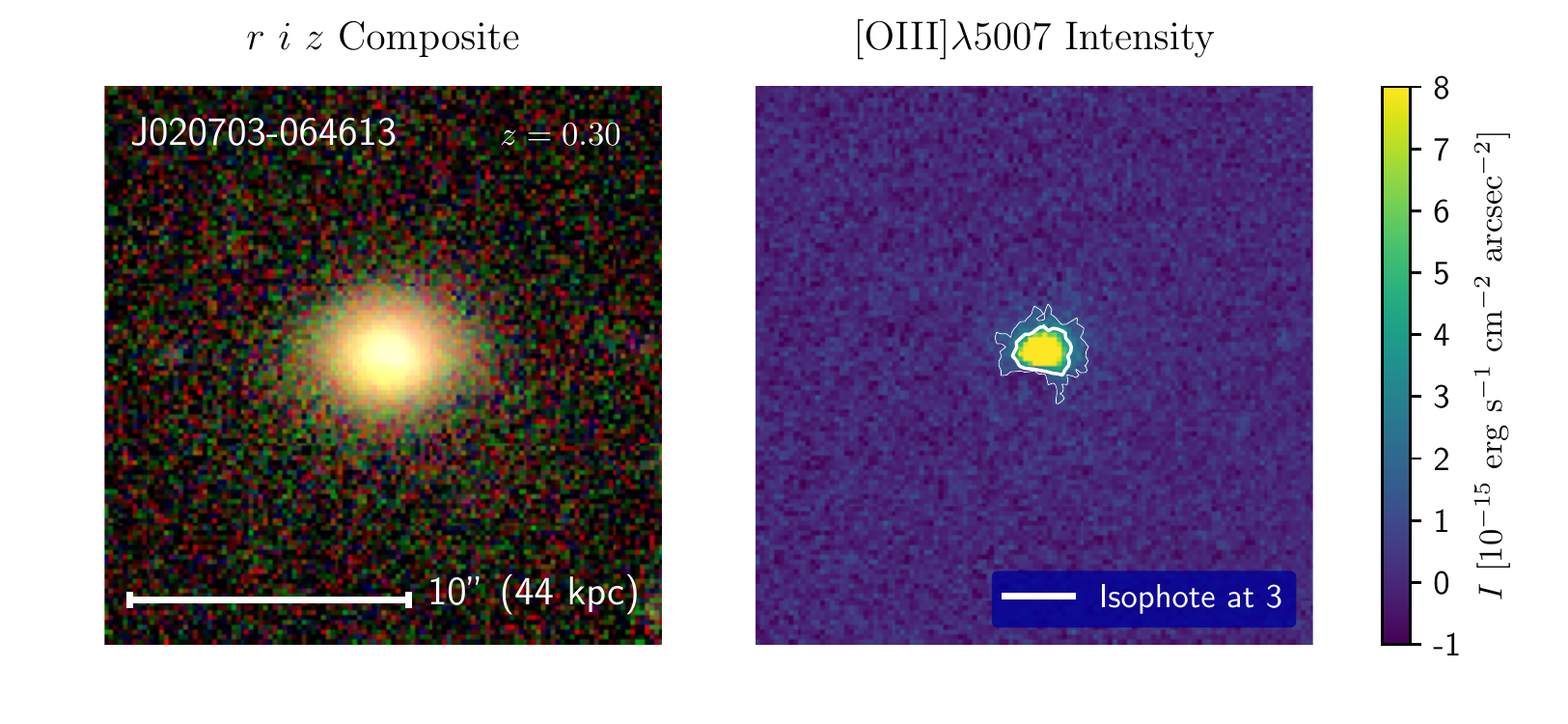}
    \includegraphics[width=3.75in]{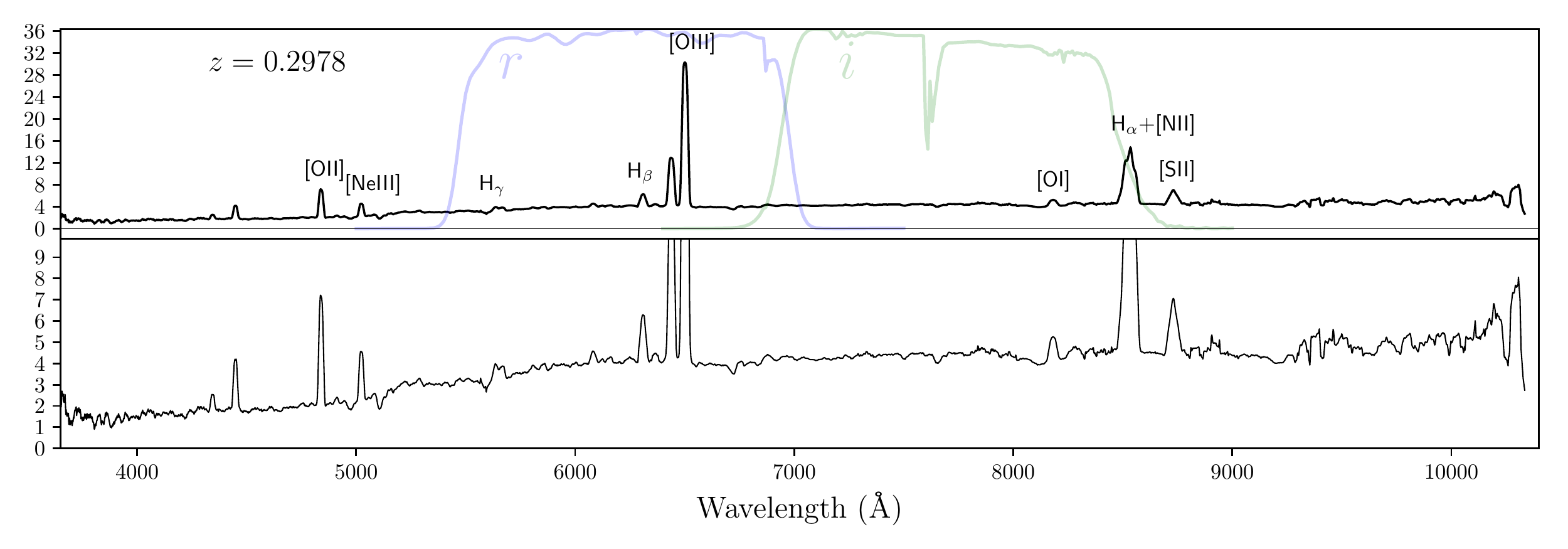}
    }
    \hbox{
    \includegraphics[width=3.in]{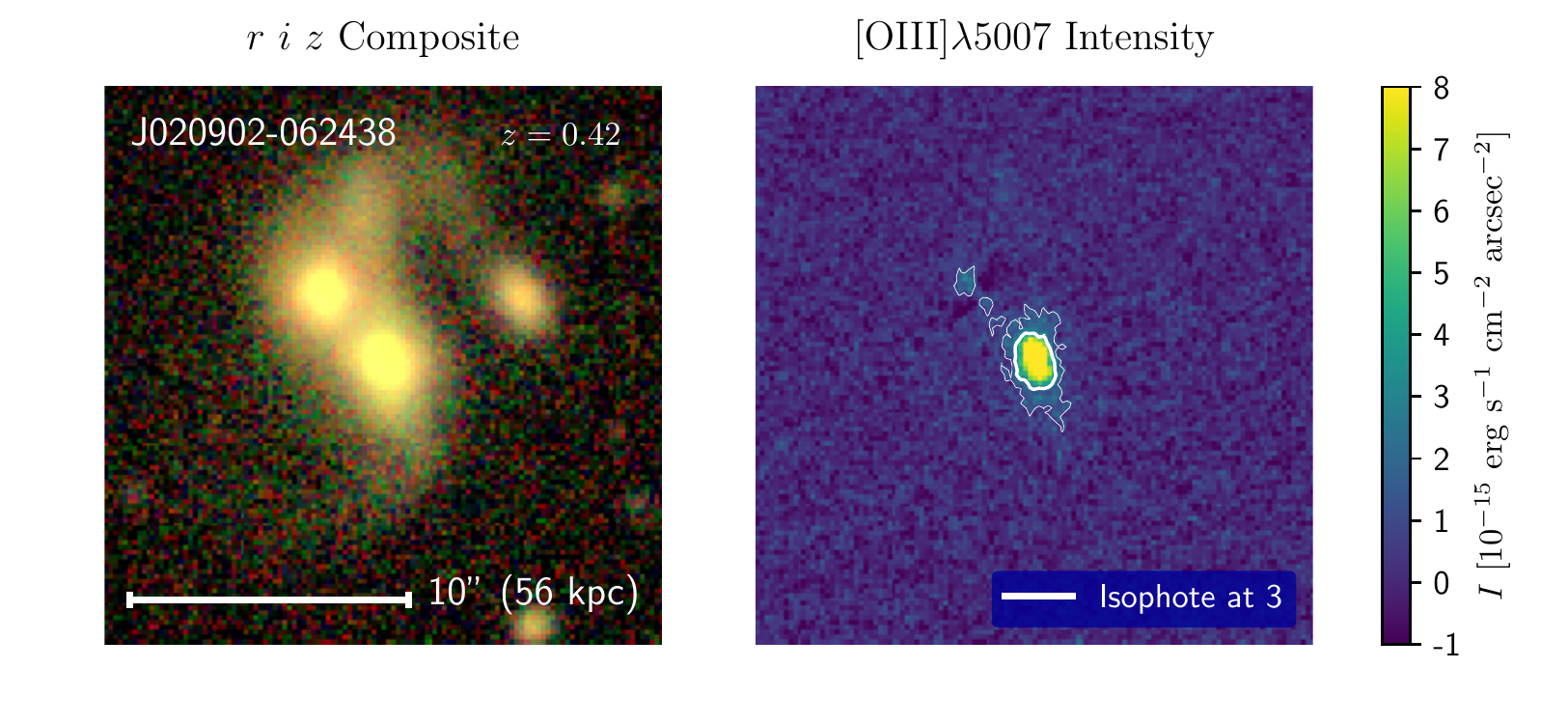}
    \includegraphics[width=3.75in]{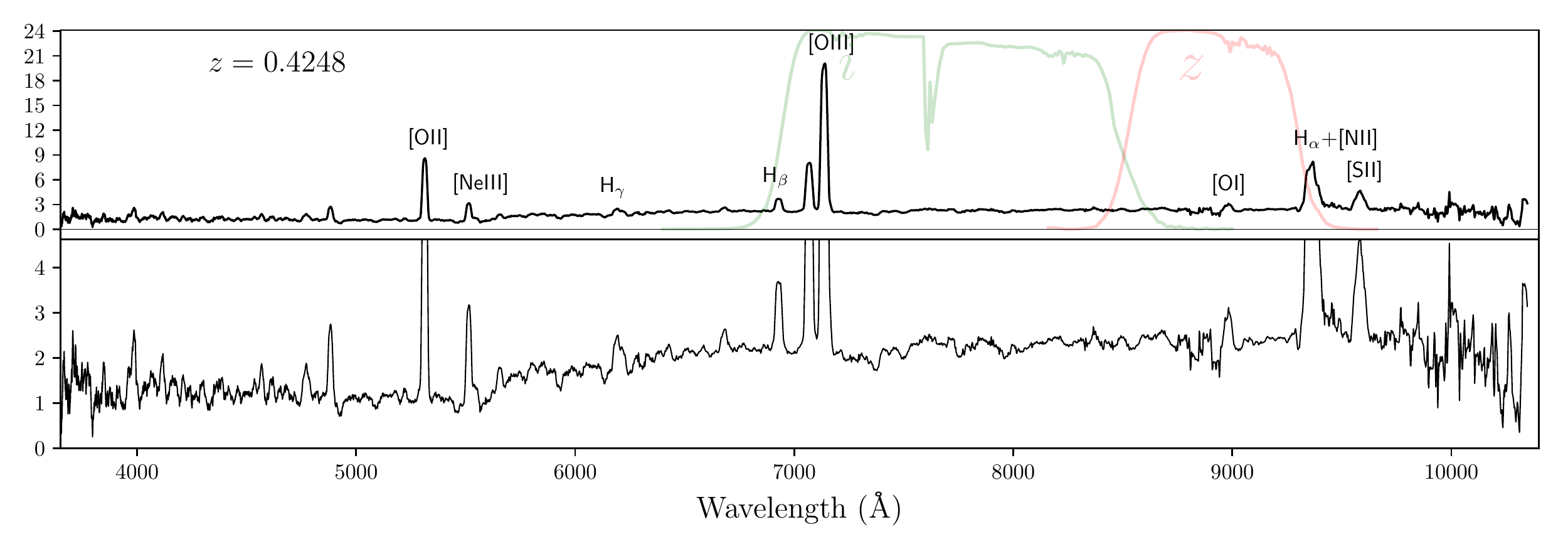}
    }
    }
    \caption{
    The HSC images, \oiii{} maps and the SDSS spectra of the primary sample. The panels are as described in Fig. \ref{fig:map_extcandi}. The full figure will be made available as online data. 
    }
    \label{fig:wholesample}
\end{figure*}

Our primary sample of 300 type 2 AGN is presented in Tab. \ref{tab:wholesample}. Their \oiii{} emission-line maps and spectra are shown in Fig. \ref{fig:wholesample}. The format is as described in Tab. \ref{tab:extcandi} and Fig. \ref{fig:map_extcandi}. The content here is intended to be presented as online data only. 

\begin{table*}
\begin{center}
\caption{The Primary Sample}
\begin{tabular}{cccccccccccc}
\hline
\hline
Name & R.A.  & Dec.  & $z$ & Bands & Source & zmag & ymag & \lbolfifteen{} & \loiii{} & Area & Radius\\
 & (J2000)  & (J2000)  &  &  &  & (mag) & (mag) & (\ergs{}) & (\ergs{}) & (kpc$^2$) & (kpc) \\
(1) & (2) & (3) & (4) & (5) & (6) & (7) & (8) & (9) & (10) & (11) & (12) \\
\hline
J015638$-$040000 & 01:56:38.0 & $-$04:00:00.4 & 0.450 & i, z & Y & $-$21.84  & $-$22.28  & 45.34   & 41.99 & 44 & 5 \\   
J020138$-$062238 & 02:01:38.5 & $-$06:22:38.0 & 0.332 & r, i & Y & $-$22.77  & $-$22.63  & 45.44   & 42.58 & 60 & 7 \\   
J020254$-$031307 & 02:02:54.2 & $-$03:13:07.7 & 0.498 & i, z & Y & $-$22.08  & $-$22.63  & 45.51   & 41.95 & 41 & 4 \\   
J020436$-$051136 & 02:04:36.9 & $-$05:11:36.9 & 0.642 & i, z & Y & $-$21.91  & $-$22.18  & 45.94   & 42.00 & 100 & 7 \\   
J020703$-$064613 & 02:07:03.0 & $-$06:46:13.5 & 0.298 & r, i & Y & $-$22.74  & $-$22.81  & 45.41   & 42.51 & 51 & 7 \\   
J020902$-$062438 & 02:09:02.0 & $-$06:24:38.9 & 0.425 & i, z & Y & $-$22.89  & $-$23.19  & 45.37   & 42.39 & 71 & 6 \\   
J021138$-$052411 & 02:11:38.7 & $-$05:24:11.7 & 0.528 & i, z & Y & $-$22.95  & $-$23.24  & 45.46   & 42.15 & 66 & 6 \\   
J021355$-$054941 & 02:13:55.4 & $-$05:49:41.3 & 0.443 & i, z & Y & $-$22.78  & $-$23.11  & 46.23   & 42.36 & 68 & 6 \\   
\hline
\end{tabular} 
\textit{Note.}
The format is as described in \ref{tab:extcandi}. The full table will be made available as online data.
\label{tab:wholesample}
\end{center}
\end{table*}

\bsp	
\label{lastpage}
\end{document}